\documentclass[lettersize,journal]{IEEEtran}
\usepackage{times}
\usepackage{hyperref}
\hypersetup{colorlinks,urlcolor=[rgb]{0.0,0.4,0.7}}
\usepackage{epsfig}
\usepackage{graphicx}
\usepackage{amsmath}
\usepackage{amssymb}
\usepackage{multicol,multirow}
\usepackage[tableposition=top,font=small]{caption}
\usepackage{cite}
\usepackage{booktabs}
\usepackage{bigstrut}
\usepackage{floatrow}
\usepackage{rotating}
\usepackage{adjustbox}
\usepackage{amsmath,url}

\def\BibTeX{{\rm B\kern-.05em{\sc i\kern-.025em b}\kern-.08em T\kern-.1667em\lower.7ex\hbox{E}\kern-.125emX}}

\begin{document}

\title{Leveraging Image Complexity in\\ Macro-Level Neural Network Design for\\ Medical Image Segmentation}

\author{Tariq M. Khan, \IEEEmembership{Member, IEEE}, Syed S. Naqvi, \IEEEmembership{Member, IEEE}, Erik Meijering, \IEEEmembership{Fellow, IEEE}
\thanks{Tariq M. Khan and Erik Meijering are with the School of Computer Science and Engineering, University of New South Wales, Sydney, NSW, Australia (email: tariq.khan@unsw.edu.au, erik.meijering@unsw.edu.au)}
\thanks{Syed S. Naqvi is with the Department of Electrical and Computer Engineering, COMSATS University Islamabad, Islamabad, Pakistan (email: saud\_naqvi@comsats.edu.pk)}}

\maketitle

\begin{abstract}
Recent progress in encoder-decoder neural network architecture design has led to significant performance improvements in a wide range of medical image segmentation tasks. However, state-of-the-art networks for a given task may be too computationally demanding to run on affordable hardware, and thus users often resort to practical workarounds by modifying various macro-level design aspects. Two common examples are downsampling of the input images and reducing the network depth to meet computer memory constraints. In this paper we investigate the effects of these changes on segmentation performance and show that image complexity can be used as a guideline in choosing what is best for a given dataset. We consider four statistical measures to quantify image complexity and evaluate their suitability on ten different public datasets. For the purpose of our experiments we also propose two new encoder-decoder architectures representing shallow and deep networks that are more memory efficient than currently popular networks. Our results suggest that median frequency is the best complexity measure in deciding about an acceptable input downsampling factor and network depth. For high-complexity datasets, a shallow network running on the original images may yield better segmentation results than a deep network running on downsampled images, whereas the opposite may be the case for low-complexity images.
\end{abstract}


\section{Introduction}
\label{sec:introduction}

\IEEEPARstart{M}{edical} image segmentation aims to delineate organs or lesions in images from computed tomography (CT), magnetic resonance imaging (MRI), optical imaging, and other medical imaging modalities, and serves as a basis for subsequent quantitative image analysis in a wide range of clinical and research applications. It is one of the most difficult tasks in medical image analysis, as it provides critical information about organ shapes and volumes, and medical images can be quite complex \cite{Hesamian2019, Tajbakhsh-2020, Liu-2021d, Fu-2021}. The challenges of obtaining a clinically applicable segmentation are multifaceted, including diverse segmentation tasks, different modalities, multiple resolutions, and varying anatomical characteristics such as shape, size, location, deformity, and texture. Recent progress in encoder-decoder architectures such as U-Net \cite{Ronneberger2015, Du-2020, Siddique-2021, Isensee-2021} has improved segmentation performance on many benchmarks. However, designing such networks requires significant effort in choosing the right network configuration.

The size of medical imaging data sets is constantly increasing \cite{Garca2017DownsamplingMF} and often it is not possible to train deep neural network architectures on a single mid-range graphics processing unit (GPU) at the native image resolution. As a result, the images are typically downsampled before training, which may cause loss or alteration of fine details that are potentially important for diagnosis. Also, in benchmarking studies, downsampling is sometimes used for both training and testing of medical image segmentation methods \cite{jcm9030871, jcm8091446}, and thus the results may not be fully representative of performance on the native images. Alternatively, shallow networks are often proposed \cite{9207668, Howard_2019_ICCV, Ma_2018_ECCV}, in an attempt to trade off image size and network size to allow training on limited computing hardware. Another common practice is iterative downsampling until training of a deeper network of choice becomes feasible on given hardware. While these approaches are understandable from a practical standpoint, we argue that the optimal choice of input size and network depth is inherently dependent upon the characteristics of the data and the segmentation task.

Recent methods in medical image segmentation adopt neural architecture search (NAS) \cite{zhu2019VNAS, kim2019scalable, weng2019NASUnet, yang2020rl, yu2020C2FNAS, he2021dints} to determine the best suitable network architecture for the task at hand. However, a computationally expensive search has to be performed for each new data set and task, and the resulting architecture may not generalize well to other data sets and tasks. Here again, the importance of the information content of the data is often ignored. We argue that we need to take a step back and base the macro-level design choices of neural networks, such as the amount of downsampling or the depth of the network, on the information complexity of the data.

Our objective in this work is to employ measures of image complexity to guide macro-level neural network design for medical image segmentation. We focus specifically on balancing input image downsampling and network depth for optimal segmentation results. To this end, we consider four statistical complexity measures: delentropy \cite{DBLP:journals/corr/Larkin16}, mean frequency \cite{10.1007/978-3-662-46578-3_81}, median frequency \cite{10.1007/978-3-662-46578-3_81}, and perimetric complexity \cite{Attneave1956}. Delentropy and perimetric complexity have been used previously as measures of data complexity in autonomous driving \cite{Ameet2020measures} and binary pattern recognition \cite{Attneave1956}, respectively, while mean and median frequency have been used in electromyography signal identification \cite{10.1007/978-3-662-46578-3_81}. Here, these measures are leveraged for the first time as complexity measures for predicting a suitable input image downsampling factor and choosing a shallow or deep neural network. We propose two new encoder-decoder architectures to represent shallow and deep networks. To find the best complexity measure, we use several data fitting models, including linear and polynomial fitting such as linear regression $\text{R}^2$, adjusted $\text{R}^2$, root mean square error (RMSE), mean absolute error (MAE), Akaike information criterion (AIC), and corrected AIC.

To demonstrate the efficacy and wide applicability of image complexity analysis for neural network based medical image segmentation, we present experiments on 10 different data sets from public challenges. The results confirm that the proposed complexity measures can indeed aid in making the said macro-level design choices and that median frequency is the best measure for this purpose. More specifically, the results show that input image size is important for data sets with high complexity and downsampling negatively affects segmentation performance in such cases, whereas downsampling does not significantly affect performance for data sets having low complexity. Also, in the case of high-complexity data sets and computational constraints, a shallow network taking the original images as input is to be preferred, whereas for low-complexity cases competitive performance with the same computational constraints is achievable by using downsampling and a deep network topology.

\section{Complexity Measures}
It has long been known that data complexity measures can be used to determine the intrinsic difficulty of a classification task on a given dataset \cite{10.5555/1177240}. In this study we consider four important complexity measures and investigate their suitability for medical image segmentation tasks.

\subsection{Delentropy}
The standard Shannon entropy of a gray-scale image is defined as \cite{DBLP:journals/corr/Larkin16}:\vspace{-0.5\baselineskip}
\begin{equation}
	H =  - \sum\limits_{i = 0}^{N - 1} {{p_i}} \log {p_i},
\end{equation}
where $N$ is the number of gray levels and $p_i$ is the probability of a pixel having gray level $i$. Delentropy (DE) is computed similarly, but using a probability density function known as deldensity \cite{DBLP:journals/corr/Larkin16}. Unlike Shannon entropy, which is entirely computed from individual pixel values, DE captures the underlying spatial image structure and pixel co-occurrence via the deldensity, which is based on gradient vectors.

\subsection{Mean Frequency}
The mean frequency (MNF) of an image is computed as the sum of the product of the power spectrum and frequency divided by the total sum of the power spectrum \cite{10.1007/978-3-662-46578-3_81}:
\begin{equation}
	\text{MNF} = \frac{\sum\limits_{i=1}^{M} f_i P_i}{\sum\limits_{i=1}^{M} P_i},
\end{equation}
where $P_i$ is the value of the power spectrum at frequency bin $i$, $f_i$ is the actual frequency of that bin, and $M$ is the total number of frequency bins. MNF can be considered as the frequency centroid or the spectral center of gravity and is also called the mean power frequency and mean spectral frequency in several works \cite{10.1007/978-3-662-46578-3_81}.

\subsection{Median Frequency}
The median frequency (MDF) of an image is the frequency at which the power spectrum of the image is divided into two regions with equal integrated power \cite{10.1007/978-3-662-46578-3_81}. In other words, at the MDF the following equality holds:
\begin{equation}
	\sum\limits_{i=1}^{\text{MDF}} P_i = \!\!
	\sum\limits_{i=\text{MDF}}^{M}\!\!\! P_i =
	\frac{1}{2}\sum\limits_{i=1}^{M} P_i.
\end{equation}

\subsection{Perimetric Complexity}
The perimetric complexity (PC) is a measure of the complexity of binary images. The general concept goes back to the early days of vision research \cite{Attneave1956} where this measure, originally called dispersion, was used to describe the perceptual complexity of visual shapes. It is defined as:
\begin{equation}
	\text{PC} = \frac{P^2}{4\pi A},
\end{equation}
where $P$ represents the perimeter of the foreground and $A$ is the foreground area. In our study this measure is computed from the annotation masks of gray-scale images.

\section{Segmentation Networks}\label{networks}
To investigate the interplay between image complexity, input downsampling, and network depth, we designed two baseline encoder-decoder networks: a light-weight (shallow) network (LW-Net) consisting of only 45 layers, and a deep (large-scale) network (LS-Net) with 10 times more layers. The reason for designing new architectures instead of using popular networks for medical image segmentation such as the U-Net variants \cite{IBTEHAZ202074,Zongwei2018} and neural architecture search approaches \cite{weng2019NASUnet,yu2020C2FNAS,he2021dints} is computational constraints. The prevalent architectures in this domain are resource hungry and training such networks on typical hardware is not always feasible. On the contrary, the proposed LW-Net and LS-Net architectures have only around 0.1 and 2 million parameters, respectively, making them 300 and 15 times smaller than U-Net \cite{Ronneberger2015}, and LW-Net is 8 times smaller than its competitor NAS based approach \cite{weng2019NASUnet}.

\begin{figure*}
	\centering
	\vspace{-1cm}
	\includegraphics[width=0.63\textwidth]{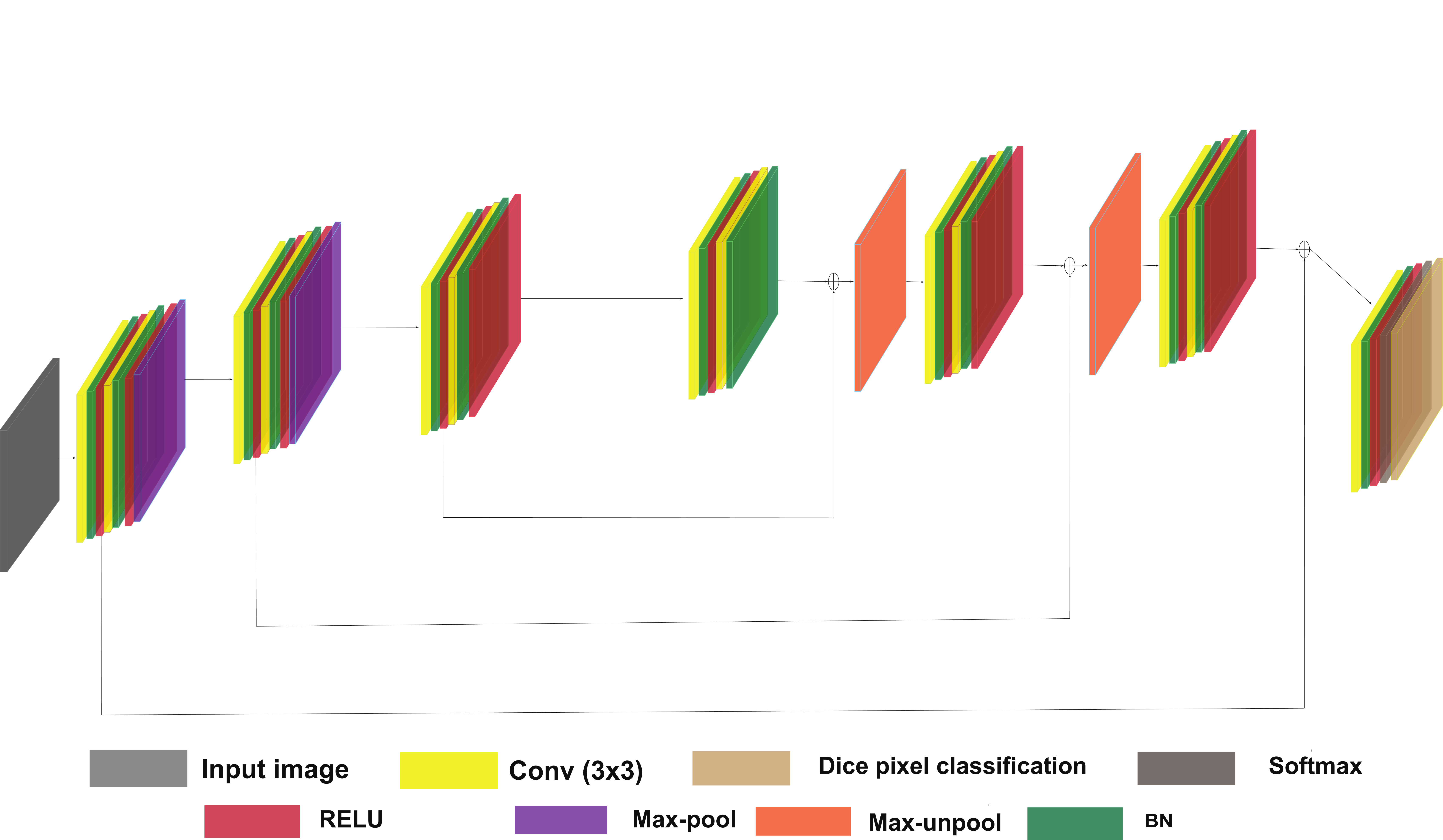} \\
	\caption{Block diagram of the proposed light-weight network LW-Net.}
	\label{LW-Net}
\end{figure*}

\begin{figure*}
	\centering
	\vspace{-1cm}
	\includegraphics[width=0.63\textwidth]{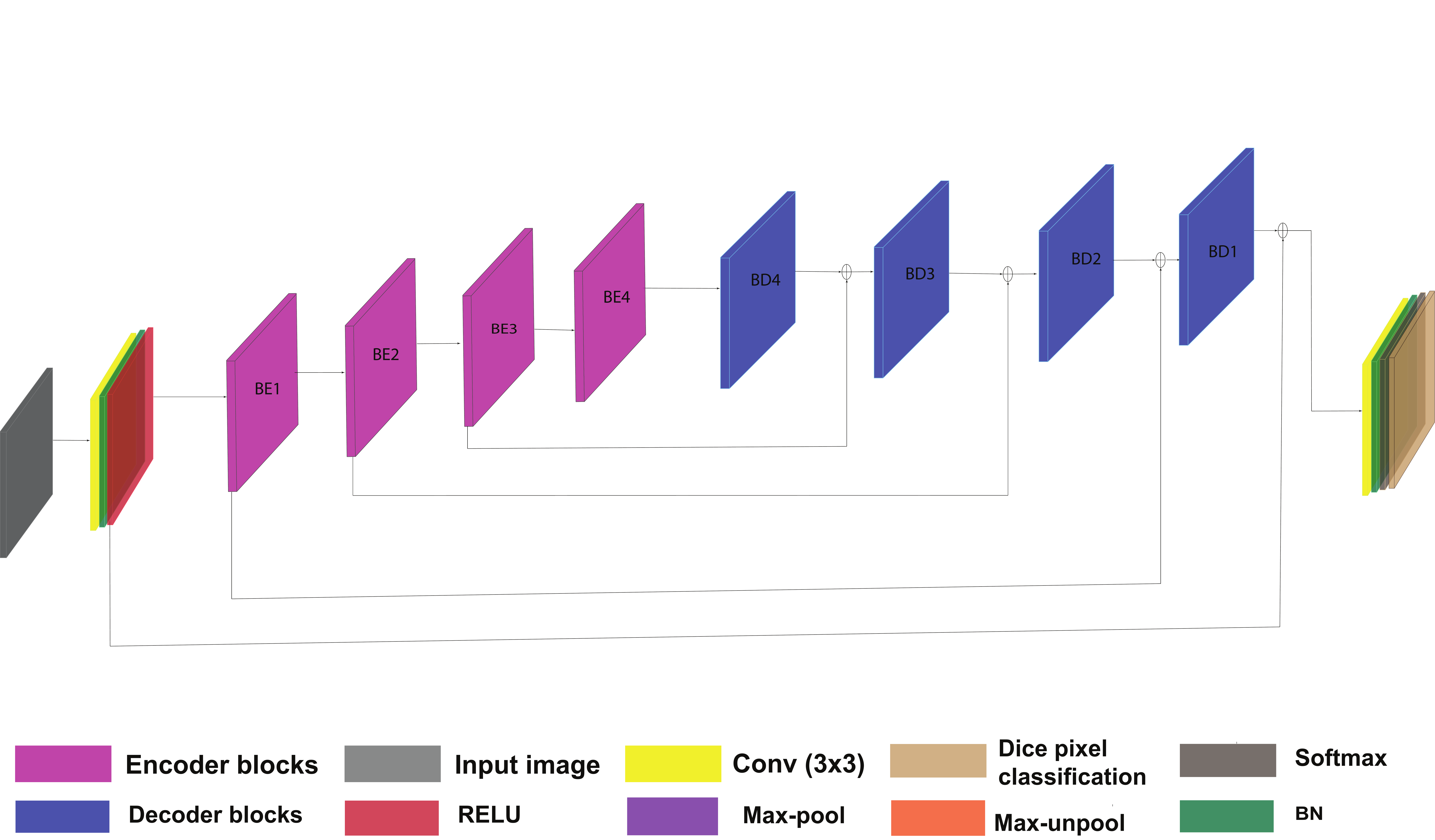} \\
	\caption{Block diagram of the proposed large-scale network LS-Net.}
	\label{LS-Net}
\end{figure*}

\subsection{Shallow Network}
LW-Net has relatively few encoding-decoding layers with few filters in each, which allows it to be trained on low-range hardware, work with larger images, and retain important high-resolution image information. It consists of six convolutional blocks (Fig.\,\ref{LW-Net}). The first/input block is followed by two downsampling (encoder) convolutional blocks, an intermediate convolutional block used as a bridge between the down- and upsampling blocks, and two upsampling (decoder) convolutional blocks, followed by the final convolutional output block which produces the segmentation map. Each block performs (a series of) convolutions of its input feature maps with its filter banks, and the output undergoes batch normalisation (BN), followed by activation using a rectified linear unit (ReLU). The convolution filters are all $3\times3$ in size. The resulting feature maps are then passed on to a max-pooling or unpooling layer, depending on whether the block is in the encoder or decoder part. The network has two max-pooling and two unpooling layers, all of which are of type 2$\times$2, non-overlapping, and use a stride of 2.

\subsection{Deep Network}
\begin{figure}
	\centering
	\includegraphics[width=0.80\textwidth]{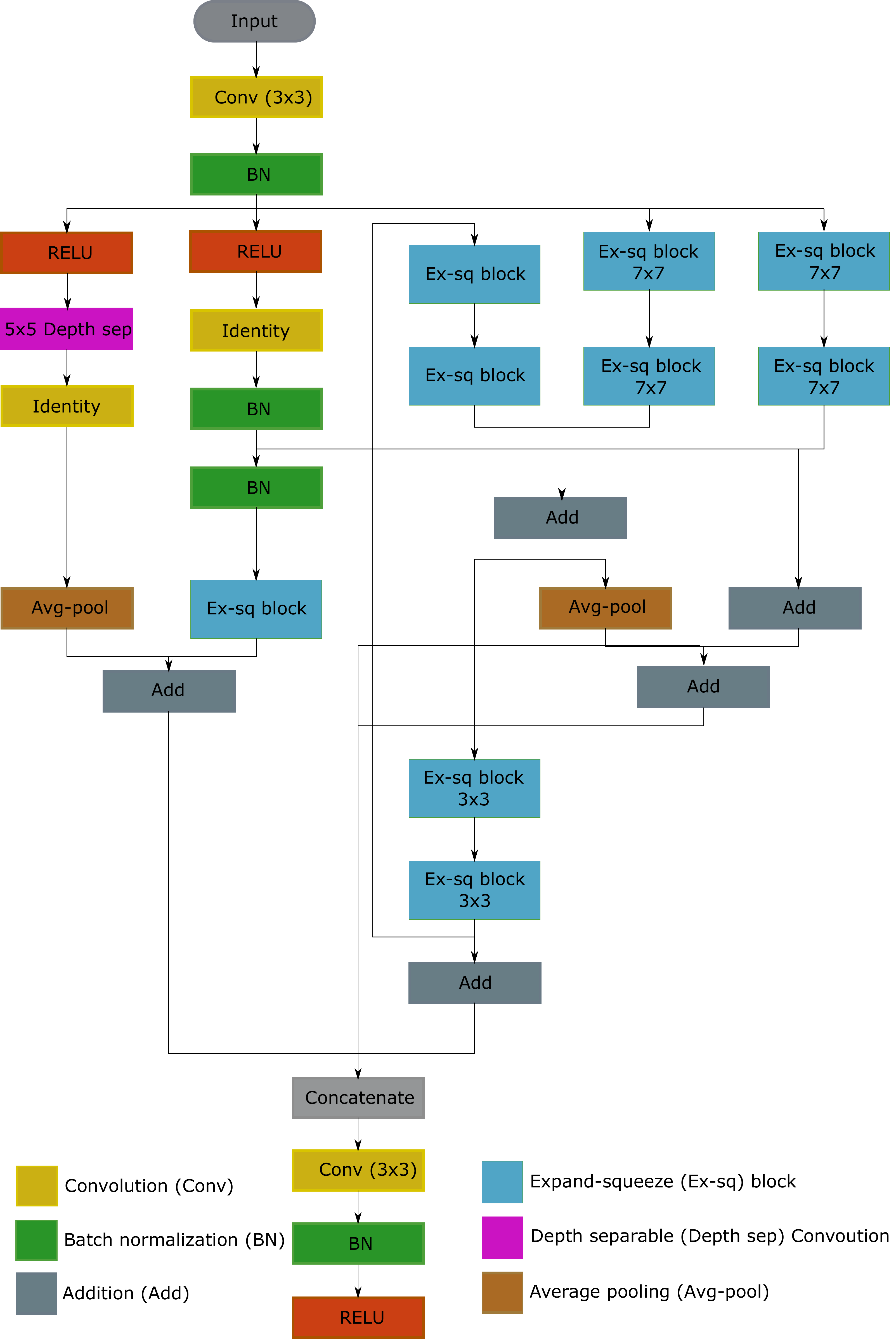}
	\caption{Architecture of the encoder block of the proposed LS-Net.}
	\label{encoderBlockLSNet}
\end{figure}

\begin{figure}
  \centering
  \includegraphics[width=0.4\textwidth]{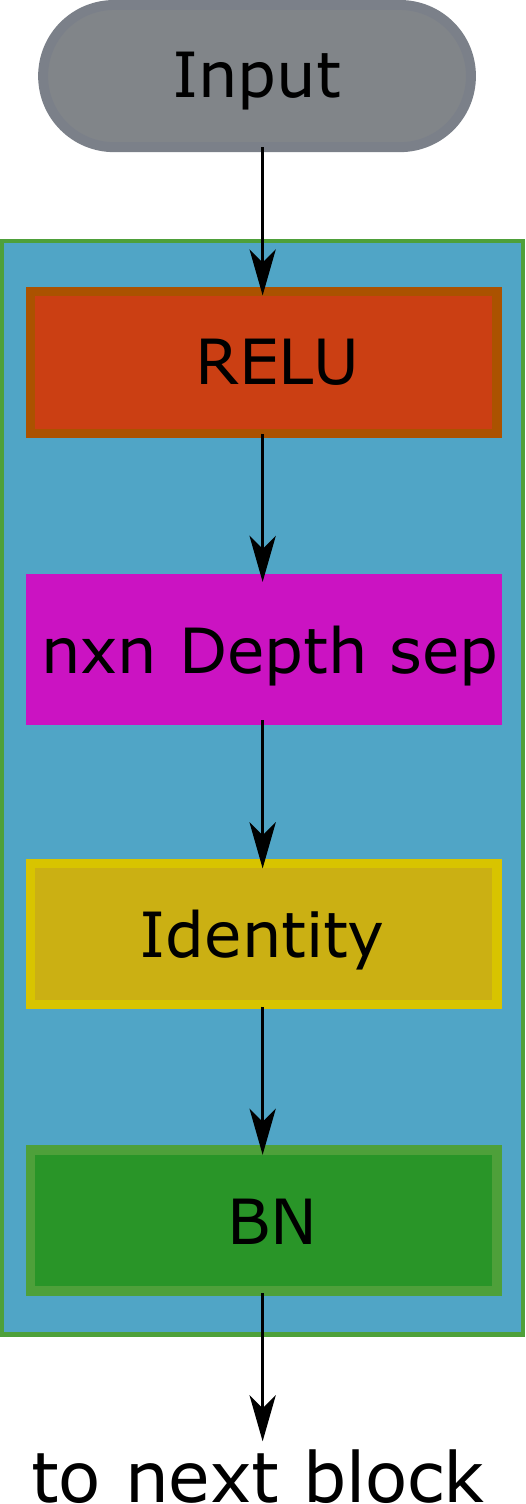}
  \caption{Architecture of the expand-squeeze module used in LS-Net. Depth-wise separable convolutions can be of size $n=3,5,7$.}
  \label{internalArch}
\end{figure}

LS-Net is based on the compound scaling principle \cite{tan2019efficientnet}, is 464 layers deep and employs a large number of branches along with depth-wise separable convolutions to make it wider. It is designed to extract rich feature information at every encoding-decoding stage while still being usable on low- to mid-range hardware. The network consists of four encoder and corresponding decoder blocks (Fig.\,\ref{LS-Net}). To capture the multiscale information in medical images, the encoder blocks (Fig.\,\ref{encoderBlockLSNet}) use filters of 3$\times$3, 5$\times$5, and 7$\times$7, along with average pooling to downsample the feature maps. To ensure that essential information is retained, the amount of downsampling in the encoding blocks is limited. The decoder blocks have the same architecture as the encoder blocks (Fig.\,\ref{encoderBlockLSNet}), except that the downsampling (average pooling) operation is replaced by bilinear upsampling. A pool size of 3$\times$3 with stride 1 is employed throughout the network. An important element in our design is the expand-squeeze module (Fig.\,\ref{internalArch}). Inspired by the existing Fire module \cite{iandola2016squeezenet}, we employ depth-wise separable convolution for the expansion operation and an identity layer for the squeeze operation. In contrast to SqueezeNet \cite{iandola2016squeezenet}, we use an expand-followed-by-squeeze strategy. We argue that the expansion operation (separable convolution) allows rich features with multiscale information to be available to the bottleneck layer that extracts all the necessary information. This enables the building block (expand-squeeze module) to simultaneously capture the completeness and the expressiveness of the features.

\section{Experimental Results}
Two experiments were performed to test the hypothesis that image complexity can and should be taken into account in making macro-level neural network design choices for medical image segmentation. In the following sections we present the network training approach, the used public datasets, segmentation performance metrics, regression analysis performance metrics, and the results of the two experiments.

\subsection{Network Training}
All experiments were carried out on an Intel(R) Core(TM) i7-8700 CPU with 64 GB RAM and a relatively low/mid-range GeForce GTX1080Ti GPU. Network training was done with adaptive moment estimation (Adam) and a fixed learning rate of 1e-3. After initial experimentation, the maximum number of epochs was set to 15 with a batch size of 8 to match the hardware constraints. Gradient clipping was employed based on the global $l_2$-norm with a gradient threshold of 3 \cite{Razvan2013jmlr}. Weighted cross-entropy loss was used as the objective function for training all models in our experiments. For the calculation of the class association weights in the loss, we used median frequency balancing \cite{Badrinarayanan2017}.

\subsection{Public Datasets}
We used 10 publicly available datasets (Table \ref{tab:datasets}) representing a range of image complexities (Table \ref{tab:downsamplingSegmentation}).

\subsubsection{STARE} The STARE (Structured Analysis of the Retina) dataset \cite{Hoover2000} consists of 20 color retinal fundus images acquired with a field of view (FOV) of $35^\circ$ and size 700$\times$605 pixels. There are various pathologies in 10 of the 20 images. For each of the 20 images, two expert manual segmentation maps are available of the retinal blood vessels, and we used the first of these as the ground truth.

\subsubsection{DRIVE} The DRIVE (Digital Retinal Images for Vessel Extraction) dataset \cite{Staal2004} is from a diabetic retinopathy screening program. It contains 20 color images for training and 20 for testing with a size of 584$\times$565 pixels and covering a wide age range of diabetic patients. Seven of the 40 images show small signs of mild early diabetic retinopathy. For each of the 40 images, an expert manual segmentation mask is available for use as ground truth.

\begin{table*}[!t]
	\centering
	\caption{Public datasets used in the experiments.}
	\resizebox{1\textwidth}{!}{
		\begin{tabular}{@{}ccccccc@{}}
		\toprule
		\textbf{Dataset} & \textbf{Organ} & \textbf{Number of Images} & \textbf{Image Size (Pixels)} & \textbf{Training } & \textbf{Testing} & \textbf{URL} \\
		\midrule
		STARE & Vessel & 20 & 700$\times$605 & 10 & 10 & \href{https://cecas.clemson.edu/\string~ahoover/stare/}{https://cecas.clemson.edu/\string~ahoover/stare/} \\
		DRIVE & Vessel & 40 & 584$\times$565 & 20 & 20 & \href{https://drive.grand-challenge.org/}{https://drive.grand-challenge.org/} \\
		CHASE-DB1 & Vessel & 28 & 999$\times$960 & 28 & 8 & \href{https://blogs.kingston.ac.uk/retinal/chasedb1/}{https://blogs.kingston.ac.uk/retinal/chasedb1/} \\
		MC & Chest & 138 & 4020$\times$4892, 4892$\times$4020 & 100 & 38 & \href{https://lhncbc.nlm.nih.gov/LHC-publications/pubs/TuberculosisChestXrayImageDataSets.html}{https://lhncbc.nlm.nih.gov/LHC-publications/pubs/TuberculosisChestXrayImageDataSets.html} \\
		PH2 & Skin & 200 & 768$\times$560 & ISIC-2016 & 200 & \href{https://www.fc.up.pt/addi/ph2 database.html}{https://www.fc.up.pt/addi/ph2 database.html} \\
		ISIC-2016 & Skin & 900 & 576-4288$\times$542-2848 & 900 & PH2 & \href{https://challenge.isic-archive.com/landing/2016}{https://challenge.isic-archive.com/landing/2016} \\
		DRISHTI-OC & Optic Cup & 101 & 2896$\times$1944 & 50 & 51 & \href{https://cvit.iiit.ac.in/projects/mip/drishti-gs/mip-dataset2/Dataset.php}{https://cvit.iiit.ac.in/projects/mip/drishti-gs/mip-dataset2/Dataset.php} \\
		DRISHTI-OD & Optic Disc & 101 & 2896$\times$1944 & 50 & 51 & \href{https://cvit.iiit.ac.in/projects/mip/drishti-gs/mip-dataset2/Dataset.php}{https://cvit.iiit.ac.in/projects/mip/drishti-gs/mip-dataset2/Dataset.php} \\
		PROMISE12 & Prostate & 274 & 512$\times$512 & 200 & 74 & \href{http://promise12.grand-challenge.org}{http://promise12.grand-challenge.org} \\
		BCSS  & Breast & 151 & 1500-3000$\times$2000-4000 & 100 & 51 & \href{https://github.com/PathologyDataScience/BCSS}{https://github.com/PathologyDataScience/BCSS} \\
		\bottomrule
		\end{tabular}
	}
	\label{tab:datasets}
\end{table*}

\subsubsection{CHASE-DB1} The CHASE-DB1 dataset \cite{Fraz2012b} (a subset of the Child Heart and Health Study in England) includes 28 color images of children. Each image is captured with a $30^\circ$ FOV centered on the optic disc and has a size of 999$\times$960 pixels. As ground truth, two different expert manual segmentation maps are available, of which we used the first for our experiments. Since there are no specific training or testing subsets, following others \cite{Fraz2012b} we used the first 20 images for training and the remaining 8 for testing.

\subsubsection{MC} The Montgomery County (MC) chest X-ray dataset \cite{Jaeger2014} contains 138 frontal chest X-ray images obtained from a tuberculosis research program and is often used as a benchmark for lung segmentation. It includes 58 tuberculosis cases and 80 normal cases with a variety of abnormalities and for which expert manual segmentations are available. The images are relatively large, either $4020\times 4892$ or $4892\times 4020$ pixels. Using stratified sampling, we selected 100 images for training and the remaining 38 for testing.

\subsubsection{PH2} The PH2 dataset \cite{Mendonca2013} (named after its provider, the Hospital Pedro Hispano in Matosinhos, Portugal) includes 200 dermoscopic images, $768\times560$ pixels each, of melanocytic skin lesions with expert annotation to be used as ground truth in evaluating both segmentation and classification methods. Following experimental protocols of others \cite{7942129, 9157193, ROY2017160, Bozorgtabar2016}, we used all images in this dataset for testing, while training was done on the ISIC-2016 training images.

\subsubsection{ISIC-2016} The ISIC-2016 dataset \cite{8363547} (named after the International Skin Imaging Collaboration who hosted the challenge at the 2016 IEEE International Symposium on Biomedical Imaging where this dataset was used) contains 900 dermoscopic training images of different sizes, from as small as $576\times768$ or $718\times542$ pixels to as large as $4288\times2848$ pixels, with expert manual annotation for benchmarking melanoma segmentation, pattern detection, and classification methods. For testing, we used the PH2 images.

\subsubsection{DRISHTI-OC} The DRISHTI-GS1 dataset \cite{Sivaswamy-2015} includes 101 retinal images for glaucoma assessment. The images were captured with a $30^\circ$ FOV centered on the optic disc (OD) and are of size 2896$\times$1944 pixels. Average boundaries of both the optic cup (OC) and the OD in all images were obtained from manual annotations by four experts. The dataset is divided into 50 images for training and 51 for testing. We refer to the OC boundaries as the DRISHTI-OC dataset.

\subsubsection{DRISHTI-OD} The DRISHTI-OD dataset refers to average boundaries of the OD regions in the 101 retinal images of the DRISHTI-GS1 dataset \cite{Sivaswamy-2015} described above.

\subsubsection{PROMISE12} The PROMISE12 (Prostate MR Image Segmentation 2012) dataset \cite{Litjens-2014} contains three-dimensional (3D) transversal T2-weighted magnetic resonance (MR) images of 50 patients scanned at various centers using various MRI scanners and imaging protocols. The size of the images varies, from 256$\times$256 pixels, to 320$\times$320, 384$\times$384, and 512$\times$512 pixels. In our experiments we used only images of patients 0-12, all of size 512$\times$512 pixels, of which we used 200 for training and 74 for testing.

\subsubsection{BCSS} The BCSS (Breast Cancer Semantic Segmentation) dataset \cite{Amgad2019} contains more than 20,000 manually segmented tissue regions in 151 whole-slide breast-cancer images from The Cancer Genome Atlas (TCGA). The images vary in size, 1500-3000$\times$2000-4000 pixels, and were annotated by 25 participants ranging in experience from senior pathologists to medical students. We used 100 images for training and the remaining 51 for testing.

\subsection{Segmentation Performance Metrics}\label{sec:seg_metrics}
To quantify segmentation performance, we used seven popular metrics \cite{Taha-2015, Yeghiazaryan-2018}. Denoting the segmented image by $S$ and the corresponding ground-truth image by $G$, each having $N$ pixels $i=1\dots N$ with value either 0 (negative $=$ background) or 1 (positive $=$ foreground), we first computed the numbers of true-positive (TP), true-negative (TN), false-positive (FP), and false-negative (FN) pixels as:
\begin{equation}\label{eq:tp}
	\text{TP} = \sum^N_{i=1} S_i\cdot G_i,
\end{equation}
\begin{equation}\label{eq:tn}
	\text{TN} = \sum^N_{i=1} (1-S_i)\cdot(1-G_i),
\end{equation}
\begin{equation}\label{eq:fp}
	\text{FP} = \sum^N_{i=1} S_i\cdot(1-G_i),
\end{equation}
\begin{equation}\label{eq:fn}
	\text{FN} = \sum^N_{i=1} (1-S_i)\cdot G_i,
\end{equation}
from which we obtained the sensitivity (Se) also known as the recall (R), the specificity (Sp), accuracy (A), balance accuracy (BA), Dice (D) coefficient which is equivalent to the F1-score, the Jaccard (J) coefficient, and the overlap error (E):
\begin{equation}\label{eq:se}
	\text{Se} = \text{R} = \frac{\text{TP}}{\text{TP}+\text{FN}},
\end{equation}
\begin{equation}\label{eq:sp}
	\text{Sp} = \frac{\text{TN}}{\text{TN}+\text{FP}},
\end{equation}
\begin{equation}\label{eq:acc}
	\text{A} = \frac{\text{TP}+\text{TN}}{\text{TP}+\text{TN}+\text{FP}+\text{FN}},
\end{equation}
\begin{equation}\label{eq:ba}
	\text{BA} = \frac{\text{Se}+\text{Sp}}{2},
\end{equation}
\begin{equation}\label{eq:f1}
	\text{D} = \text{F1} = \frac{2|S\cap G|}{|S|+|G|} = \frac{2\text{TP}}{2\text{TP}+\text{FP}+\text{FN}},
\end{equation}
\begin{equation}\label{eq:jaccard}
	\text{J} = \frac{|S\cap G|}{|S\cup G|} = \frac{\text{TP}}{\text{TP}+\text{FP}+\text{FN}},
\end{equation}
\begin{equation}\label{eq:error}
	\text{E} = 1 - \text{J} = \frac{\text{FP}+\text{FN}}{\text{TP}+\text{FP}+\text{FN}}.
\end{equation}
The values of all metrics are in the range $[0,1]$, where 0 means worst and 1 means best performance, except for E, where 0 means best and 1 means worst performance.

\subsection{Regression Analysis Performance Metrics}\label{Regression Analysis}
To evaluate the performance of the linear regression models, we used the most common regression performance metrics, including the coefficient of determination $\text{R}^2$, adjusted $\text{R}^2$, RMSE, MAE, and important unbiased metrics, namely AIC and its corrected version AICc \cite{kassambara2018machine}.

The first is a statistical measure of proportional variance in the outcome that is explained by the independent variables \cite{glantz2001primer} and is computed as:
\begin{equation}\label{eq:r2}
	\text{R}^2 = 1-\frac{\text{RSS}}{\text{TSS}},
\end{equation}
with the total sum of squares
\begin{equation}
	\text{TSS} = \sum_{i=1}^n \left( y_i - \bar{y} \right)^2
\end{equation}
and the residual sum of squares
\begin{equation}\label{Eq SSres}
	\text{RSS} = \sum_{i=1}^n \left( y_i - m_i \right)^2
\end{equation}
computed from the observed values $y_i$ and the values $m_i$ predicted by the model \cite{glantz2001primer}. The regression model having a higher $\text{R}^2$ value is considered to be better. To account for the numbers of independent variables, $k$, and observations, $n$, the adjusted $\text{R}^2$ is also employed \cite{miles2005r}:
\begin{equation}\label{Eq AdjR2}
	\text{AR}^2 = 1-\frac{\left ( 1-R^2 \right )\left ( n-1 \right )}{\left ( n-k-1 \right )}.
\end{equation}

To measure the average error of the models in predicting the observations, we computed the RMSE, defined as:
\begin{equation}\label{Eq RMSE}
	\text{RMSE} = \sqrt{\frac{1}{n}\sum_{i=1}^{n}\left ( y_i-m_i \right )^2},
\end{equation}
as well as the MAE, defined as:
\begin{equation}\label{Eq MAE}
	\text{MAE} = \frac{1}{n}\sum_{i=1}^{n}\left | y_i-m_i \right |.
\end{equation}

Finally, to get an unbiased estimate of a model's performance, we computed the AIC metric:
\begin{equation}\label{Eq AIC}
	\text{AIC} = n\,\text{log} \left ( \frac{\text{RSS}}{n} \right )+2k,
\end{equation}
and because our sample size is relatively small ($n=10$ data\-sets), we also employed the AICc metric:
\begin{equation}\label{Eq AICc}
	\text{AICc} = \text{AIC} + \frac{2k^2+2k}{n-k-1}.
\end{equation}

\begin{figure}[t]
	\centering
	\begin{tabular}{@{}c@{}}
	\includegraphics[width=\textwidth]{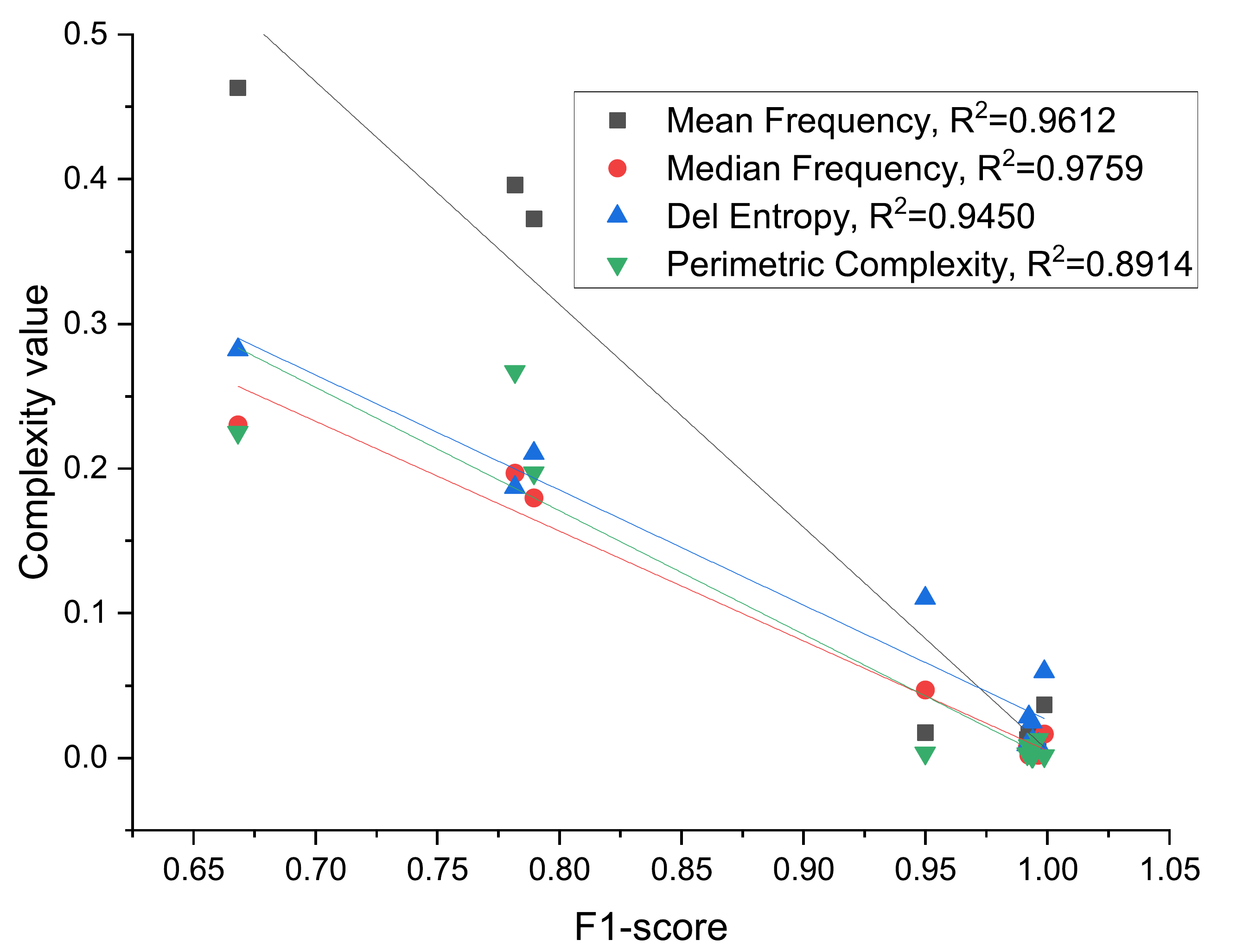} \\
	\includegraphics[width=\textwidth]{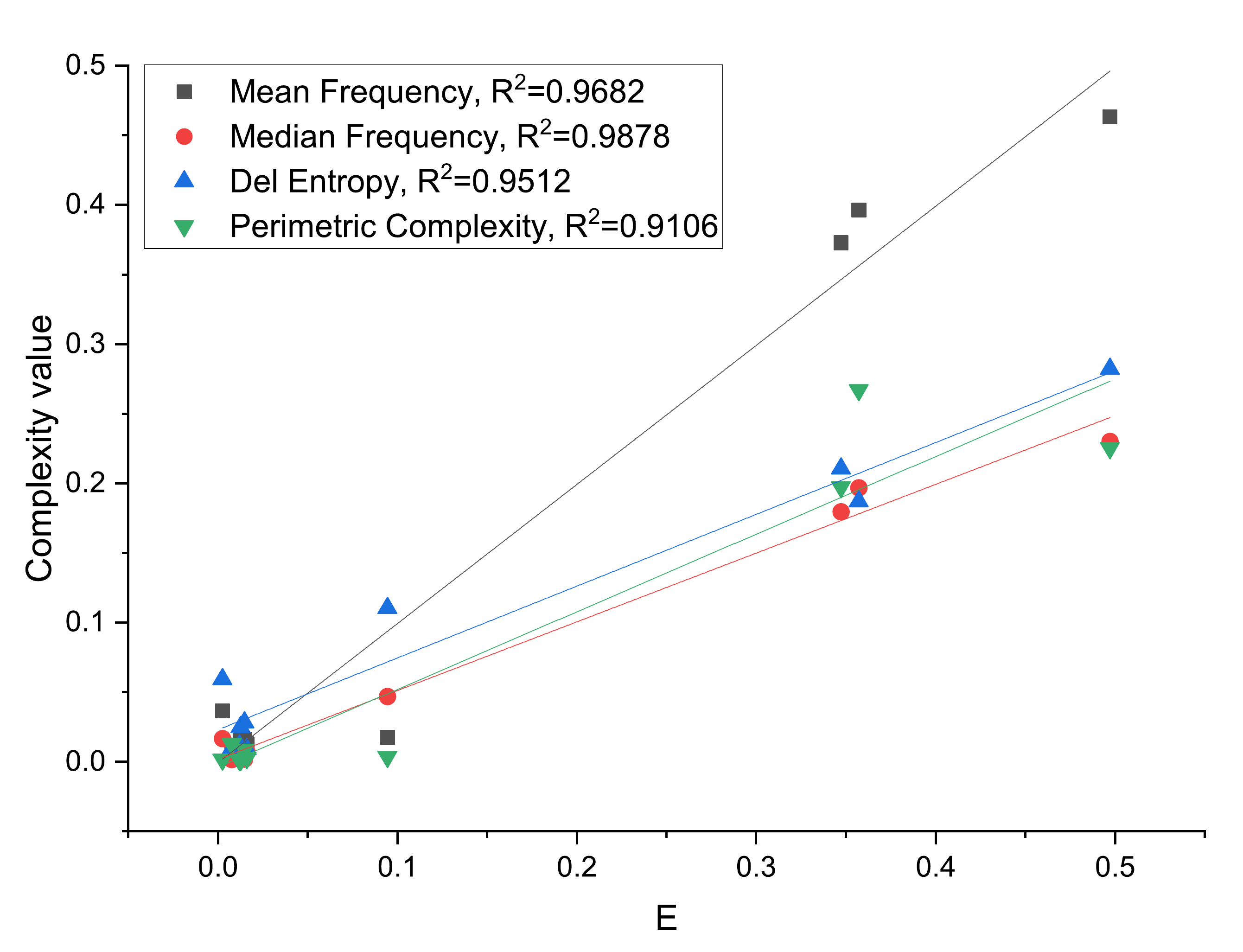} \\
	\end{tabular}
	\caption{Comparison of complexity measures in terms of their predictive performance.}
	\label{regressionPlots}
\end{figure}

\subsection{Experiment I: Image Complexity as a Guide for Input Downsampling}
This experiment was designed to investigate the effect of input downsampling on medical image segmentation performance and how, precisely, the proposed complexity measures predict the corresponding information loss. We considered three downsampling factors: 2, 3, and 4. For this experiment, we did not employ the proposed networks LW-Net and LS-Net, as the goal was to study the effect of input downsampling alone. To this end, the binary annotation masks of the images of all considered datasets were downsampled by a given factor, and then upsampled with the same factor to restore their size for comparison with the original masks using the segmentation performance metrics (Section~\ref{sec:seg_metrics}). Bilinear interpolation was employed in our implementation for both downsampling and upsampling. To minimize aliasing artifacts in the reconstructions, we removed all frequency components above the resampling Nyquist frequency using a low-pass filter \cite{SCHUMACHER19928} before downsampling, and after upsampling we applied optimal thresholding to get binary masks maximizing the Dice/F1-measure \cite{Lipton-2014}. From the results of this experiment (Table~\ref{tab:downsamplingSegmentation}) we observe two important trends: 1) the segmentation quality is consistently decreasing with increasing downsampling, and 2) this effect is less severe for datasets with relatively low image complexity. These trends clearly support our hypothesis that the proposed complexity measures are indicative of the information loss caused by downsampling and therefore can be employed as a guideline to determine the amount of acceptable downsampling.

\begin{table*}[!t]
	\centering
	\resizebox{0.8\textwidth}{!}{
		\begin{tabular}{@{}cccccccc|cccc@{}}
		\toprule
		\textbf{Downsampling} & \textbf{Se} & \textbf{Sp} & \textbf{A} & \textbf{BA} & \textbf{D} & \textbf{J} & \textbf{E} & \textbf{DE} & \textbf{MNF} & \textbf{MDF} & \textbf{PC} \\
		\toprule
		& \multicolumn{7}{c}{\textbf{STARE}} & 0.2105 & 0.3725 & 0.1796 & 0.1971 \\
		\midrule
		2 & 0.8741 & 0.9915 & 0.9826 & 0.9328 & 0.8833 & 0.7913 & 0.2087 \\
		3 & 0.8086 & 0.9872 & 0.9738 & 0.8979 & 0.8226 & 0.6991 & 0.3009 \\
		4 & 0.7586 & 0.9868 & 0.9696 & 0.8727 & 0.7895 & 0.6527 & 0.3473 \\
		\midrule
		& \multicolumn{7}{c}{\textbf{DRIVE}} & 0.2821 & 0.4632 & 0.2301 & 0.2253 \\
		\midrule
		2 & 0.8077 & 0.9872 & 0.9715 & 0.8975 & 0.8317 & 0.7121 & 0.2879 \\
		3 & 0.6931 & 0.9799 & 0.9549 & 0.8365 & 0.7282 & 0.5731 & 0.4269 \\
		4 & 0.6242 & 0.9767 & 0.9460 & 0.8005 & 0.6683 & 0.5027 & 0.4973 \\
		\midrule
		& \multicolumn{7}{c}{\textbf{CHASE-DB1}} & 0.1869 & 0.3961 & 0.1967 & 0.2670 \\
		\midrule
		2 & 0.8785 & 0.9911 & 0.9837 & 0.9355 & 0.8755 & 0.7788 & 0.2212 \\
		3 & 0.8280 & 0.9872 & 0.9769 & 0.9076 & 0.8233 & 0.7002 & 0.2998 \\
		4 & 0.7655 & 0.9866 & 0.9723 & 0.8761 & 0.7818 & 0.6427 & 0.3573 \\
		\midrule
		& \multicolumn{7}{c}{\textbf{MC}} & 0.0594 & 0.0367 & 0.0166 & 0.0016 \\
		\midrule
		2 & 0.9996 & 0.9997 & 0.9997 & 0.9996 & 0.9993 & 0.9986 & 0.0014 \\
		3 & 0.9991 & 0.9996 & 0.9995 & 0.9991 & 0.9990 & 0.9980 & 0.0020 \\
		4 & 0.9990 & 0.9995 & 0.9994 & 0.9990 & 0.9987 & 0.9975 & 0.0025 \\
		\midrule
		& \multicolumn{7}{c}{\textbf{PH2}} & 0.0248 & 0.0181 & 0.0049 & 0.0014 \\
		\midrule
		2 & 0.9985 & 0.9975 & 0.9984 & 0.9985 & 0.9965 & 0.9931 & 0.0069 \\
		3 & 0.9966 & 0.9974 & 0.9980 & 0.9966 & 0.9955 & 0.9910 & 0.0090 \\
		4 & 0.9958 & 0.9962 & 0.9971 & 0.9958 & 0.9936 & 0.9873 & 0.0127 \\
		\midrule
		& \multicolumn{7}{c}{\textbf{ISIC-2016}} & 0.0093 & 0.0106 & 0.0017 & 0.0128 \\
		\midrule
		2 & 0.9979 & 0.9994 & 0.9995 & 0.9979 & 0.9976 & 0.9953 & 0.0047 \\
		3 & 0.9968 & 0.9990 & 0.9992 & 0.9968 & 0.9962 & 0.9925 & 0.0075 \\
		4 & 0.9964 & 0.9990 & 0.9991 & 0.9964 & 0.9961 & 0.9922 & 0.0078 \\
		\midrule
		& \multicolumn{7}{c}{\textbf{DRISHTI-OC}} & 0.0090 & 0.0128 & 0.0072 & 0.0029 \\
		\midrule
		2 & 0.9943 & 1.0000 & 0.9990 & 0.9971 & 0.9961 & 0.9922 & 0.0078 \\
		3 & 0.9943 & 0.9999 & 0.9998 & 0.9971 & 0.9950 & 0.9901 & 0.0099 \\
		4 & 0.9901 & 0.9999 & 0.9997 & 0.9950 & 0.9918 & 0.9838 & 0.0162 \\
		\midrule
		& \multicolumn{7}{c}{\textbf{DRISHTI-OD}} & 0.0117 & 0.0104 & 0.0045 & 0.0013 \\
		\midrule
		2 & 0.9957 & 1.0000 & 0.9998 & 0.9978 & 0.9972 & 0.9943 & 0.0057 \\
		3 & 0.9955 & 0.9999 & 0.9998 & 0.9977 & 0.9963 & 0.9925 & 0.0075 \\
		4 & 0.9924 & 0.9998 & 0.9996 & 0.9961 & 0.9939 & 0.9880 & 0.0120 \\
		\midrule
		& \multicolumn{7}{c}{\textbf{PROMISE12}} & 0.1104 & 0.0469 & 0.0175 & 0.0035 \\
		\midrule
		2 & 0.9623 & 0.9988 & 0.9978 & 0.9805 & 0.9654 & 0.9336 & 0.0664 \\
		3 & 0.9453 & 0.9988 & 0.9969 & 0.9722 & 0.9568 & 0.9178 & 0.0822 \\
		4 & 0.9398 & 0.9985 & 0.9963 & 0.9692 & 0.9499 & 0.9054 & 0.0946 \\
		\midrule
		& \multicolumn{7}{c}{\textbf{BCSS}} & 0.0282 & 0.0163 & 0.0018 & 0.0085 \\
		\midrule
		2 & 0.9950 & 0.9988 & 0.9977 & 0.9969 & 0.9963 & 0.9927 & 0.0073 \\
		3 & 0.9944 & 0.9971 & 0.9966 & 0.9957 & 0.9946 & 0.9894 & 0.0106 \\
		4 & 0.9914 & 0.9964 & 0.9953 & 0.9939 & 0.9924 & 0.9851 & 0.0149 \\
		\bottomrule
		\end{tabular}
    }
	\caption{Effect of input image downsampling on segmentation performance for the considered datasets. The proposed complexity measures computed on the original images are also reported.\vspace{-0.2\baselineskip}}
	\label{tab:downsamplingSegmentation}
\end{table*}

To compare the predictive power of the different complexity measures on segmentation performance, we performed linear regression for the two most common segmentation performance metrics: Dice (F1) and Jaccard (expressed via E). The results (Fig.~\ref{regressionPlots}) indicate that the MDF measure outperforms the other measures in predicting segmentation quality, as confirmed by its highest $\text{R}^2$ values. As both MNF and MDF are higher than DE and PC, it can be concluded that frequency information is most predictive of segmentation performance in the datasets considered in our experiments. The other measures capture different types of complexity and may prove useful in other medical image segmentation tasks.

\begin{table}[!t]
	\centering
	\resizebox{0.945\textwidth}{!}{
	\begin{tabular}{@{}lccccccc@{}}
	\toprule
	\textbf{Measure} & \textbf{DoF} & $\mathbf{\text{R}^2}$ & $\mathbf{\text{AR}^2}$ & \textbf{RMSE} & \textbf{MAE} & \textbf{AIC} & \textbf{AICc} \\
	\midrule
	\multicolumn{8}{c}{\textbf{Downsampled by 2}} \\
	\midrule
	\multicolumn{1}{@{}l@{}}{\multirow{6}[1]{*}{DE}}
		& 1 & 0.964387 & 0.958452 & 0.021752 & 0.016354 & -59.2487 & -58.5821 \\
		& 2 & 0.964775 & 0.950684 & 0.021634 & 0.016246 & -57.3362 & -54.9362 \\
		& 3 & 0.987878 & 0.978787 & 0.012691 & 0.008659 & -63.8702 & -57.8702 \\
		& 4 & 0.994052 & 0.986121 & 0.008890 & 0.007013 & -67.5658 & -54.2325 \\
		& 5 & 0.999930 & 0.999754 & 0.000967 & 0.000626 & -101.062 & \textbf{-71.0617} \\
		& 6 & \textbf{0.999958} & \textbf{0.999707} & \textbf{0.000746} & \textbf{0.000399} & \textbf{-103.223} & -19.2228 \\
	\midrule
	\multicolumn{1}{@{}l@{}}{\multirow{6}[1]{*}{MNF}}
		& 1 & 0.992983 & 0.991814 & 0.009655 & 0.008384 & -72.2439 & -71.5772 \\
		& 2 & 0.998666 & 0.998133 & 0.004209 & 0.003328 & -83.5268 & -81.1268 \\
		& 3 & 0.999446 & 0.999030 & 0.002714 & 0.002272 & -88.5520 & -82.5520 \\
		& 4 & 0.999833 & 0.999610 & 0.001490 & 0.001202 & -96.1454 & \textbf{-82.8121} \\
		& 5 & 0.999972 & 0.999903 & 0.000606 & 0.000369 & -108.533 & -78.5332 \\
		& 6 & \textbf{0.999992} & \textbf{0.999942} & \textbf{0.000331} & \textbf{0.000177} & \textbf{-116.238} & -32.2378 \\
	\midrule
	\multicolumn{1}{@{}l@{}}{\multirow{6}[1]{*}{MDF}}
		& 1 & 0.993635 & 0.992574 & 0.009196 & 0.007828 & -73.0241 & -72.3574 \\
		& 2 & 0.997561 & 0.996585 & 0.005693 & 0.004479 & -78.6963 & -76.2963 \\
		& 3 & 0.998331 & 0.997079 & 0.004709 & 0.003594 & -79.7330 & -73.7330 \\
		& 4 & 0.999695 & 0.999288 & 0.002014 & 0.001472 & -91.3206 & -77.9873 \\
		& 5 & 0.999989 & 0.999961 & 0.000382 & 0.000236 & -115.902 & \textbf{-85.9019} \\
		& 6 & \textbf{0.999996} & \textbf{0.999971} & \textbf{0.000235} & \textbf{0.000126} & \textbf{-121.711} & -37.7109 \\
	\midrule
	\multicolumn{1}{@{}l@{}}{\multirow{6}[1]{*}{PC}}
		& 1 & 0.942353 & 0.932746 & 0.027675 & 0.016837 & -55.3957 & -54.7290 \\
		& 2 & 0.966339 & 0.952875 & 0.021148 & 0.015842 & -57.6996 & -55.2996 \\
		& 3 & 0.982266 & 0.968966 & 0.015350 & 0.011684 & -60.8266 & -54.8266 \\
		& 4 & 0.999788 & 0.999506 & 0.001677 & 0.001038 & -94.2488 & \textbf{-80.9155} \\
		& 5 & 0.999806 & 0.999322 & 0.001604 & 0.000961 & -92.9601 & -62.9601 \\
		& 6 & \textbf{0.999966} & \textbf{0.999763} & \textbf{0.000670} & \textbf{0.000377} & \textbf{-104.927} & -20.9274 \\
	\midrule
	\multicolumn{8}{c}{\textbf{Downsampled by 3}} \\
	\midrule
	\multicolumn{1}{@{}l@{}}{\multirow{6}[1]{*}{DE}}
		& 1 & 0.971305 & 0.966522 & 0.028192 & 0.020556 & -55.0995 & -54.4329 \\
		& 2 & 0.974376 & 0.964127 & 0.026640 & 0.019587 & -54.0052 & -51.6052 \\
		& 3 & 0.992955 & 0.987671 & 0.013969 & 0.009600 & -62.3344 & -56.3344 \\
		& 4 & 0.996215 & 0.991169 & 0.010238 & 0.008100 & -65.3059 & -51.9726 \\
		& 5 & 0.999964 & 0.999875 & 0.000994 & 0.000635 & -100.623 & \textbf{-70.6233} \\
		& 6 & \textbf{0.999989} & \textbf{0.999924} & \textbf{0.000548} & \textbf{0.000293} & \textbf{-108.159} & -24.1591 \\
	\midrule
	\multicolumn{1}{@{}l@{}}{\multirow{6}[1]{*}{MNF}}
		& 1 & 0.985851 & 0.983493 & 0.019796 & 0.015826 & -60.7561 & -60.0894 \\
		& 2 & 0.996766 & 0.995473 & 0.009464 & 0.006754 & -70.5639 & -68.1639 \\
		& 3 & 0.996932 & 0.994631 & 0.009218 & 0.006069 & -68.9855 & -62.9855 \\
		& 4 & 0.999801 & 0.999537 & 0.002345 & 0.001862 & -88.8881 & -75.5548 \\
		& 5 & 0.999991 & 0.999970 & 0.000486 & 0.000303 & -112.080 & \textbf{-82.0802} \\
		& 6 & \textbf{1.000000} & \textbf{0.999999} & \textbf{0.000008} & \textbf{0.000004} & \textbf{-139.762} & -55.7619 \\
	\midrule
	\multicolumn{1}{@{}l@{}}{\multirow{6}[1]{*}{MDF}}
		& 1 & 0.986262 & 0.983972 & 0.019507 & 0.015018 & -60.9920 & -60.3254 \\
		& 2 & 0.994309 & 0.992032 & 0.012555 & 0.009130 & -66.0420 & -63.6420 \\
		& 3 & 0.994451 & 0.990290 & 0.012397 & 0.008372 & -64.2447 & -58.2447 \\
		& 4 & 0.999674 & 0.999240 & 0.003003 & 0.002189 & -84.9303 & -71.5970 \\
		& 5 & 0.999988 & 0.999958 & 0.000574 & 0.000346 & -109.406 & \textbf{-79.4060} \\
		& 6 & \textbf{0.999996} & \textbf{0.999975} & \textbf{0.000314} & \textbf{0.000169} & \textbf{-117.049} & -33.0493 \\
	\midrule
	\multicolumn{1}{@{}l@{}}{\multirow{6}[1]{*}{PC}}
		& 1 & 0.918892 & 0.905374 & 0.047397 & 0.027916 & -46.7871 & -46.1204 \\
		& 2 & 0.954481 & 0.936273 & 0.035507 & 0.026147 & -49.4083 & -47.0083 \\
		& 3 & 0.974561 & 0.955482 & 0.026544 & 0.019937 & -52.0630 & -46.0630 \\
		& 4 & 0.999846 & 0.999641 & 0.002064 & 0.001275 & -90.9320 & \textbf{-77.5987} \\
		& 5 & 0.999852 & 0.999480 & 0.002028 & 0.001215 & -89.2110 & -59.2110 \\
		& 6 & \textbf{0.999974} & \textbf{0.999819} & \textbf{0.000847} & \textbf{0.000476} & \textbf{-101.187} & -17.1869 \\
	\midrule
	\multicolumn{8}{c}{\textbf{Downsampled by 4}} \\
	\midrule
	\multicolumn{1}{@{}l@{}}{\multirow{6}[1]{*}{DE}} & 1 & 0.968211 & 0.962913 & 0.034523 & 0.025492 & -51.8580 & -51.1913 \\
		& 2 & 0.970937 & 0.959316 & 0.033010 & 0.024596 & -50.5752 & -48.1752 \\
		& 3 & 0.990215 & 0.982877 & 0.019153 & 0.013639 & -57.2843 & -51.2843 \\
		& 4 & 0.994972 & 0.988268 & 0.013730 & 0.010878 & -60.6105 & -47.2772 \\
		& 5 & 0.999867 & 0.999536 & 0.002231 & 0.001323 & \textbf{-87.6878} & \textbf{-57.6878} \\
		& 6 & \textbf{0.999885} & \textbf{0.999193} & \textbf{0.002079} & \textbf{0.001114} & -86.8103 & -2.81032 \\
	\midrule
	\multicolumn{1}{@{}l@{}}{\multirow{6}[1]{*}{MNF}} & 1 & 0.986812 & 0.984614 & 0.022236 & 0.018458 & -58.8966 & -58.2299 \\
		& 2 & 0.997850 & 0.996990 & 0.008978 & 0.006973 & -71.4084 & -69.0084 \\
		& 3 & 0.998171 & 0.996799 & 0.008281 & 0.006226 & -70.7005 & -64.7005 \\
		& 4 & 0.999691 & 0.999278 & 0.003406 & 0.002665 & -82.9140 & -69.5806 \\
		& 5 & 0.999909 & 0.999682 & 0.001847 & 0.001183 & -90.7113 & \textbf{-60.7113} \\
		& 6 & \textbf{0.999958} & \textbf{0.999709} & \textbf{0.001248} & \textbf{0.000668} & \textbf{-94.9813} & -10.9813 \\
	\midrule
	\multicolumn{1}{@{}l@{}}{\multirow{6}[1]{*}{MDF}} & 1 & 0.987747 & 0.985705 & 0.021433 & 0.017538 & -59.4849 & -58.8182 \\
		& 2 & 0.995947 & 0.994325 & 0.012328 & 0.009407 & -66.3344 & -63.9344 \\
		& 3 & 0.996180 & 0.993315 & 0.011968 & 0.008978 & -64.8085 & -58.8085 \\
		& 4 & 0.999414 & 0.998633 & 0.004686 & 0.003487 & -77.8094 & -64.4761 \\
		& 5 & 0.999990 & 0.999964 & 0.000622 & 0.000438 & -108.118 & -78.1177 \\
		& 6 & \textbf{1.000000} & \textbf{1.000000} & \textbf{6.54E-06} & \textbf{3.52E-06} & \textbf{-179.008} & \textbf{-95.0082} \\
	\midrule
	\multicolumn{1}{@{}l@{}}{\multirow{6}[1]{*}{PC}} & 1 & 0.924150 & 0.911508 & 0.053327 & 0.032039 & -44.9009 & -44.2342 \\
		& 2 & 0.954159 & 0.935823 & 0.041457 & 0.030712 & -46.9295 & -44.5295 \\
		& 3 & 0.974786 & 0.955876 & 0.030746 & 0.023560 & -49.7118 & -43.7118 \\
		& 4 & 0.999678 & 0.999250 & 0.003473 & 0.002166 & -82.6055 & \textbf{-69.2721} \\
		& 5 & 0.999721 & 0.999023 & 0.003235 & 0.001967 & -81.7402 & -51.7402 \\
		& 6 & \textbf{0.999977} & \textbf{0.999836} & \textbf{0.000937} & \textbf{0.000526} & \textbf{-99.5720} & -15.5720 \\
		\bottomrule
	\end{tabular}
	}
	\caption{Performance comparison of image complexity measures in terms of regression performance in predicting the error measure E for various downsampling factors and degrees of freedom (DoF) of the regression model. Best values are indicated in bold.\vspace{-2\baselineskip}}
	\label{tab:regressionMetricsF1}
\end{table}

To evaluate the trade-off between the goodness-of-fit and model complexity in terms of number of independent variables (or the degree of freedom), we compared the regression performance of models by varying the degree of freedom (DoF) and using the regression performance metrics (Section~\ref{Regression Analysis}). The metrics were computed for the three considered downsampling factors: 2, 3, and 4. The DoF is the number of independent variables in the polynomial function (or the degree of the polynomial) that best fits the data. In our experiments, models with $\text{DoF}>5$ did not improve the regression performance in general (Table~\ref{tab:regressionMetricsF1}). More specifically, while performance further improved in terms of the other metrics, according to the AICc metric optimal performance was reached for $\text{DoF}=4$ or $5$ in most cases. Given our small sample size, we considered AICc to be decisive owing to its unbiased nature.

\begin{figure}[!t]
	\centering
	\begin{tabular}{@{}c@{\hspace{0.0125\textwidth}}c@{\hspace{0.0125\textwidth}}c@{\hspace{0.0125\textwidth}}c@{\hspace{0.0125\textwidth}}c@{}}
		\includegraphics[width=0.19\textwidth]{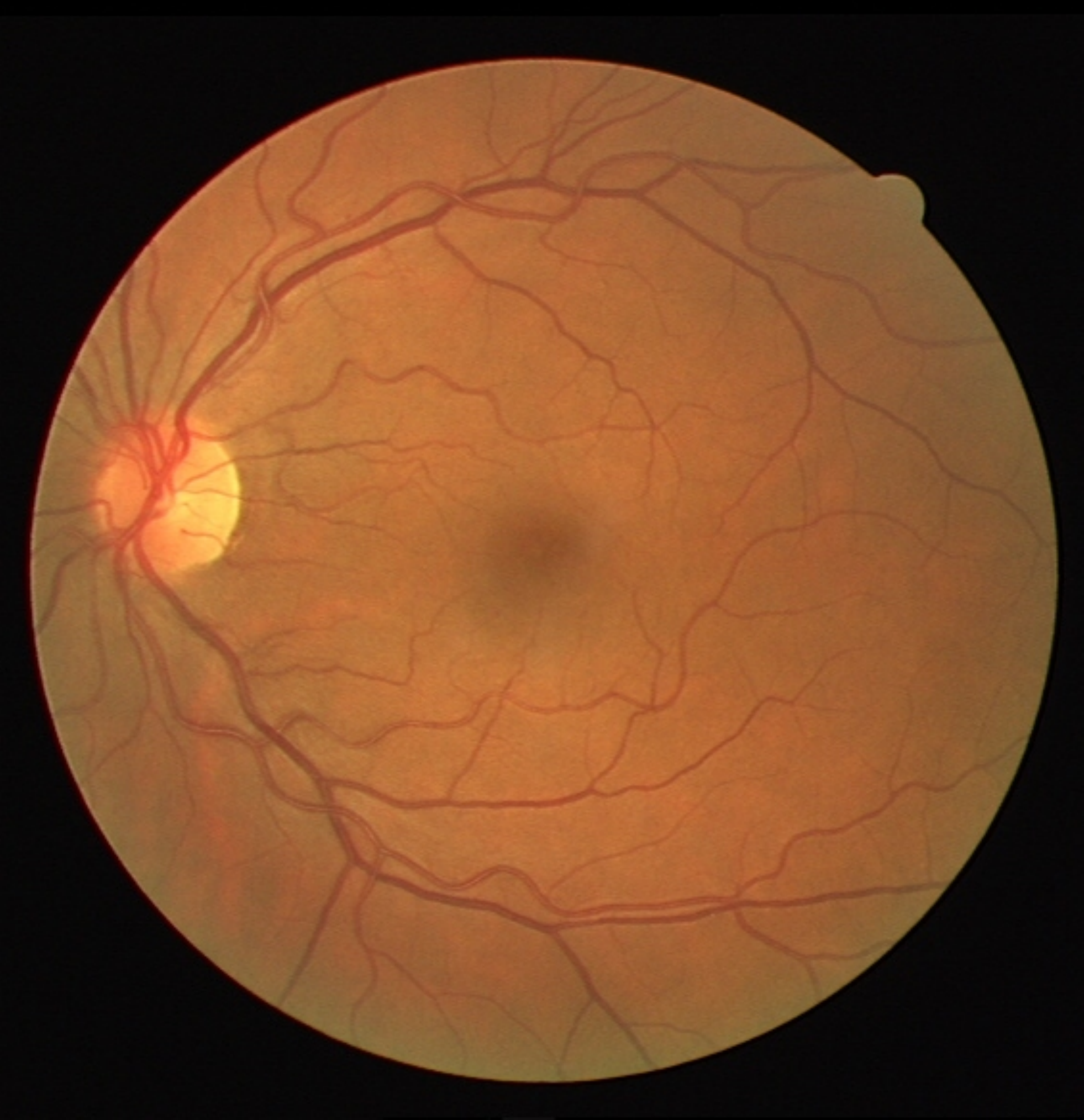} &
		\includegraphics[width=0.19\textwidth]{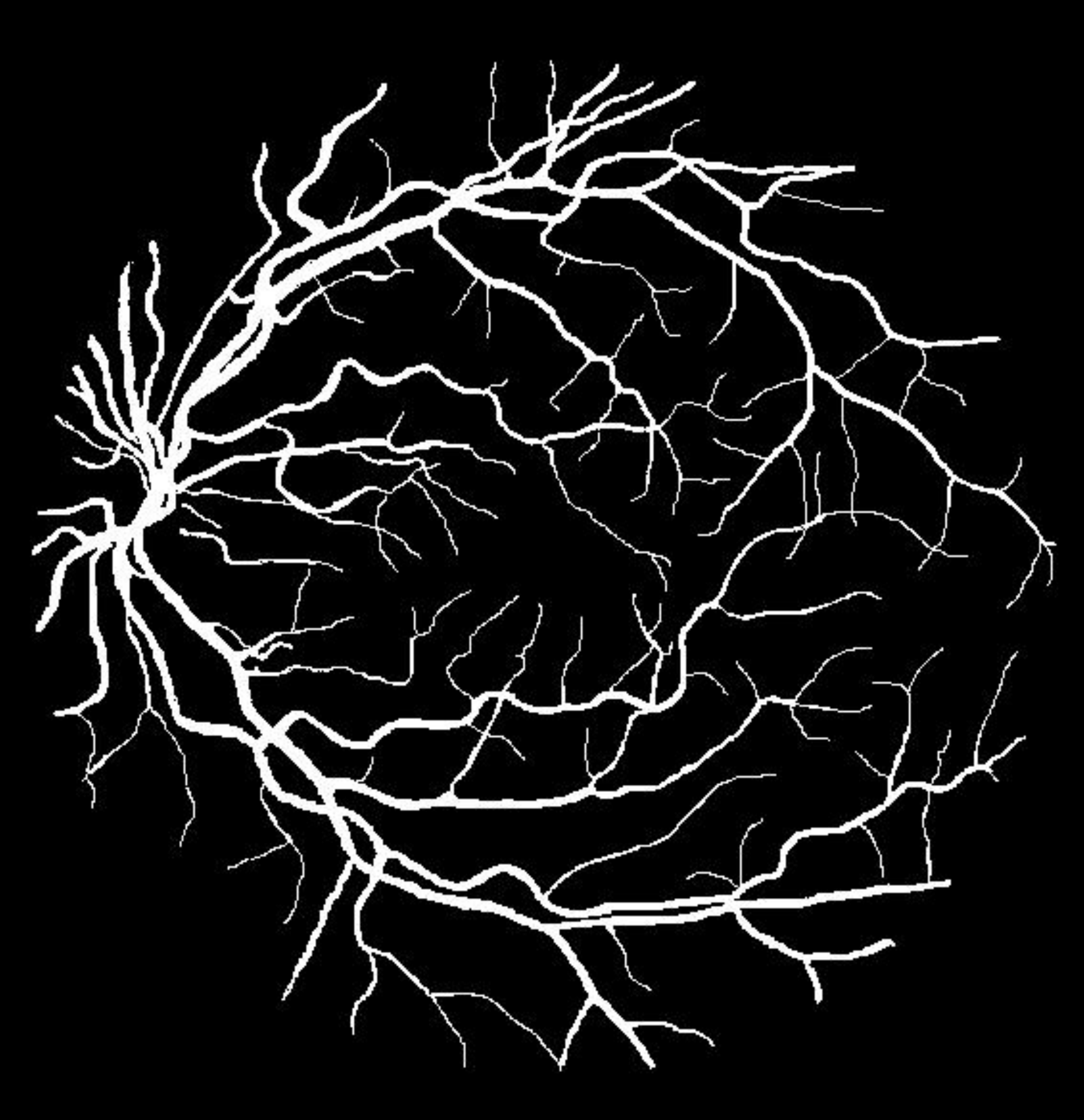} &
		\includegraphics[width=0.19\textwidth]{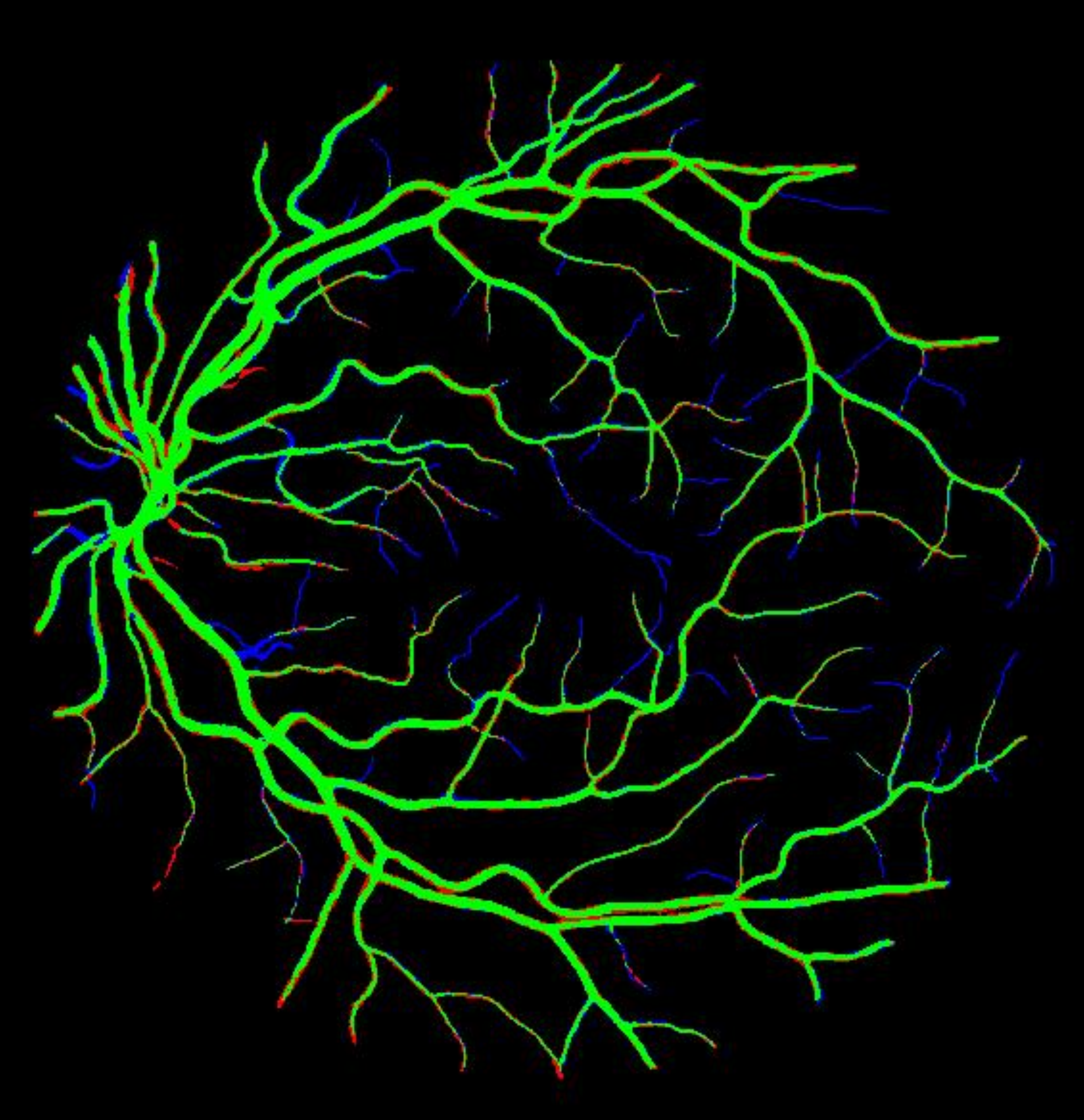} &
		\includegraphics[width=0.19\textwidth]{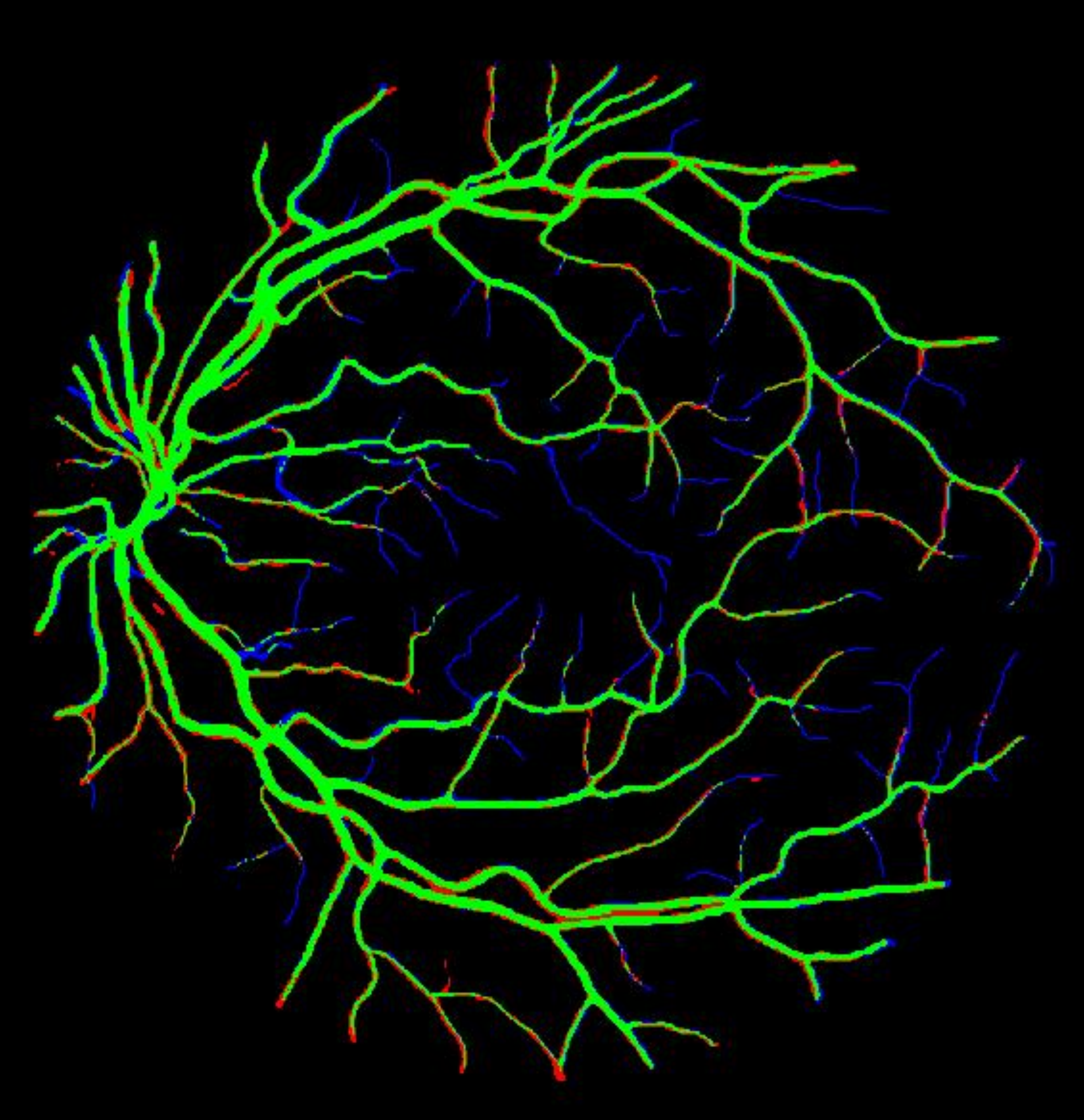} &
		\includegraphics[width=0.19\textwidth]{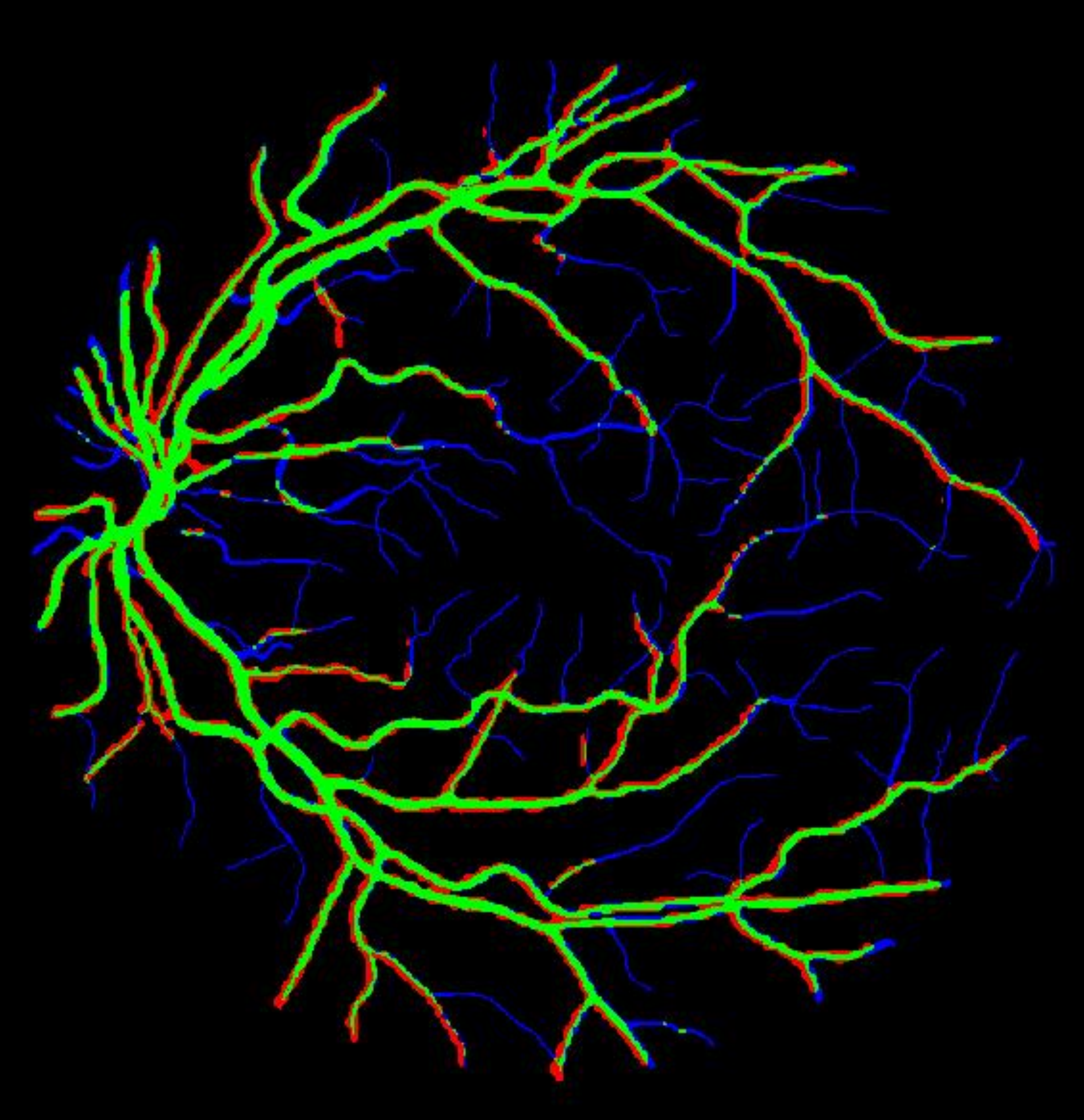} \\
		\includegraphics[width=0.19\textwidth]{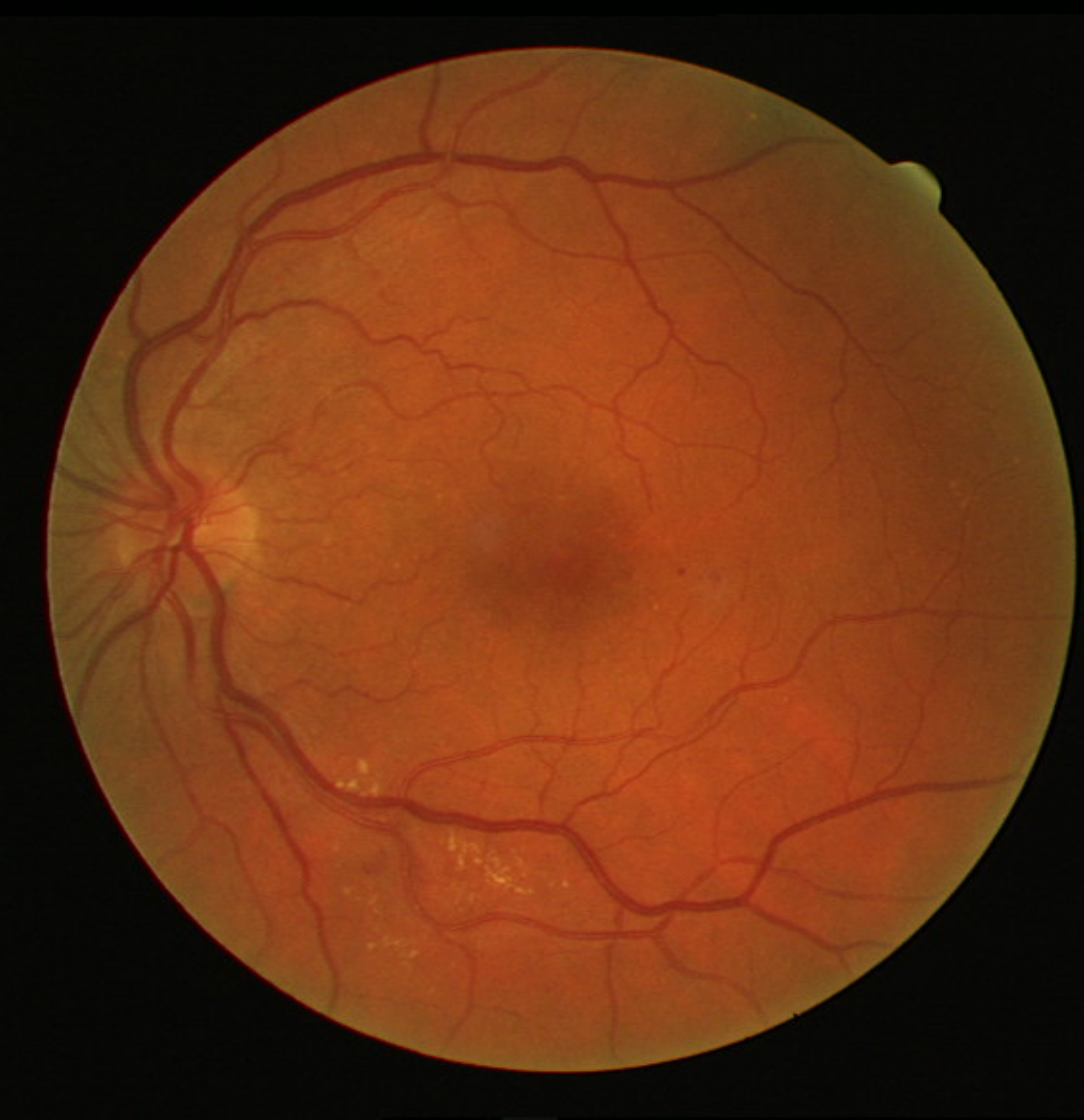} &
		\includegraphics[width=0.19\textwidth]{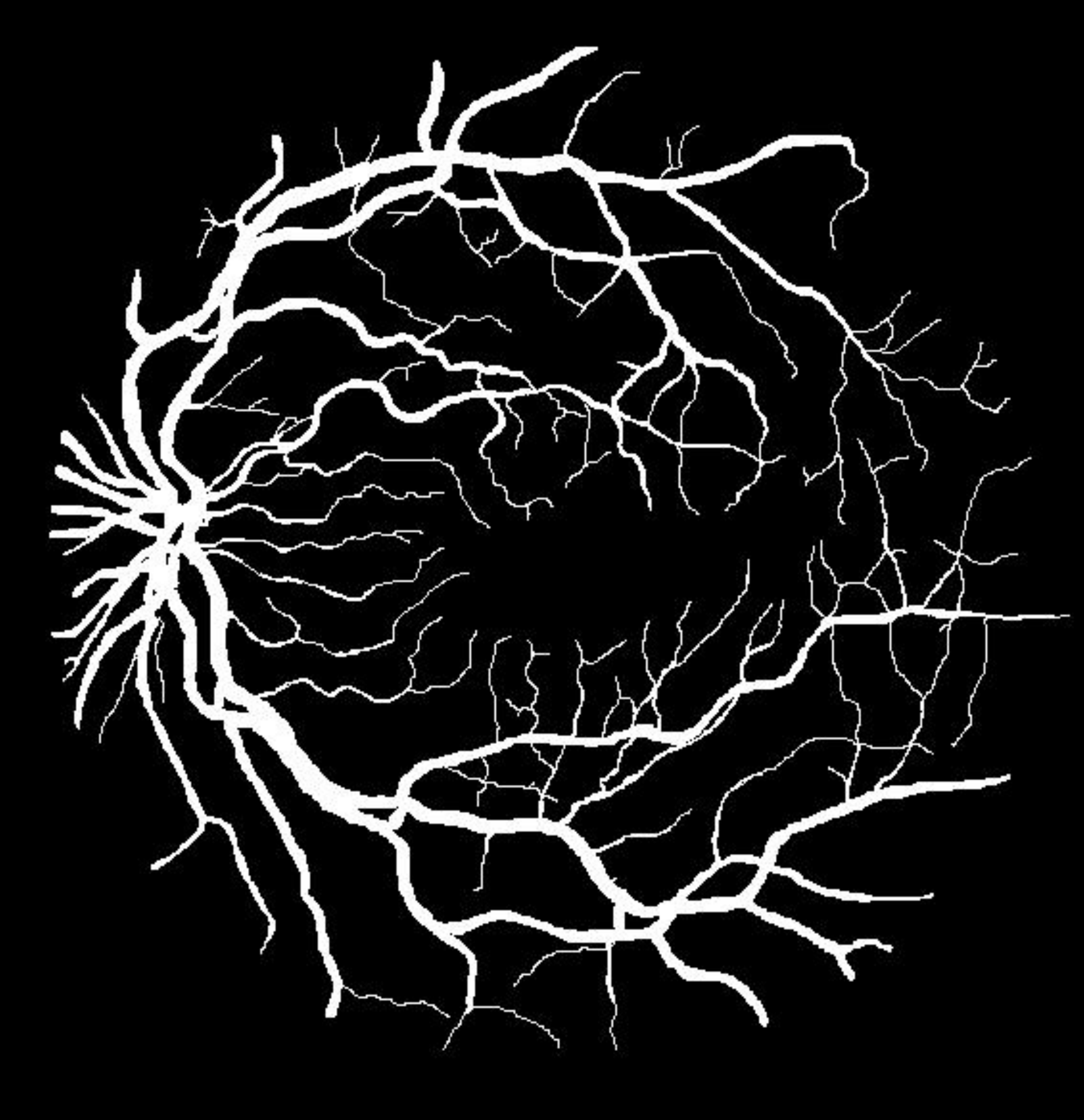} &
		\includegraphics[width=0.19\textwidth]{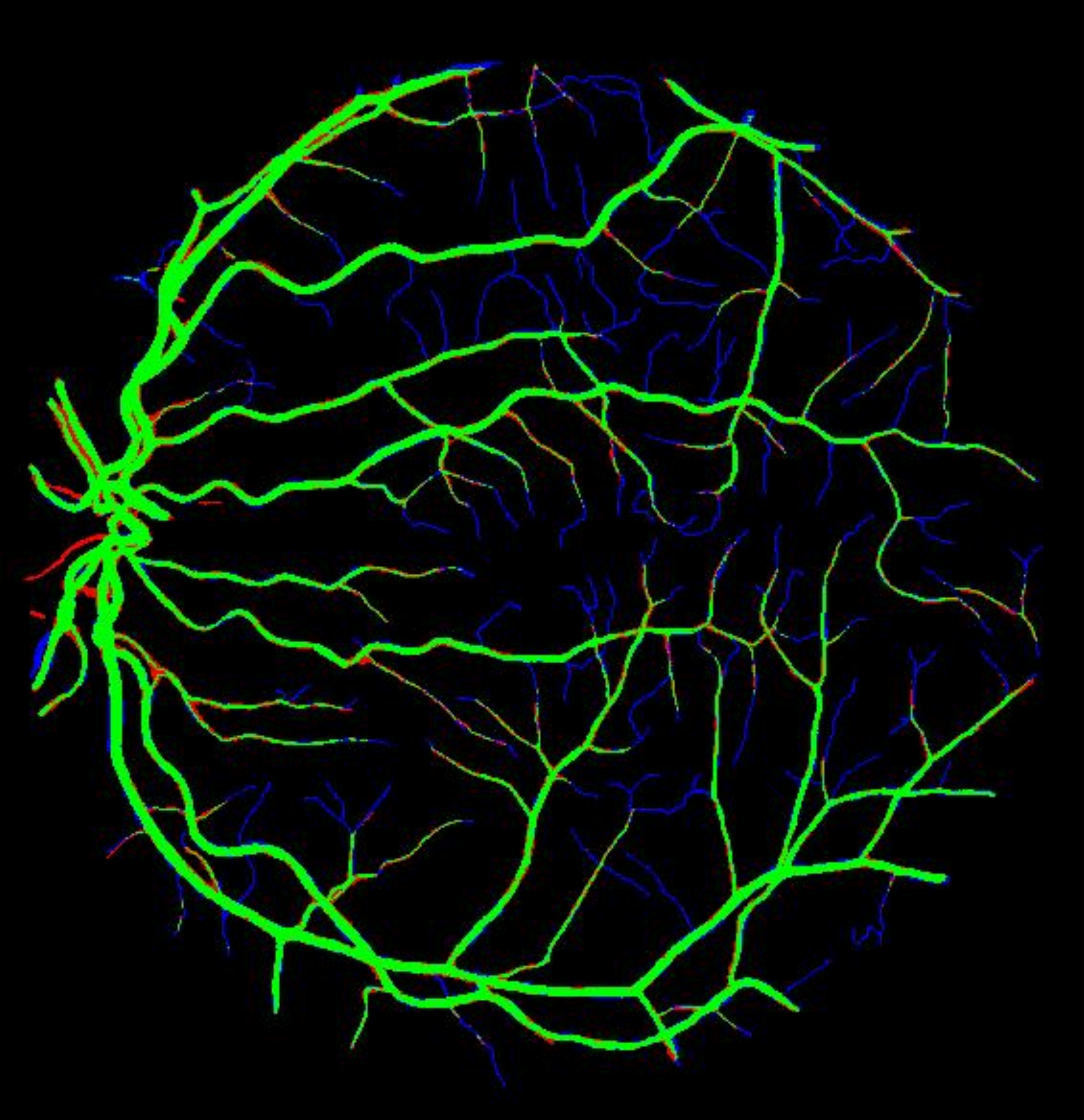} &
		\includegraphics[width=0.19\textwidth]{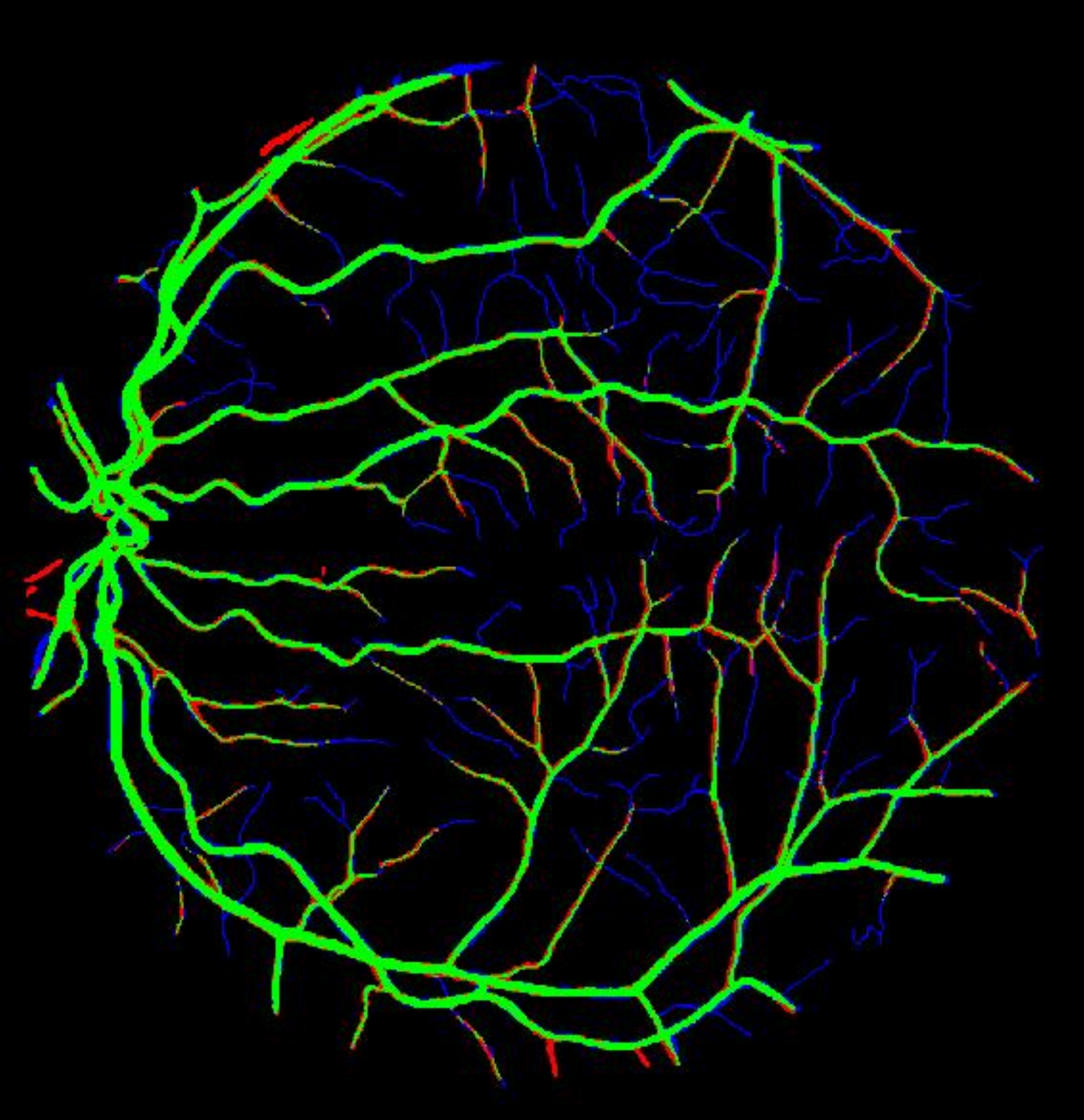} &
		\includegraphics[width=0.19\textwidth]{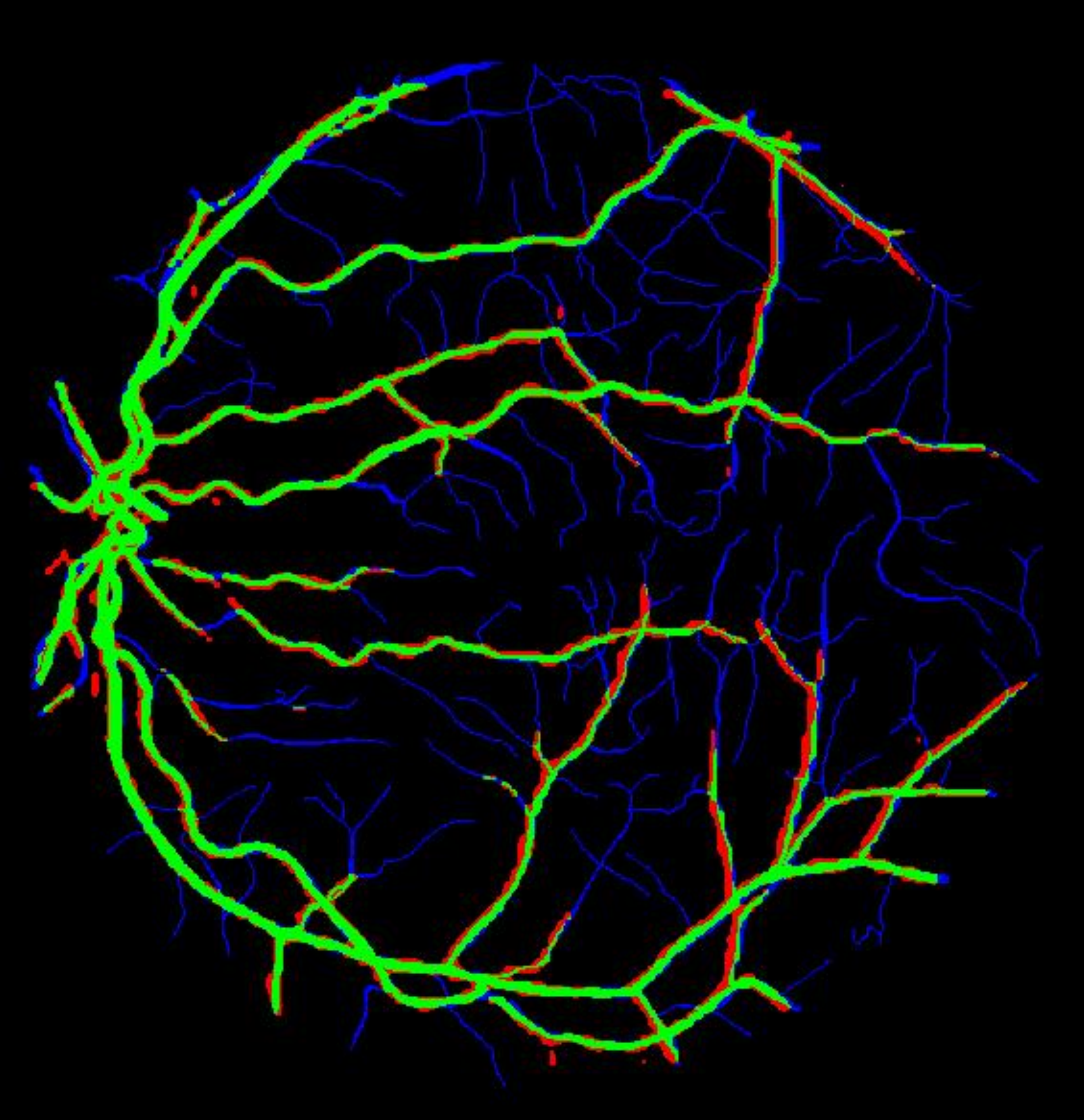} \\
		\includegraphics[width=0.19\textwidth]{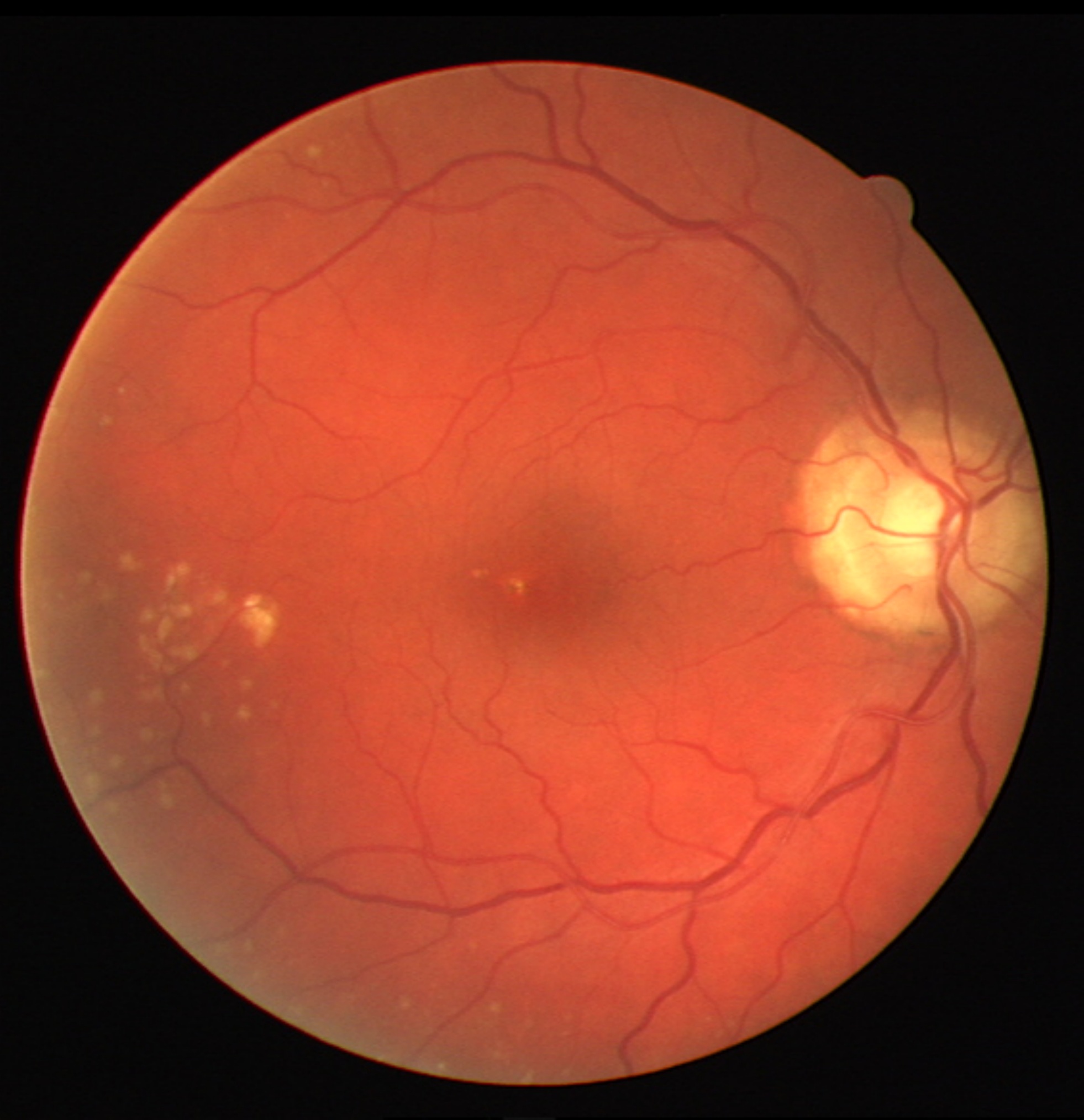} &
		\includegraphics[width=0.19\textwidth]{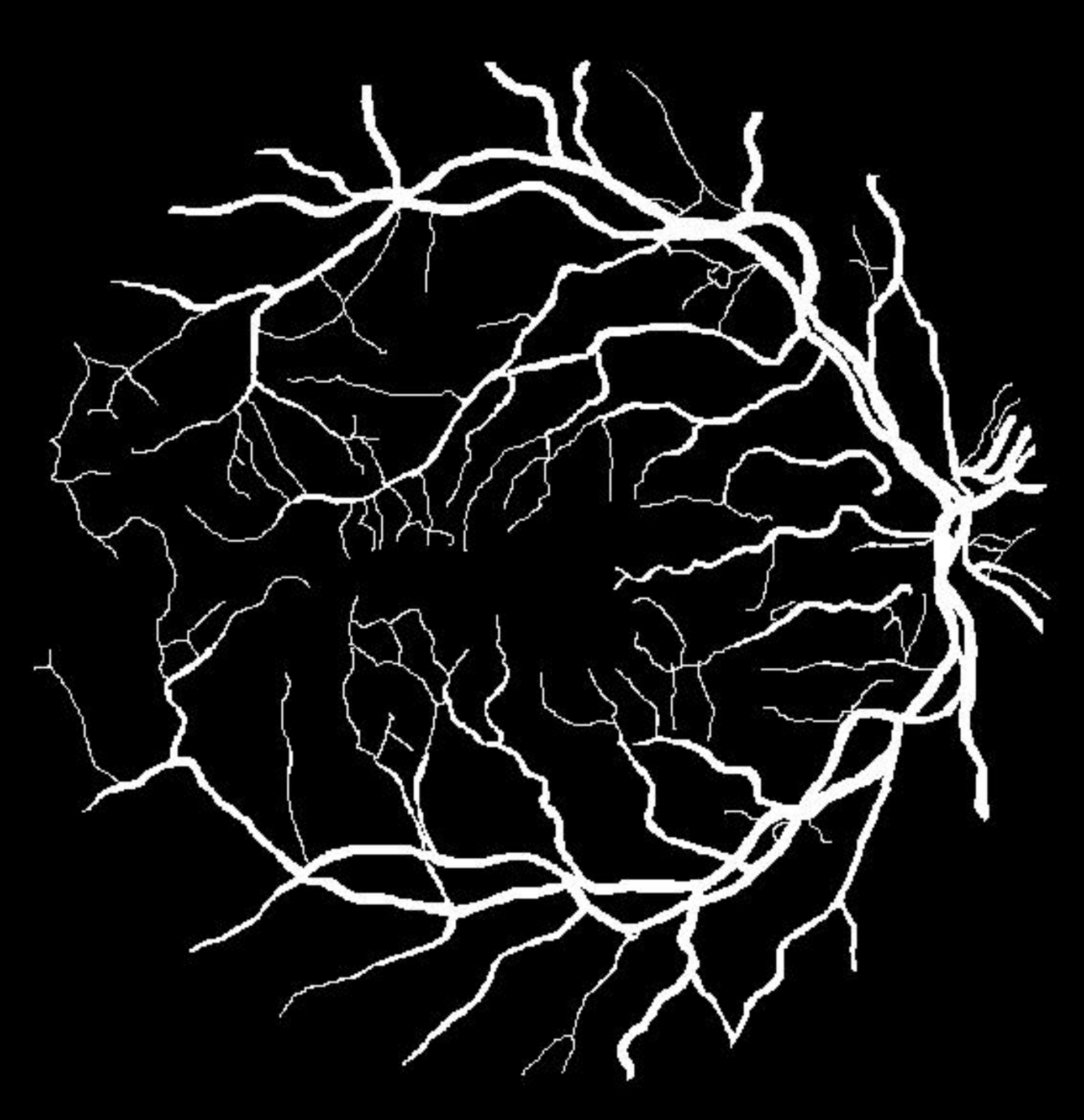} &
		\includegraphics[width=0.19\textwidth]{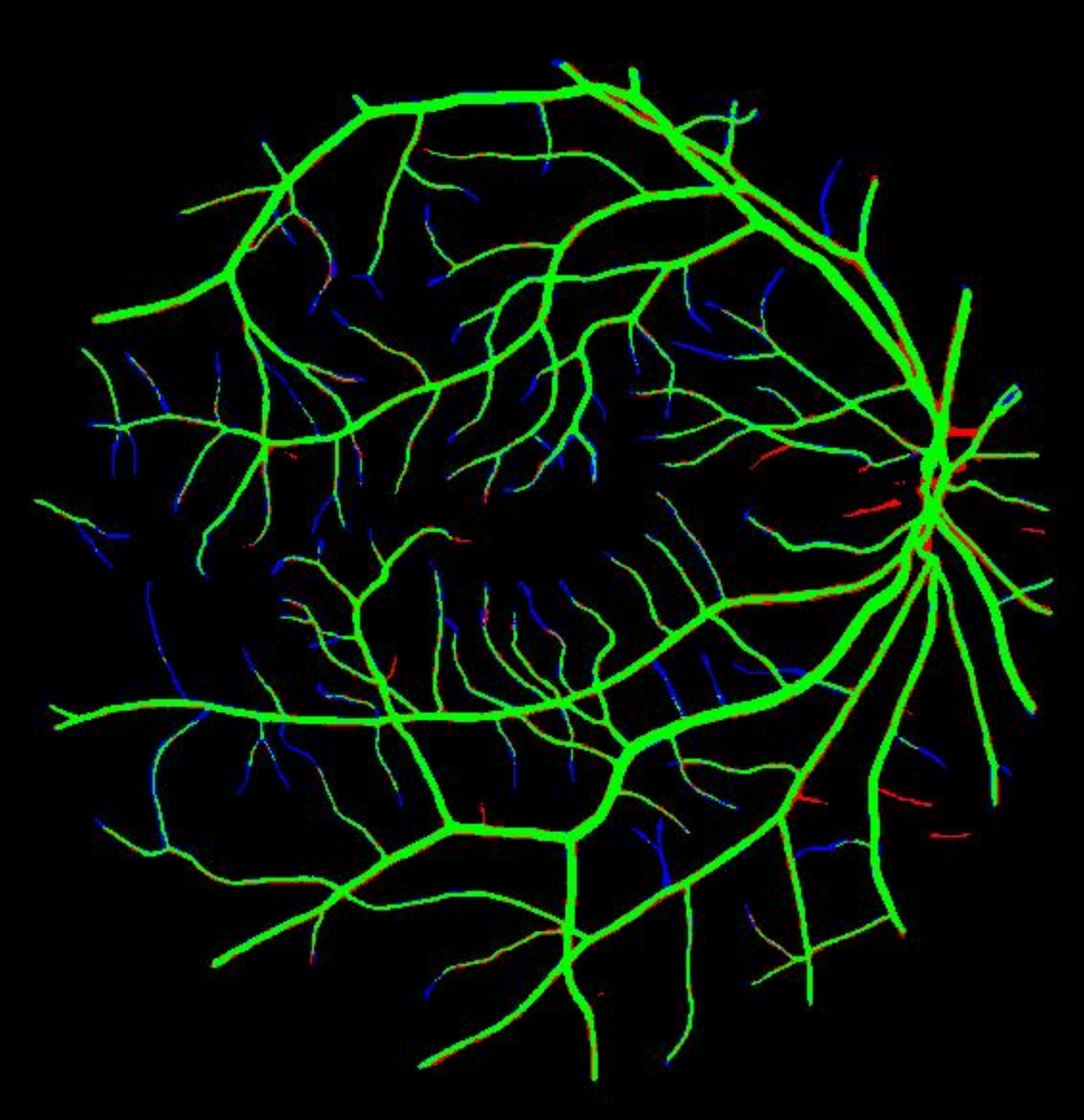} &
		\includegraphics[width=0.19\textwidth]{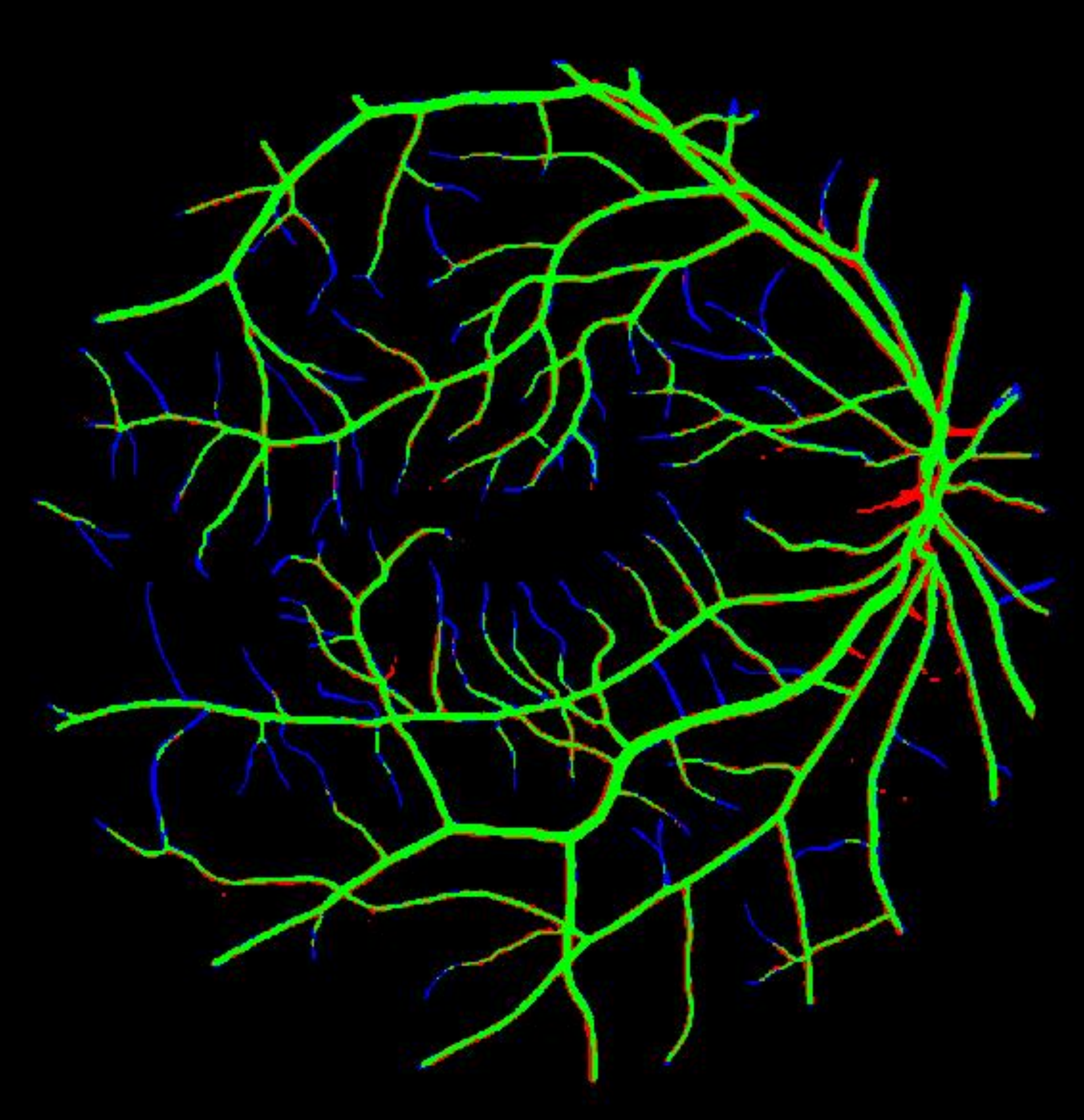} &
		\includegraphics[width=0.19\textwidth]{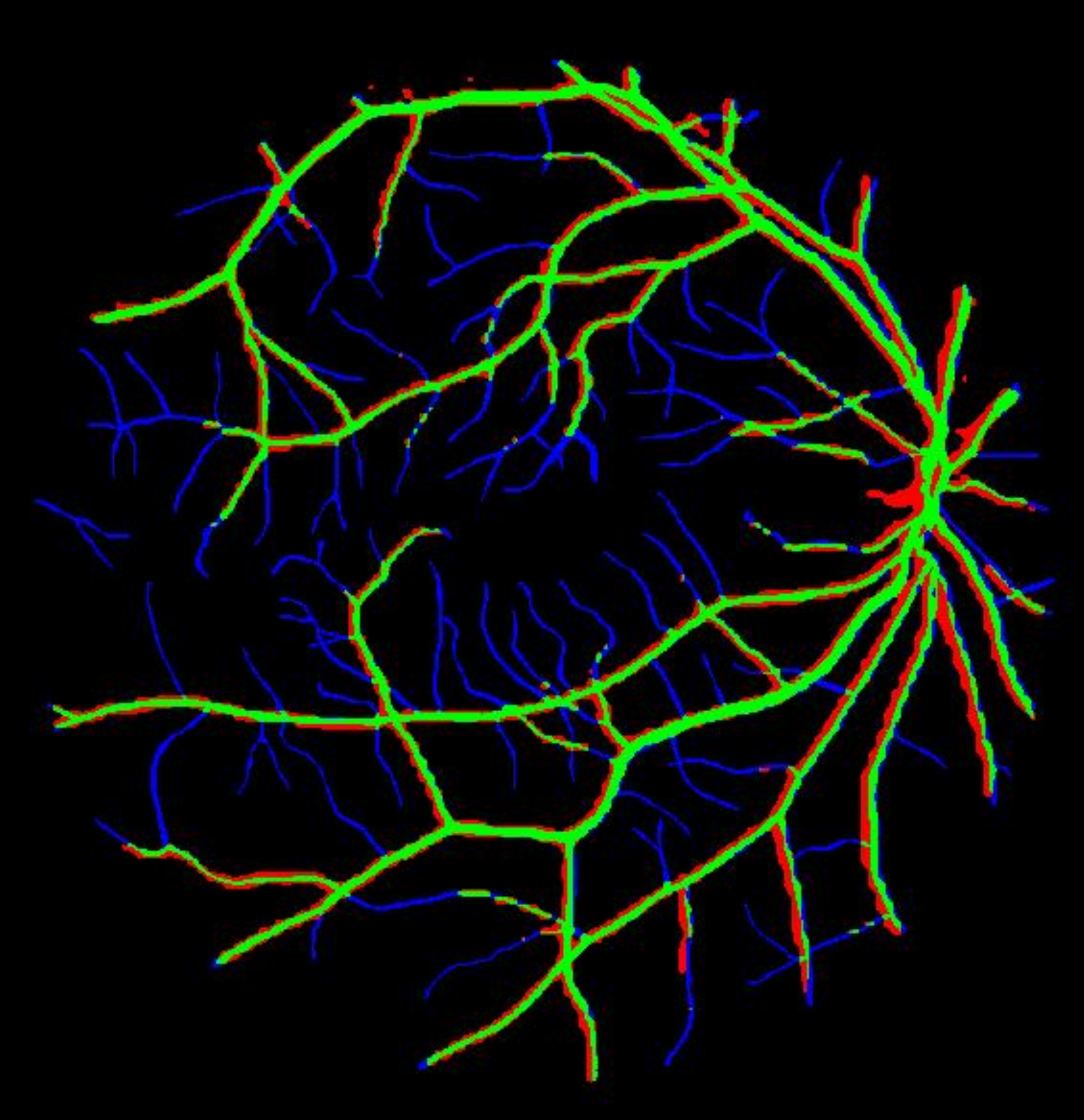} \\
		\includegraphics[width=0.19\textwidth]{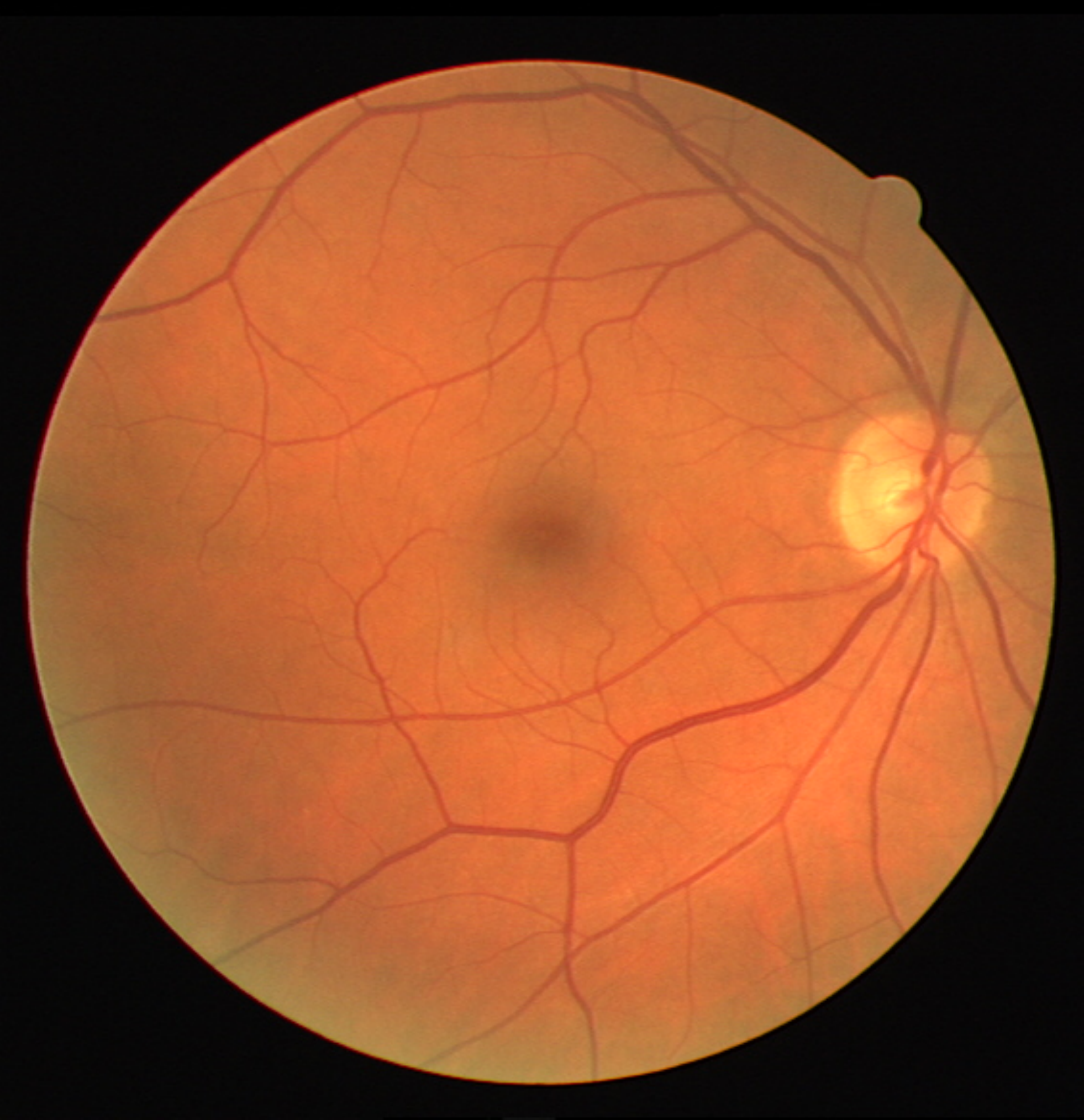} &
		\includegraphics[width=0.19\textwidth]{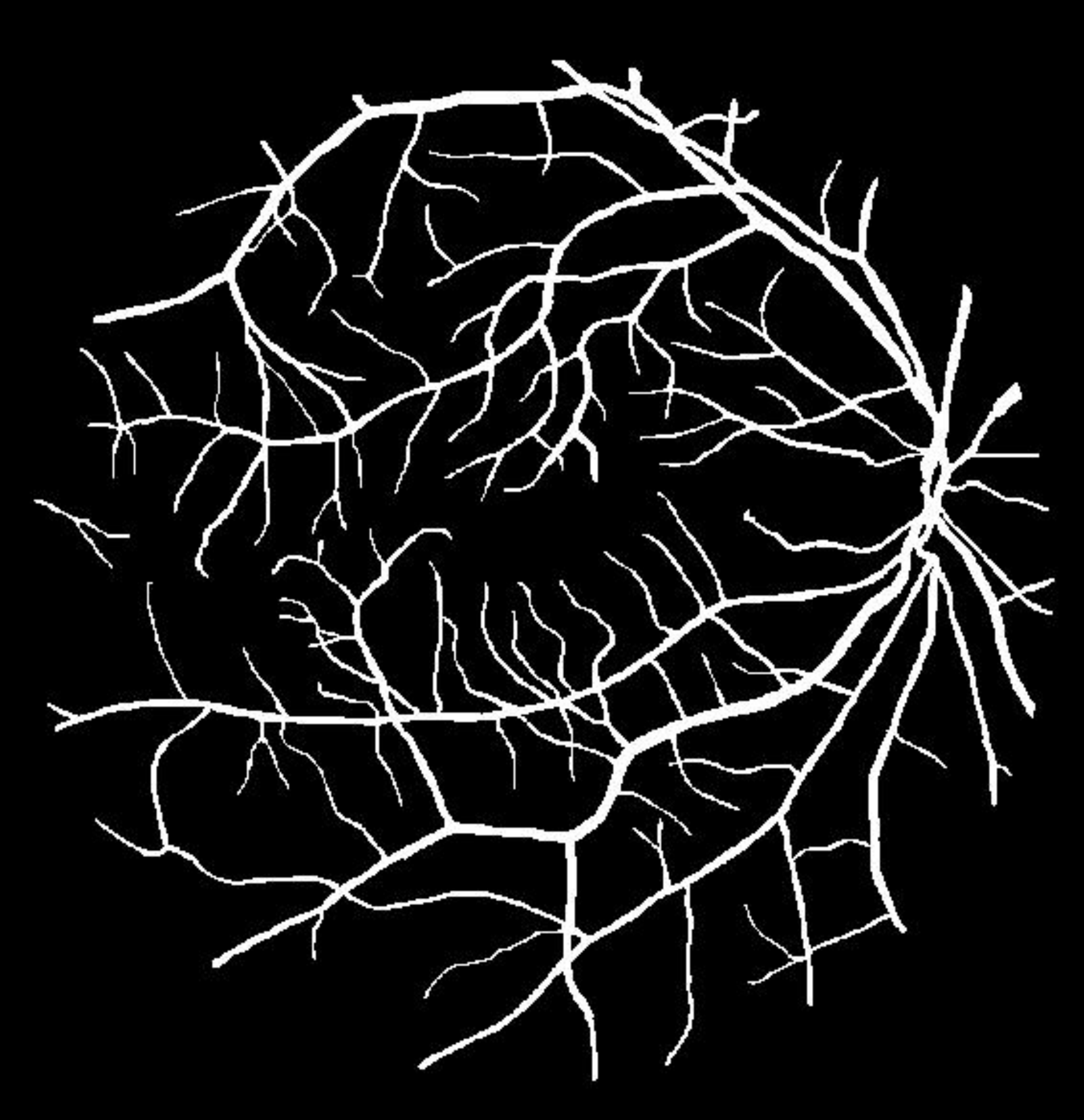} &
		\includegraphics[width=0.19\textwidth]{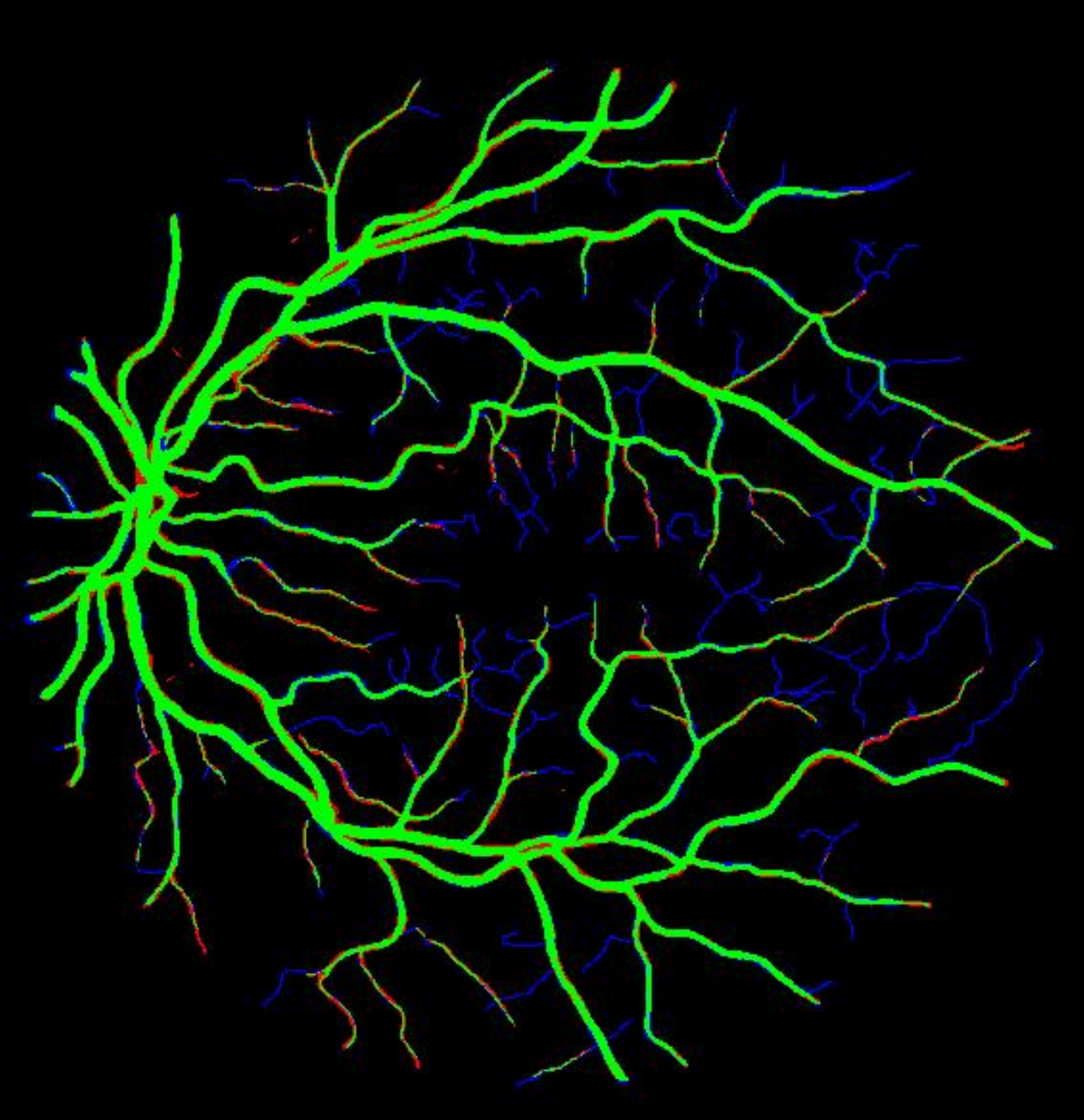} &
		\includegraphics[width=0.19\textwidth]{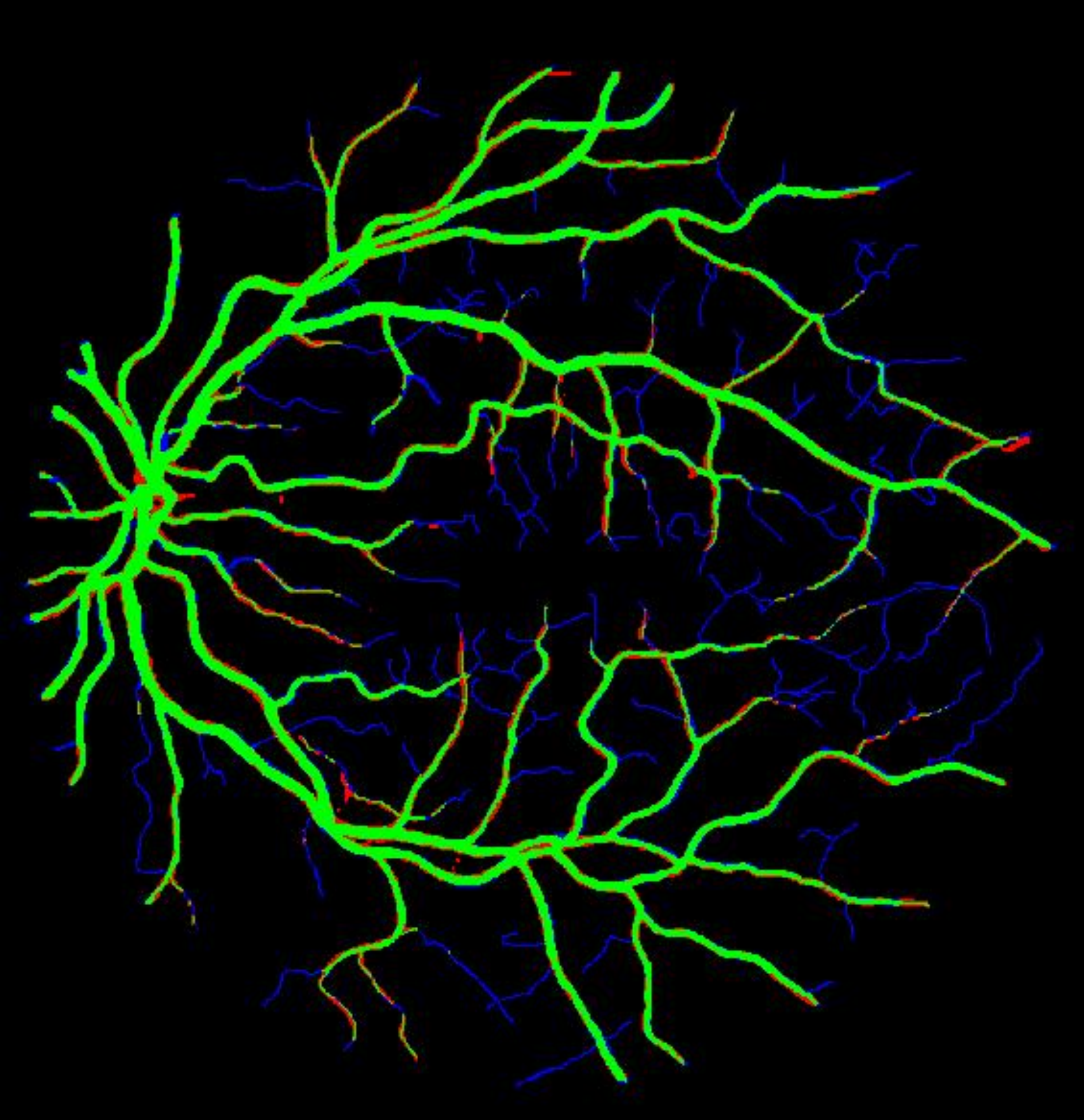} &
		\includegraphics[width=0.19\textwidth]{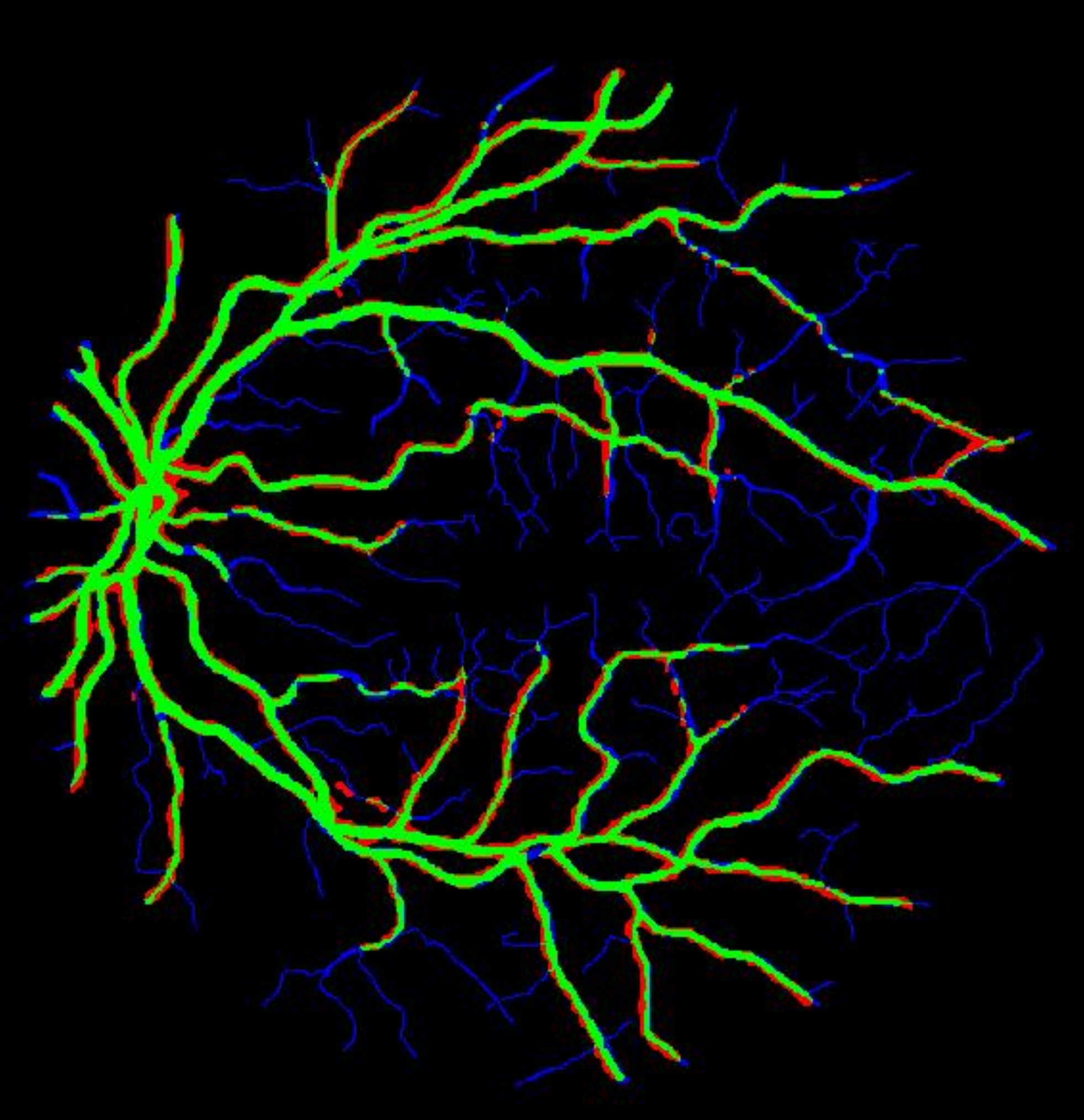} \\
	\end{tabular}
	\caption{Sample segmentation results with LS-Net on the DRIVE dataset. Four examples are shown from top to bottom. From left to right: the input images, the ground truth manually annotated by an expert, and the results on $2\times$, $3\times$, and $4\times$ downsampled input images. Correctly segmented foreground and background pixels are shown in, respectively, green and black. False positive and false negative pixels are shown in, respectively, red and blue.}
	\label{visualdrive}
\end{figure}

\begin{figure}[!t]
	\centering
	\begin{tabular}{@{}c@{\hspace{0.0125\textwidth}}c@{\hspace{0.0125\textwidth}}c@{\hspace{0.0125\textwidth}}c@{\hspace{0.0125\textwidth}}c@{}}
		\includegraphics[width=0.19\textwidth]{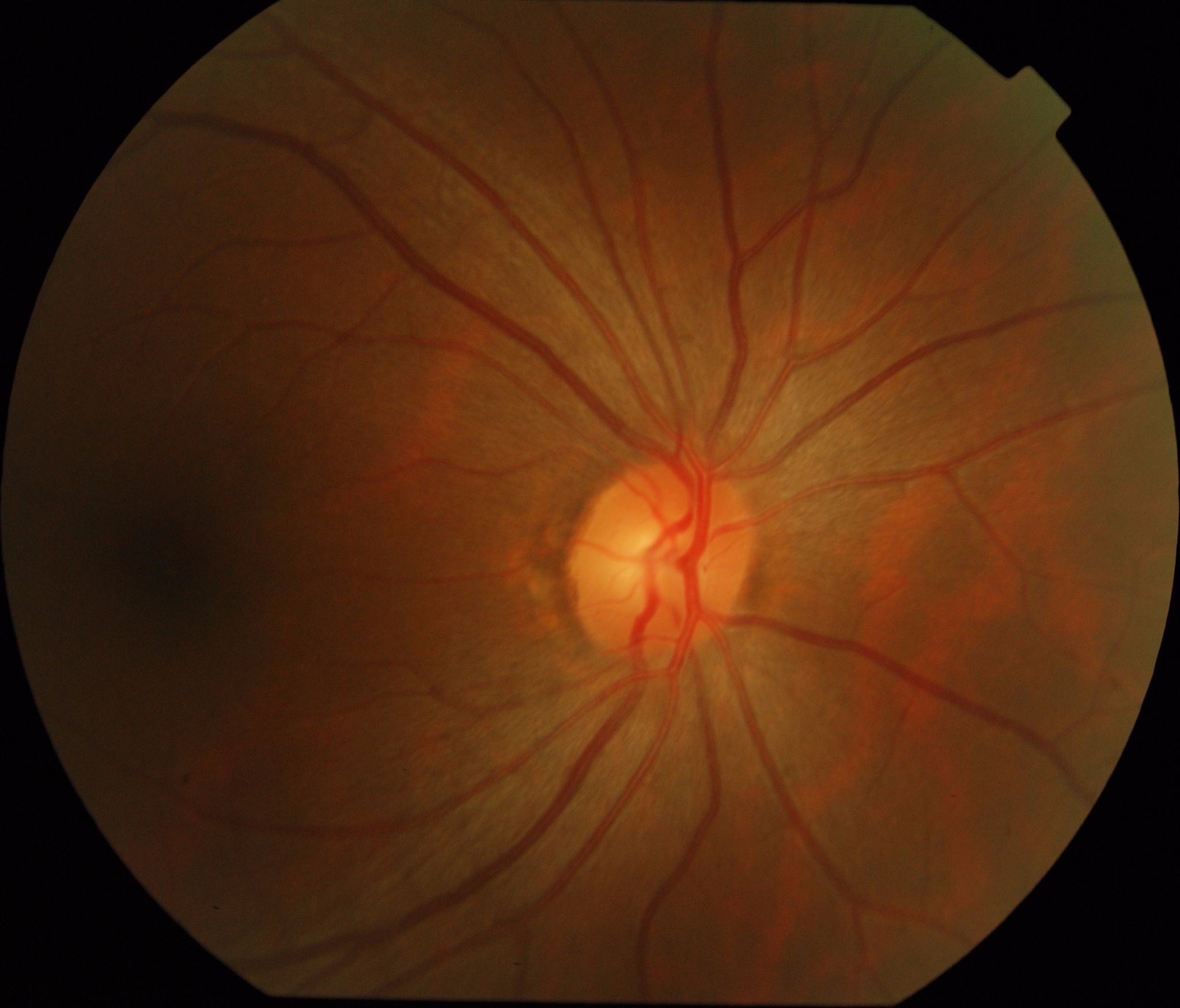} &
		\includegraphics[width=0.19\textwidth]{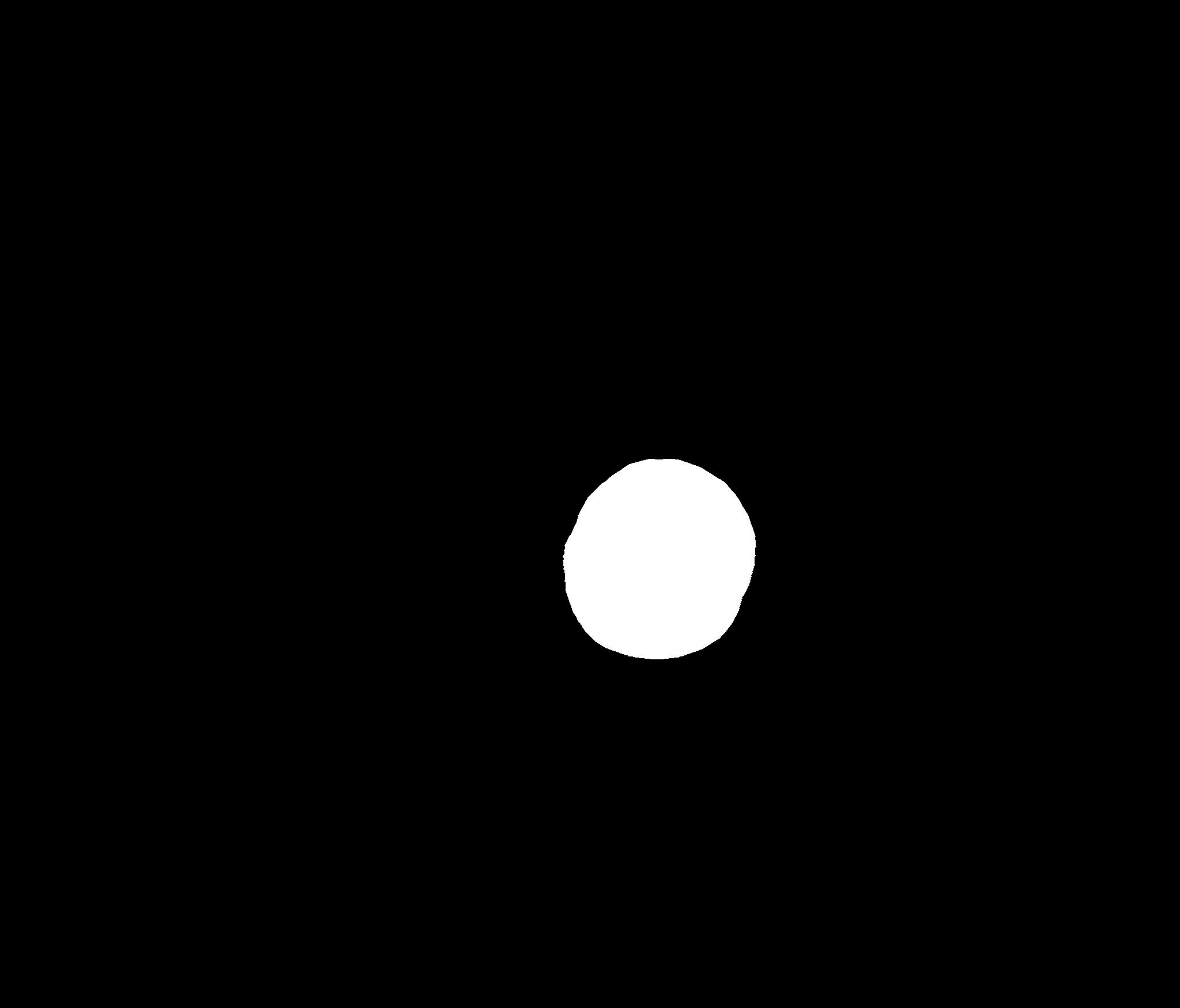} &
		\includegraphics[width=0.19\textwidth]{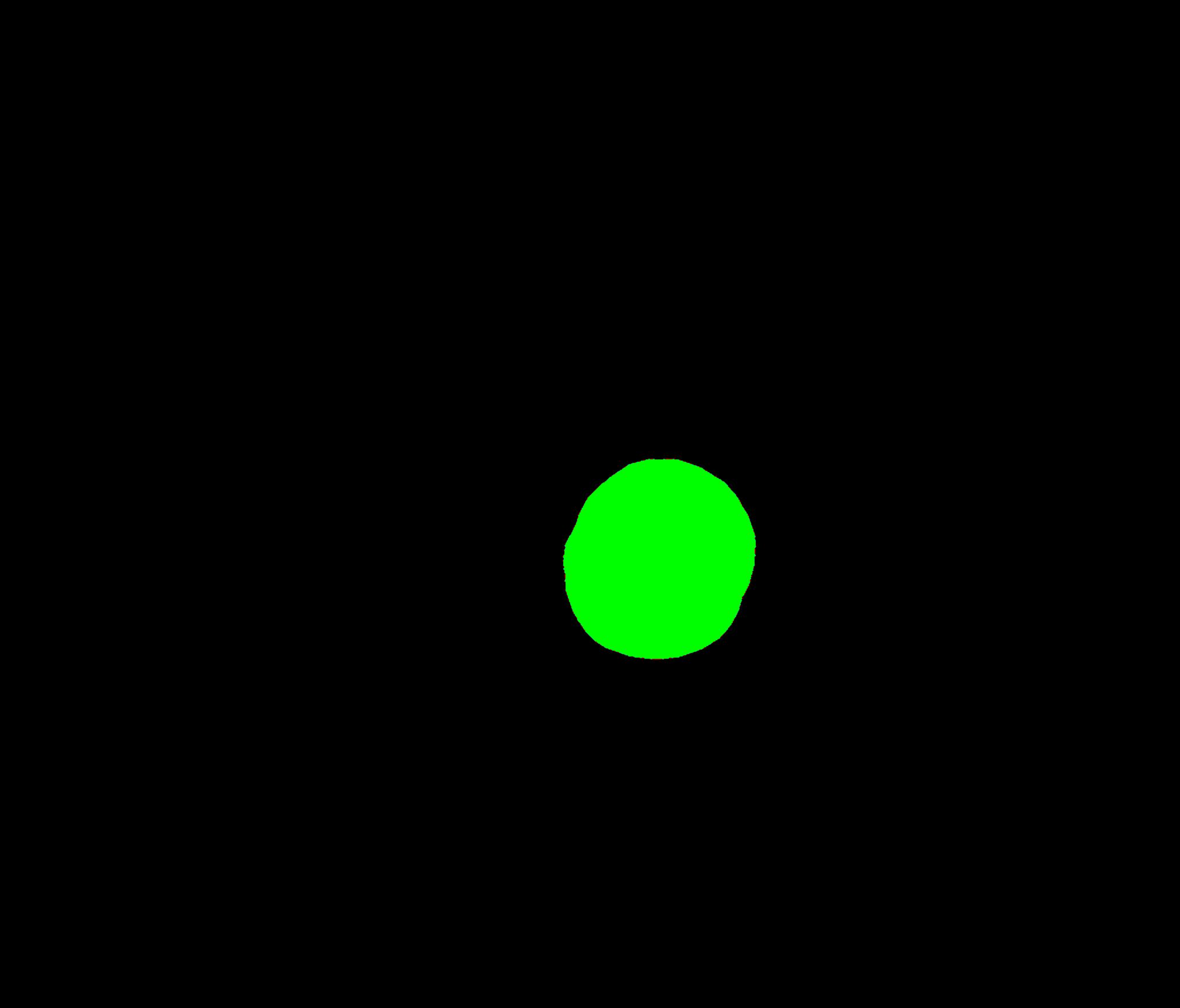} &
		\includegraphics[width=0.19\textwidth]{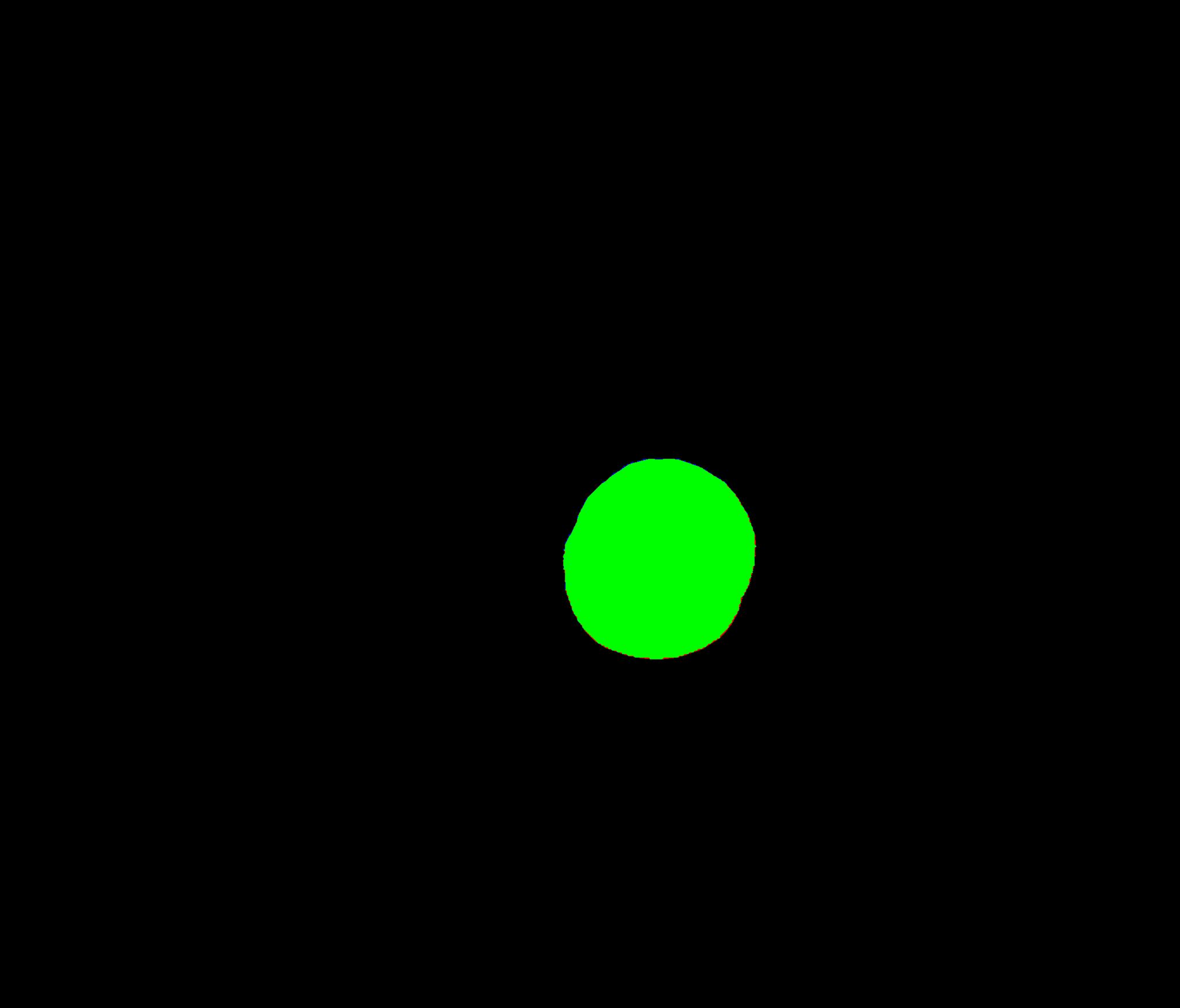} &
		\includegraphics[width=0.19\textwidth]{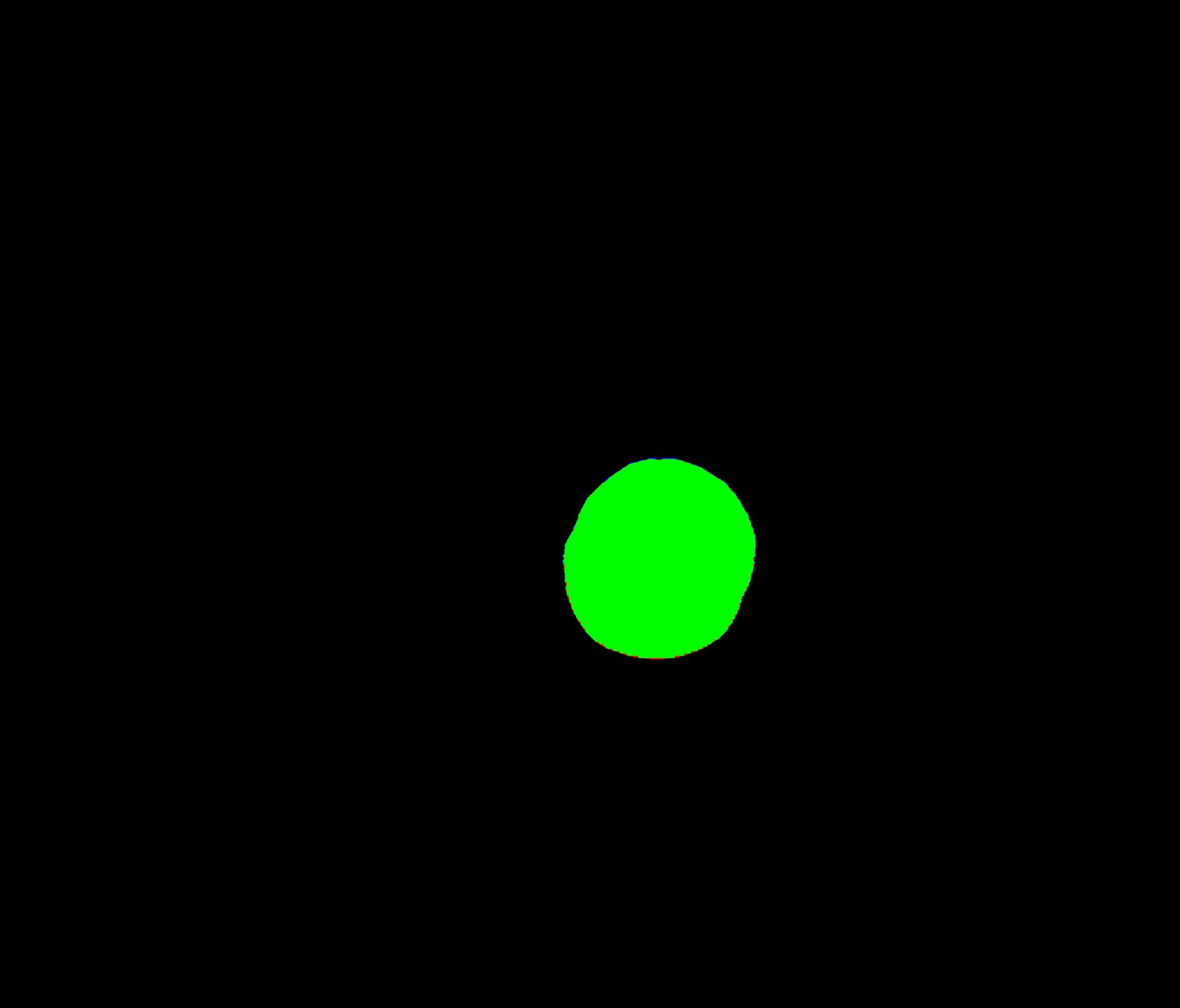} \\
		\includegraphics[width=0.19\textwidth]{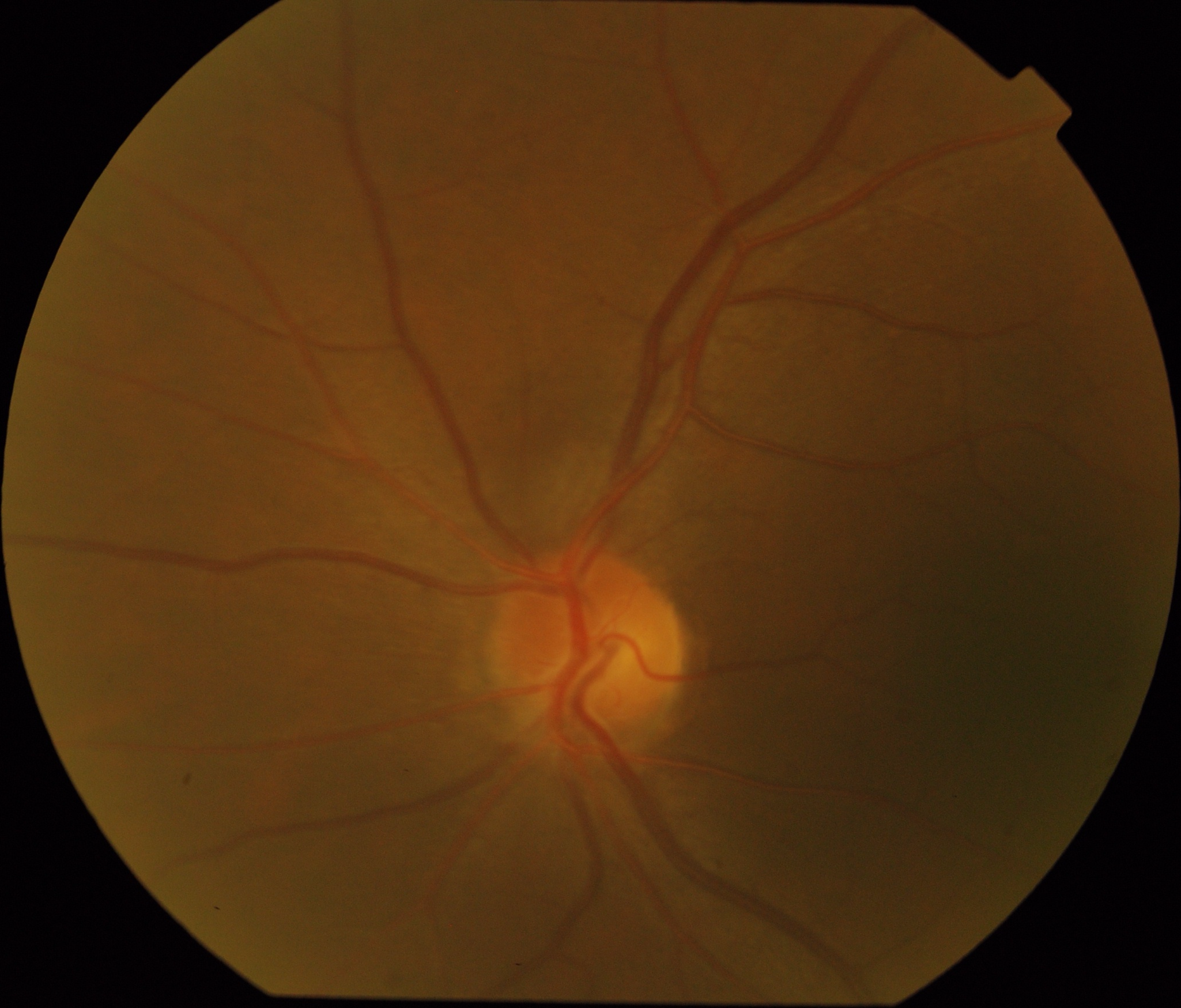} &
		\includegraphics[width=0.19\textwidth]{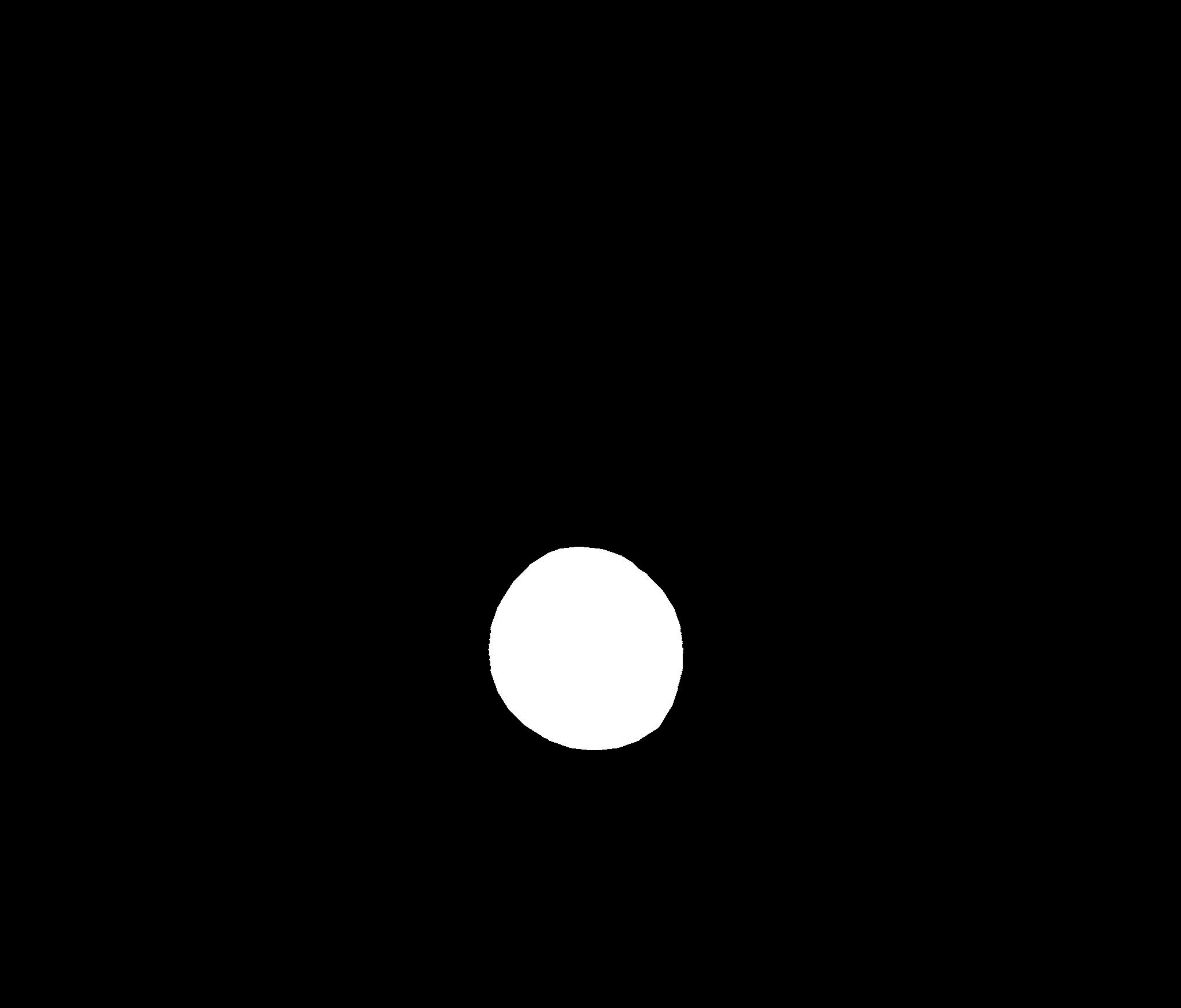} &
		\includegraphics[width=0.19\textwidth]{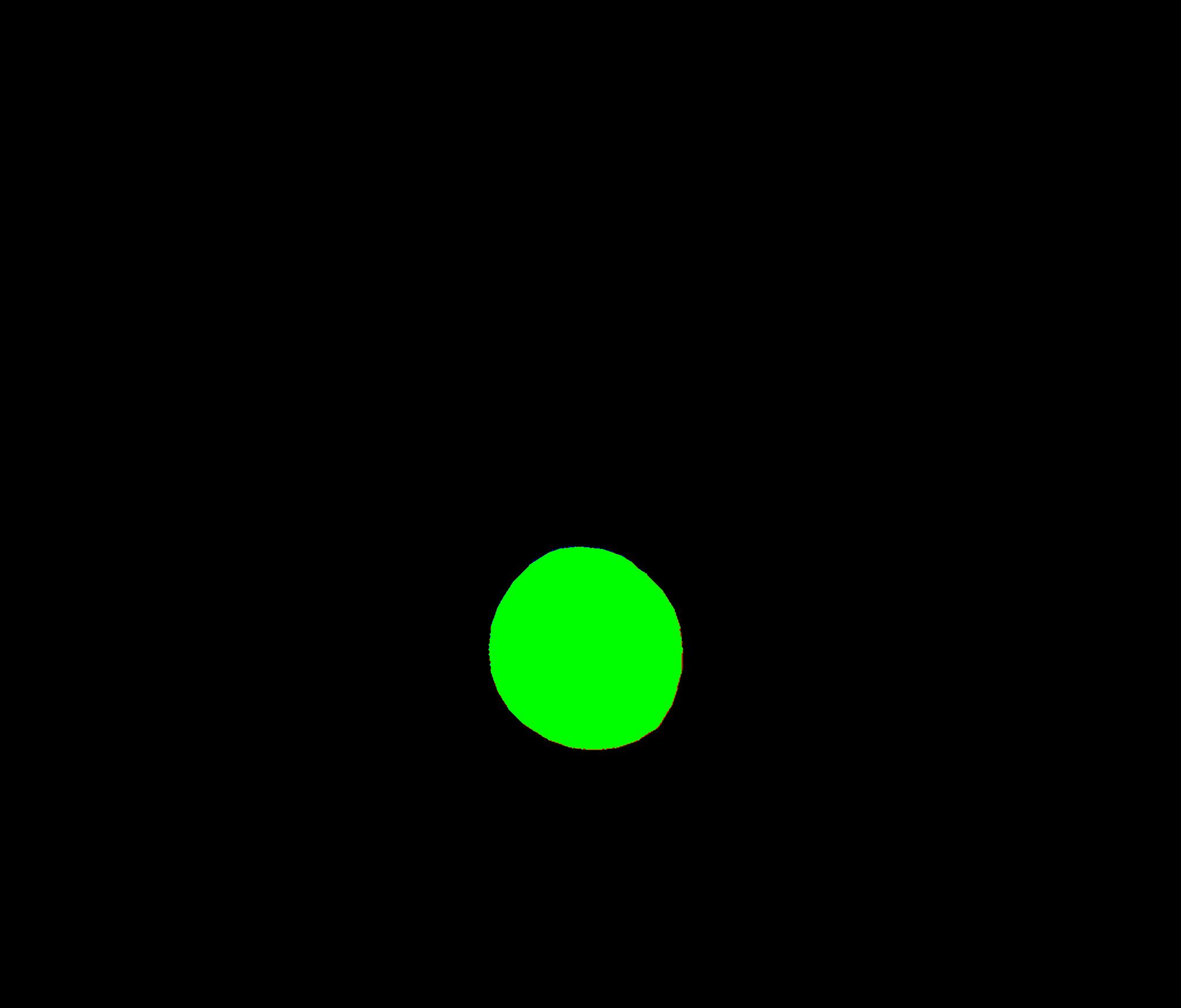} &
		\includegraphics[width=0.19\textwidth]{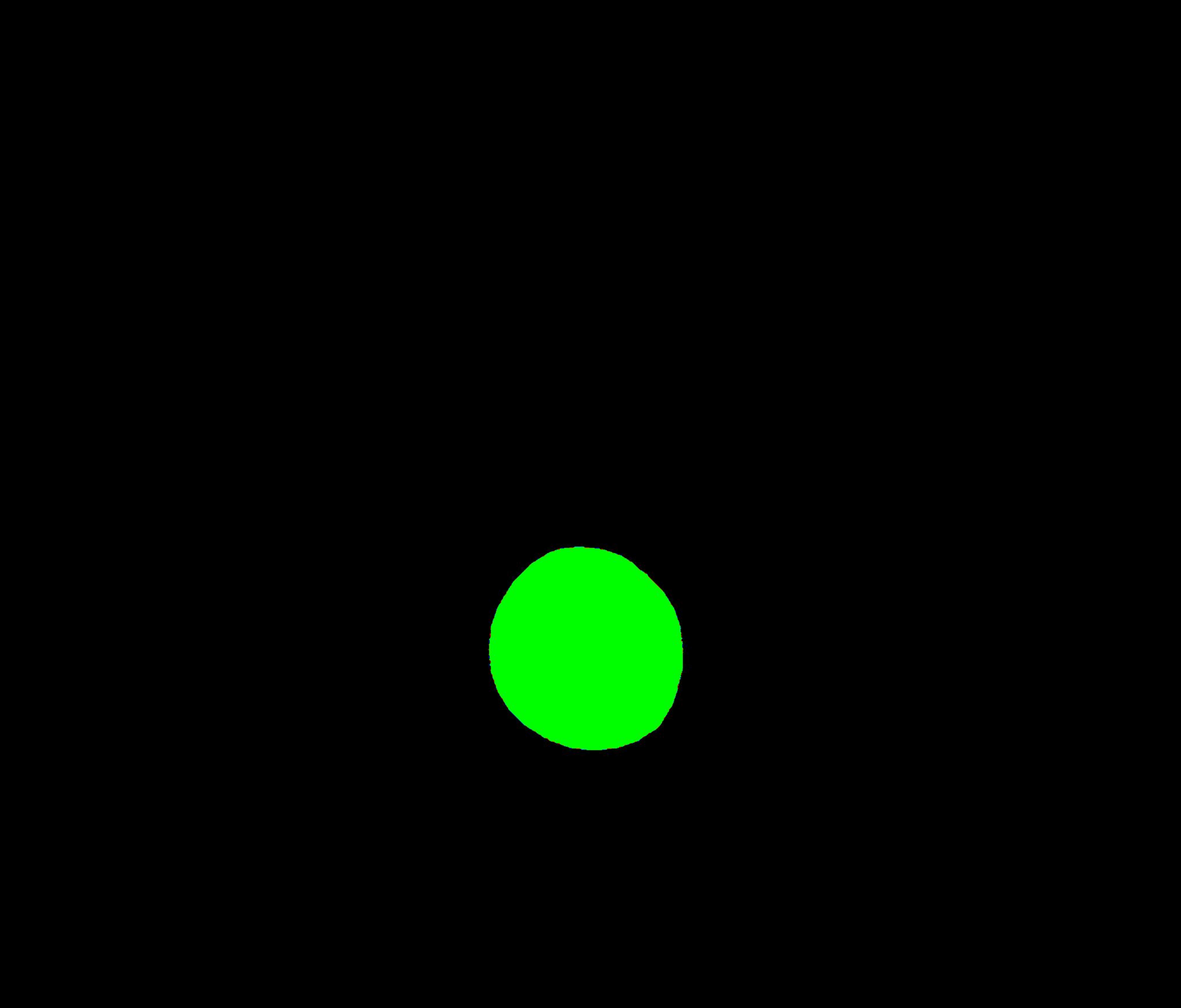} &
		\includegraphics[width=0.19\textwidth]{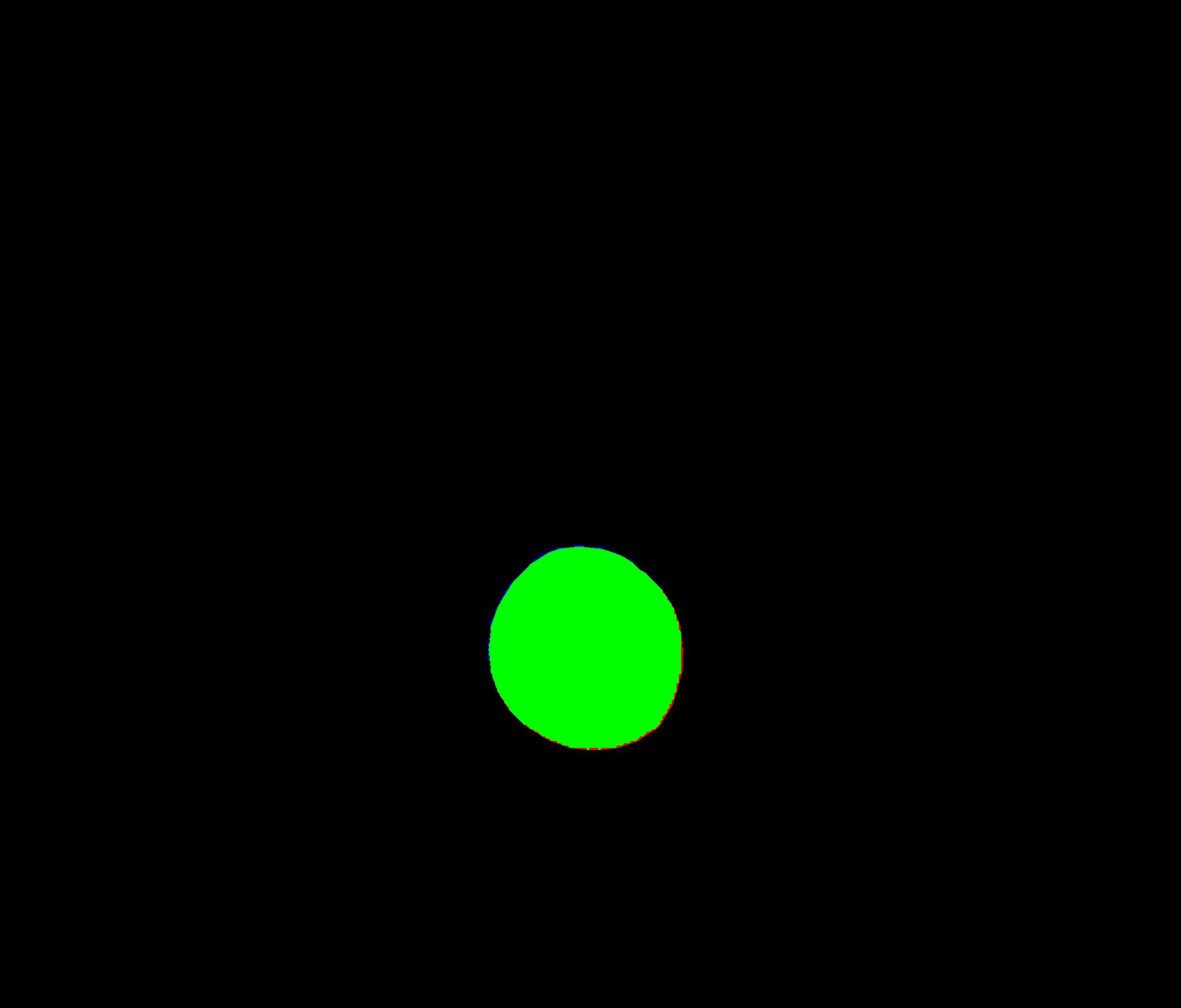} \\
		\includegraphics[width=0.19\textwidth]{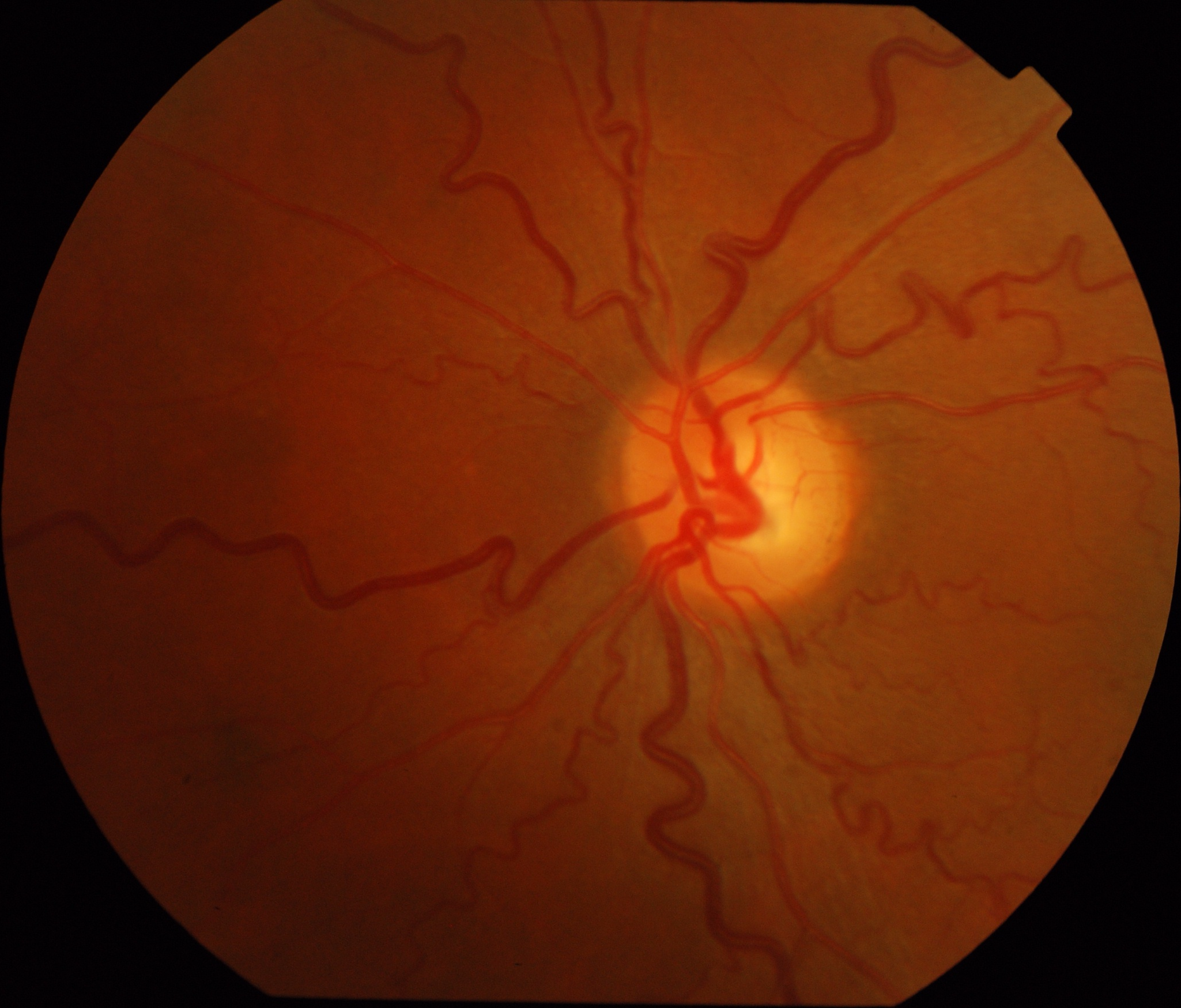} &
		\includegraphics[width=0.19\textwidth]{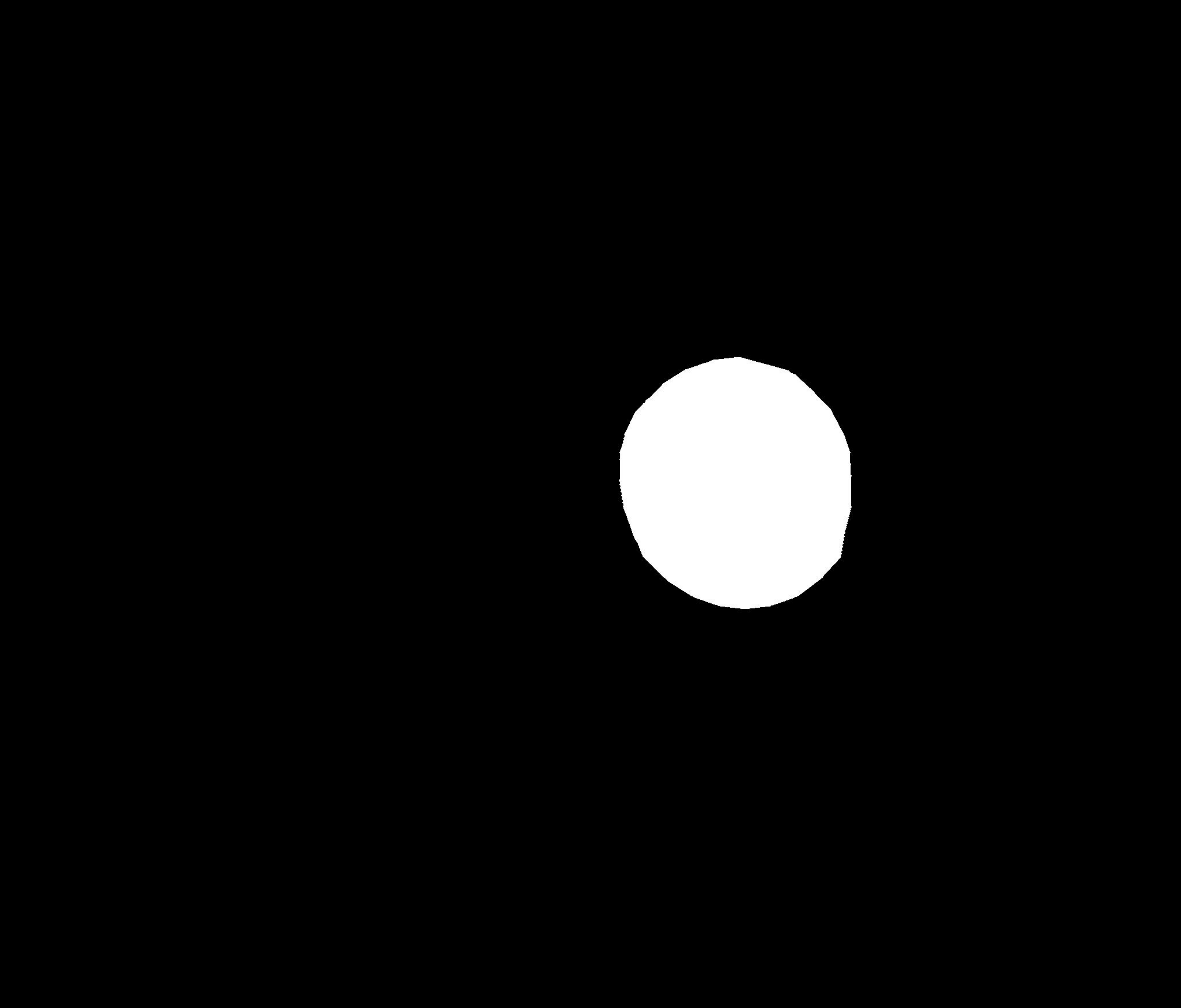} &
		\includegraphics[width=0.19\textwidth]{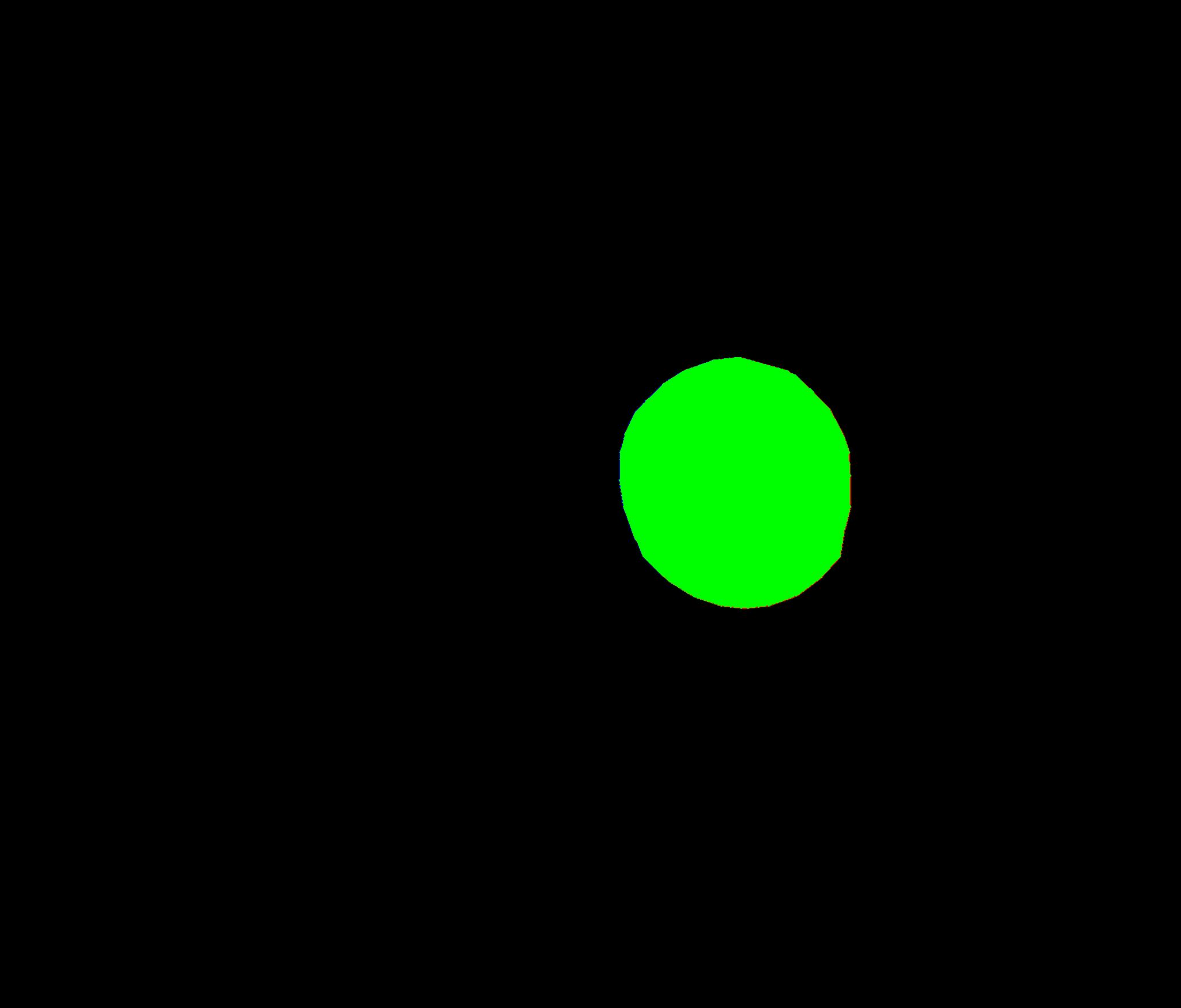} &
		\includegraphics[width=0.19\textwidth]{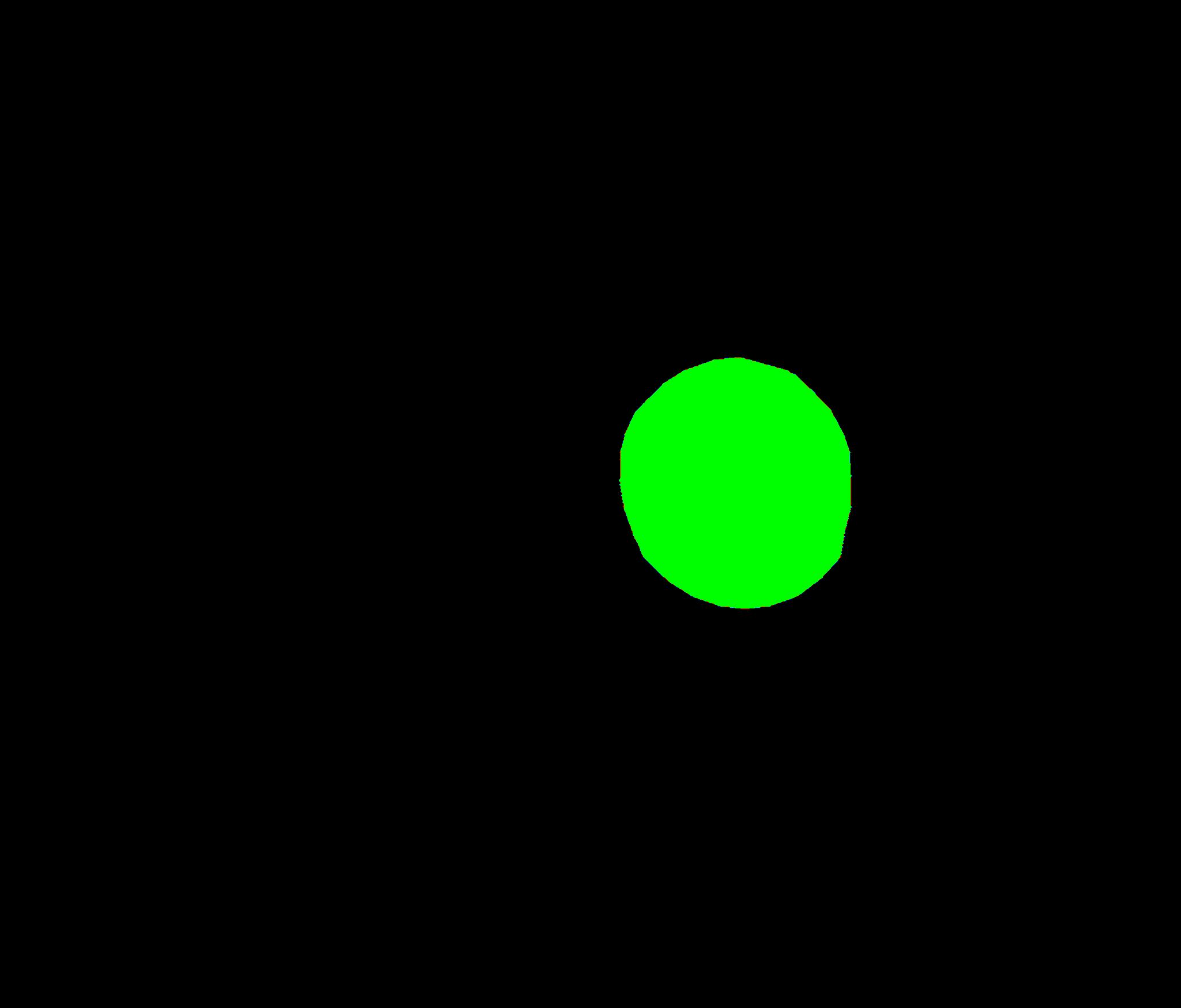} &
		\includegraphics[width=0.19\textwidth]{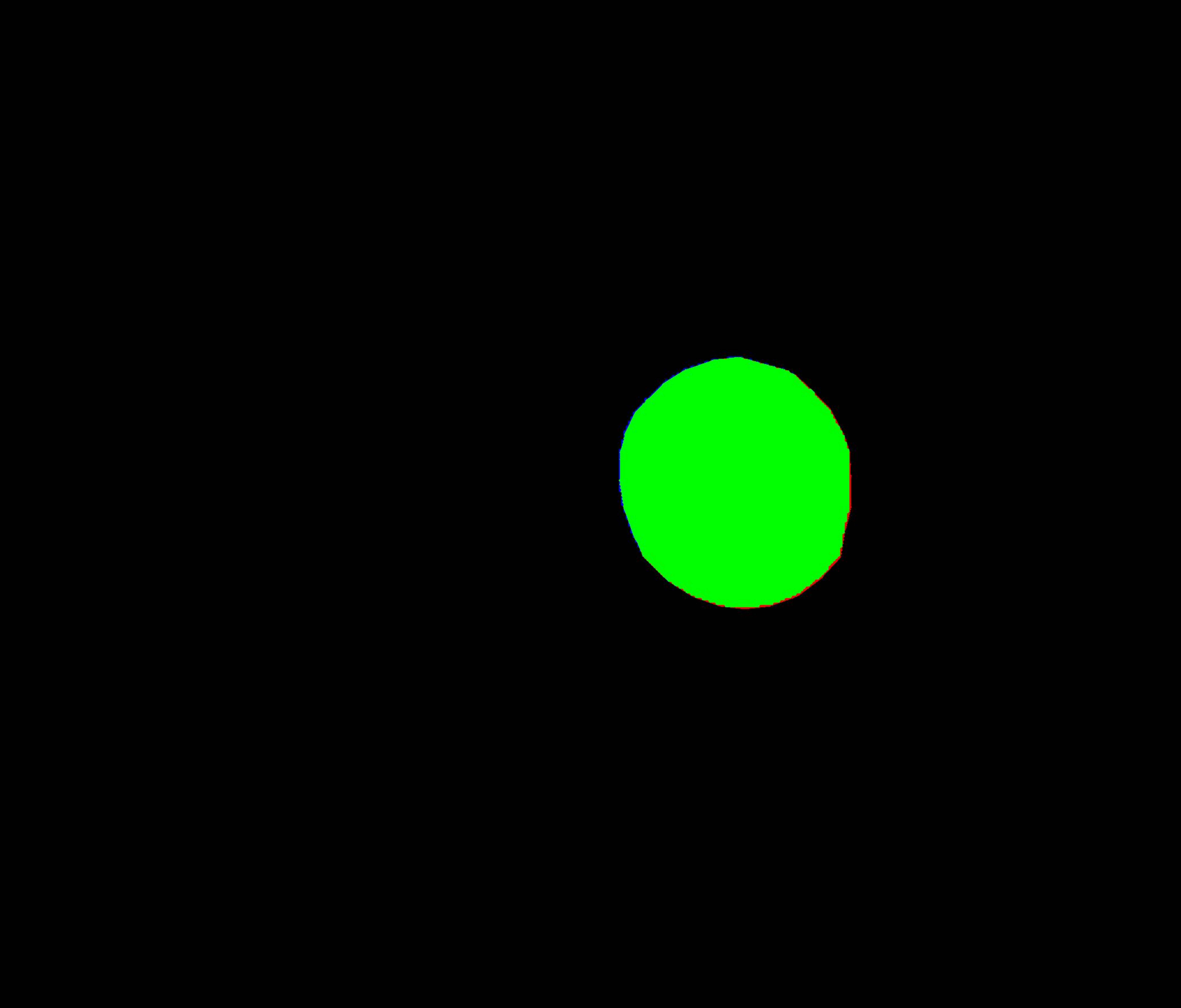} \\
		\includegraphics[width=0.19\textwidth]{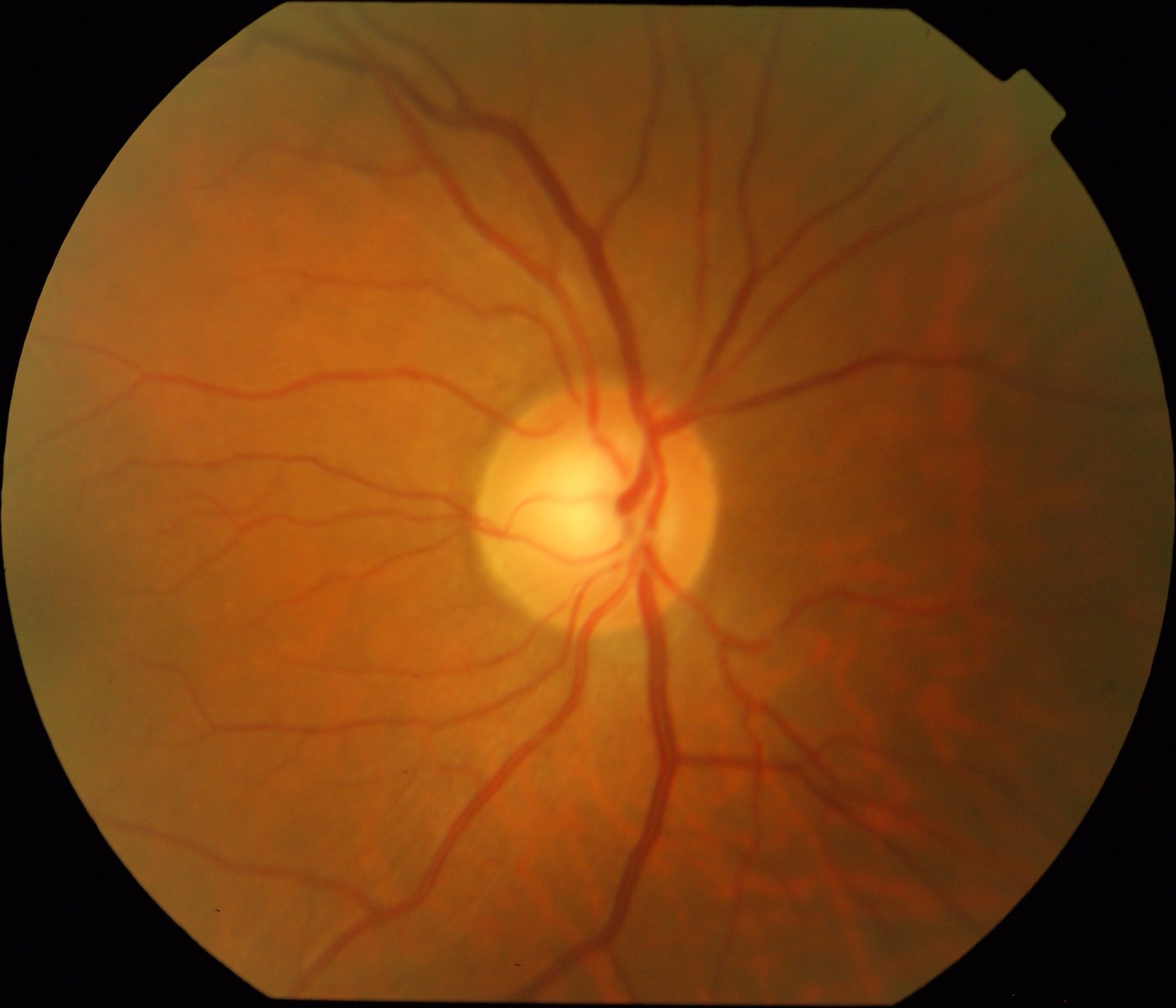} &
		\includegraphics[width=0.19\textwidth]{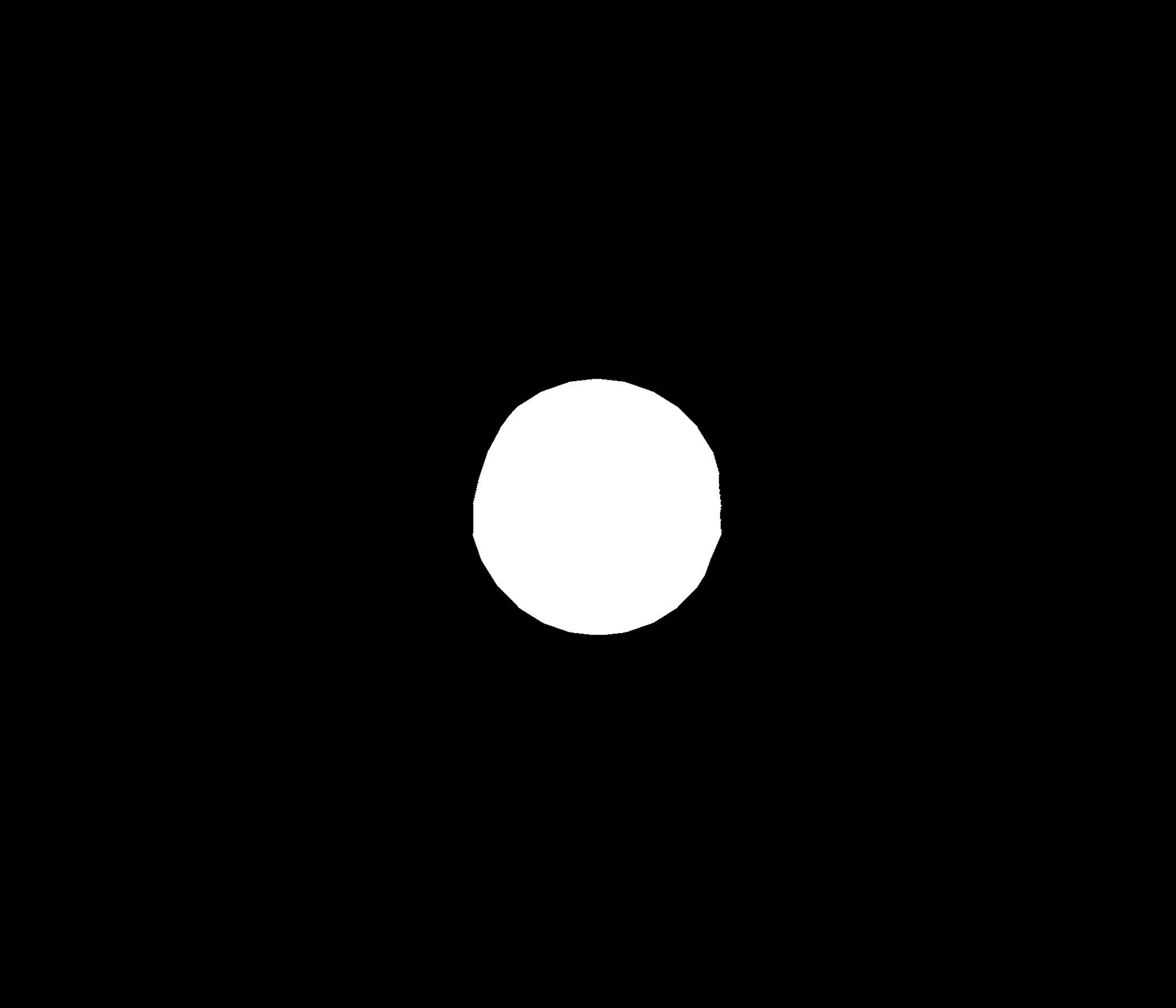} &
		\includegraphics[width=0.19\textwidth]{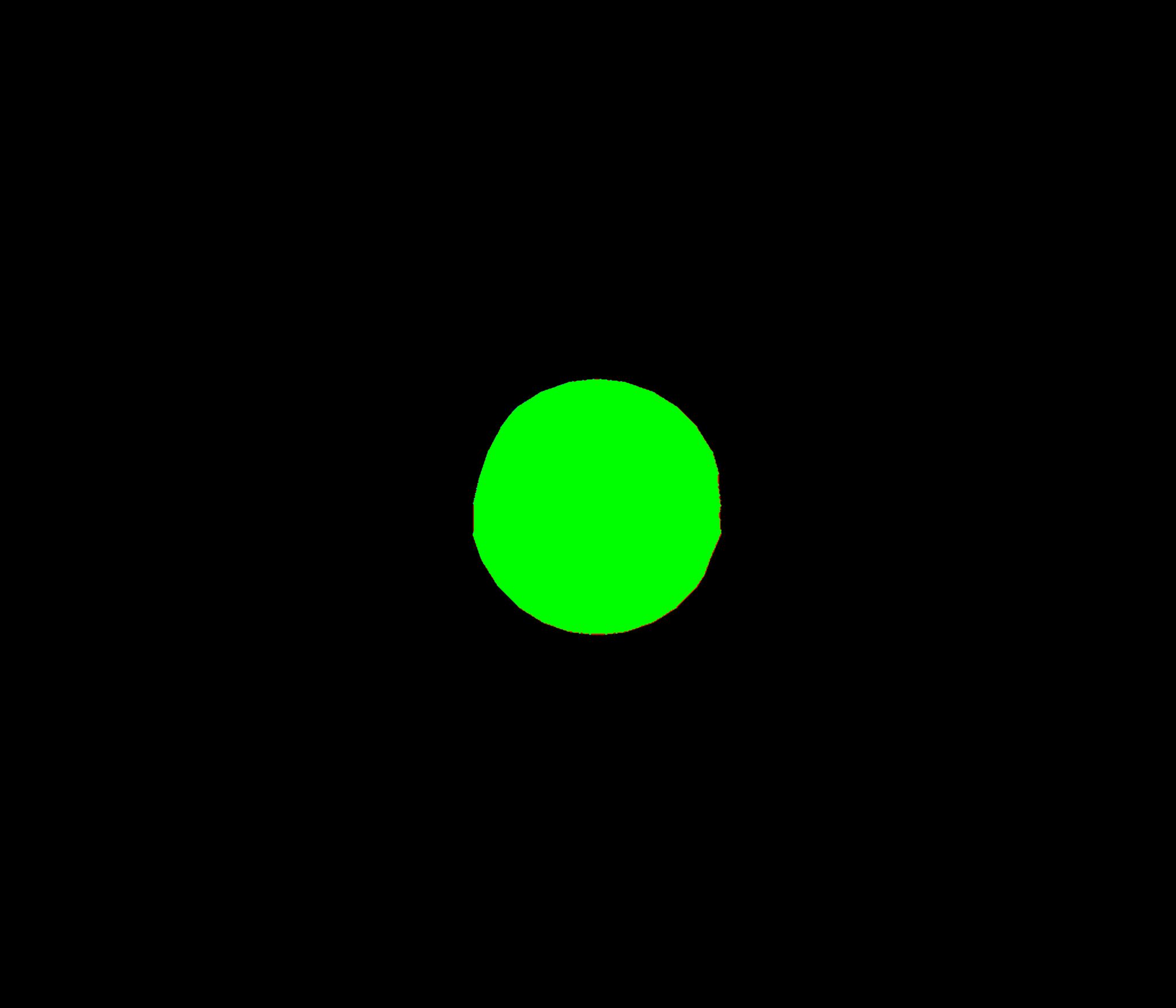} &
		\includegraphics[width=0.19\textwidth]{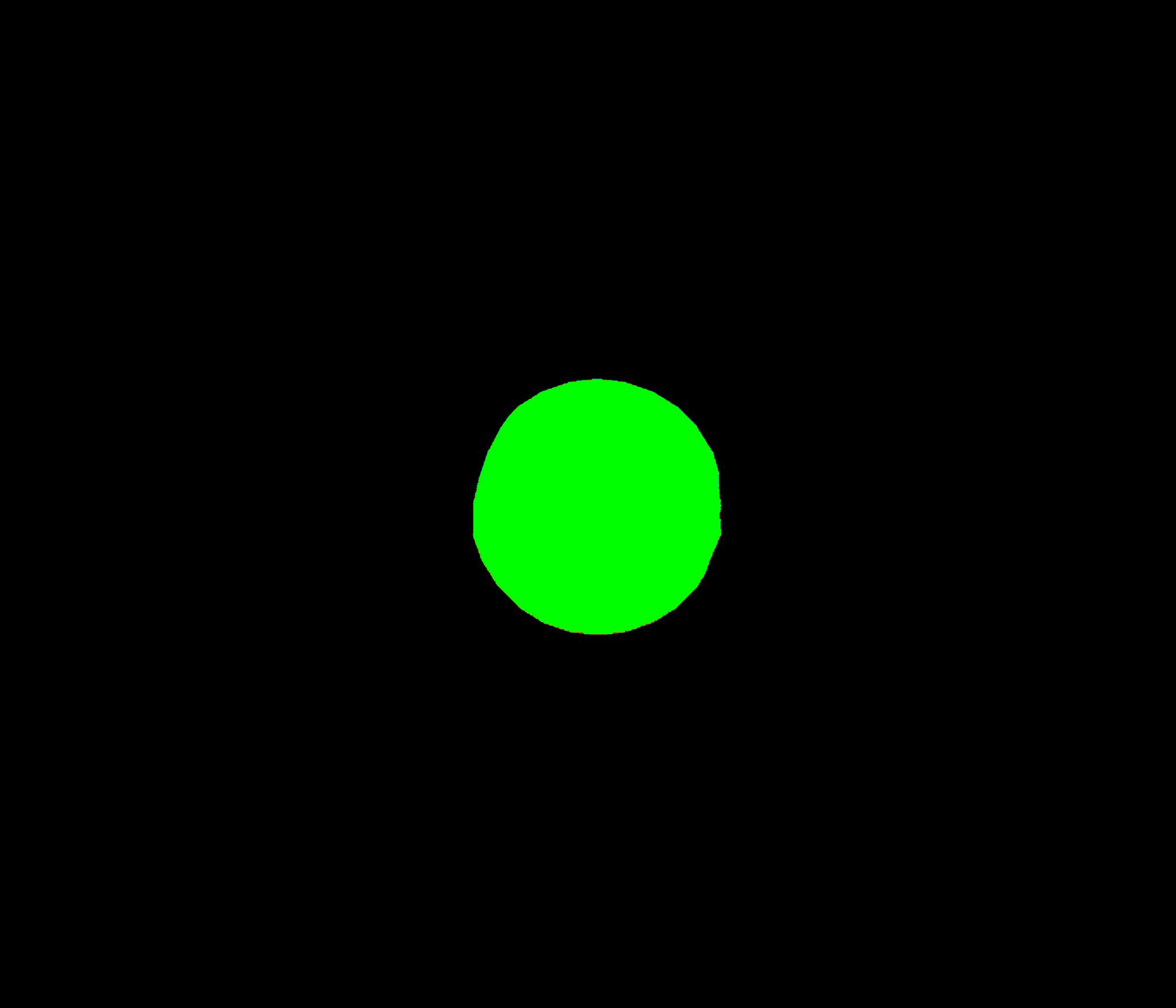} &
		\includegraphics[width=0.19\textwidth]{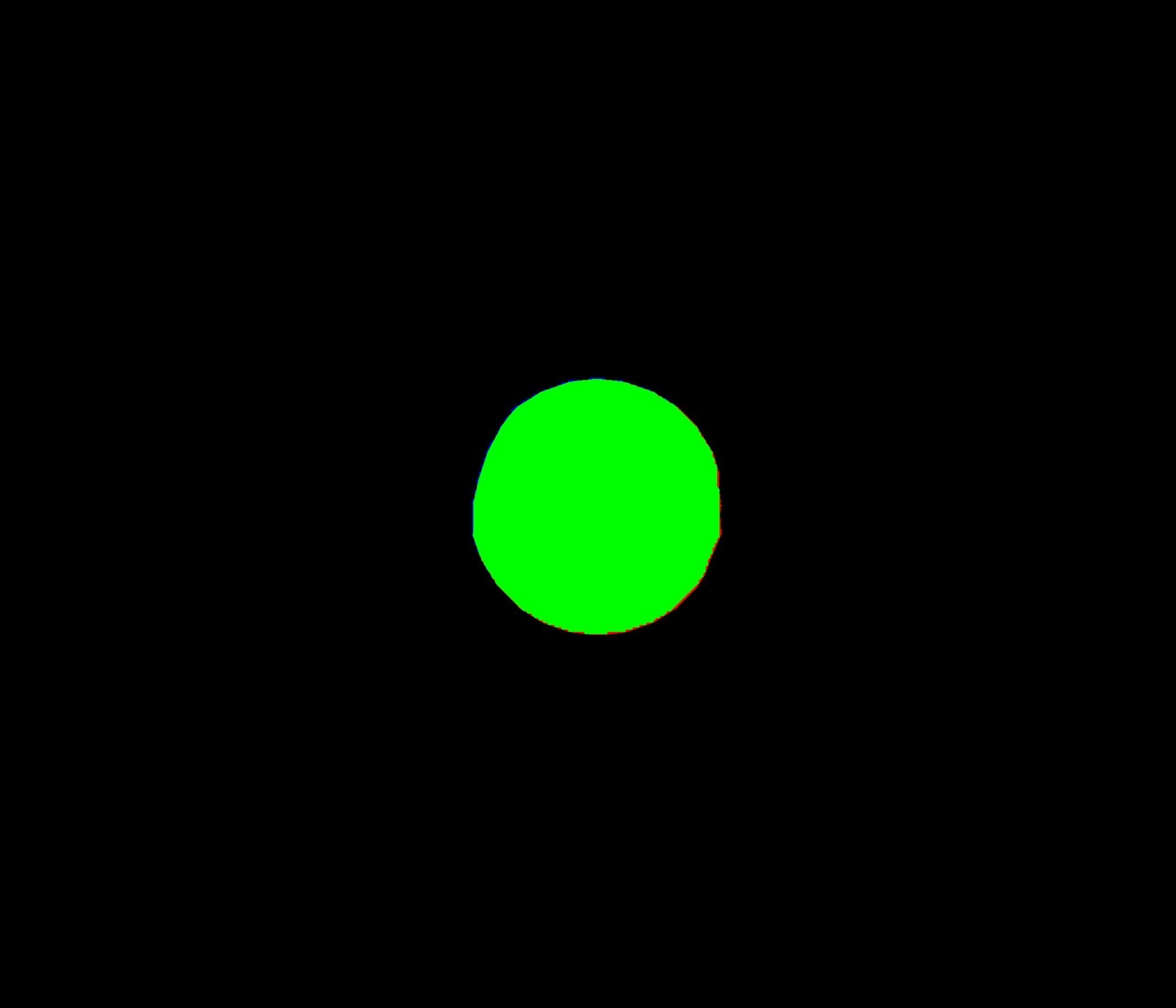} \\
	\end{tabular}
	\caption{Sample segmentation results with LS-Net on the DRISHTI-OC dataset. Four examples are shown from top to bottom. From left to right: the input images, the ground truth manually annotated by an expert, and the results on $2\times$, $3\times$, and $4\times$ downsampled input images. Correctly segmented foreground and background pixels are shown in, respectively, green and black. False positive and false negative pixels are shown in, respectively, red and blue (visible around the object edges only at very high magnification).}
	\label{visualchase}
\end{figure}

To reaffirm the predictive power of the proposed image complexity measures for segmentation performance, we trained the LS-Net (Section~\ref{networks}) with the original images and separately with downsampled images (factors 2, 3, 4) from two relatively high-complexity datasets (DRIVE and CHASE-DB1) and two relatively low-complexity datasets (DRISHTI-OC and DRISHTI-OD). From the quantitative results (Table \ref{tab:reaffirm}) we again observe that segmentation performance consistently decreases with increasing downsampling factor, and the loss is more pronounced for the high-complexity datasets. For example, in this experiment the performance loss was 19\% in Dice/F1 and 28\% in Jaccard, with an increase of 66\% in E, for a downsampling factor of 4 on the DRIVE dataset. Similarly, a decrease of 6\% in Dice/F1 and 10\% in Jaccard, and an increase of 21\% was seen in the CHASE-DB1 dataset for the same downsampling factor. By contrast, as expected, no noteworthy loss in segmentation performance was observed in either of the DRISHTI datasets, due to their low complexity. This is confirmed by visual inspection (Figs.\ \ref{visualdrive} and \ref{visualchase}). We also notice that with increasing downsampling, the number of false negatives increased more than the number of false positives in the DRIVE dataset. This was to be expected, as it is increasingly harder for the LS-Net to capture the tiny vessels, which tend to get lost in the downsampling process. In the DRISHTI dataset, on the other hand, the loss due to downsampling is negligible.

\begin{table}
	\centering
	\caption{Effect of input image downsampling on the segmentation performance of LS-Net compared to no downsampling for selected high- and low-complexity datasets.}
	\resizebox{1\textwidth}{!}{
		\begin{tabular}{@{}lccccccc@{}}
		\toprule
		\textbf{Image Size} & \textbf{Se} & \textbf{Sp} & \textbf{A} & \textbf{BA} & \textbf{D} & \textbf{J} & \textbf{E} \\
		\midrule
		\multicolumn{8}{c}{\textbf{DRIVE} ($\text{MDF}=0.2301$)} \\
		\midrule
		Original         & 0.8259 & 0.9826 & 0.9689 & 0.9043 & 0.8227 & 0.6993 & 0.3007 \\
		Downsampled by 2 & 0.7954 & 0.9779 & 0.9619 & 0.8867 & 0.7852 & 0.6467 & 0.3533 \\
		Downsampled by 3 & 0.7397 & 0.9712 & 0.9563 & 0.8600 & 0.7713 & 0.5789 & 0.4211 \\
		Downsampled by 4 & 0.6843 & 0.9648 & 0.9402 & 0.8245 & 0.6665 & 0.5002 & 0.4998 \\
		\midrule
		\multicolumn{8}{c}{\textbf{CHASE-DB1} ($\text{MDF}=0.1967$)} \\
		\midrule
		Original         & 0.8200 & 0.9845 & 0.9732 & 0.9022 & 0.8073 & 0.6775 & 0.3225 \\
		Downsampled by 2 & 0.8119 & 0.9827 & 0.9711 & 0.8973 & 0.7931 & 0.6593 & 0.3407 \\
		Downsampled by 3 & 0.7877 & 0.9816 & 0.9667 & 0.8850 & 0.7721 & 0.6297 & 0.3703 \\
		Downsampled by 4 & 0.7700 & 0.9804 & 0.9660 & 0.8752 & 0.7556 & 0.6099 & 0.3901 \\
		\midrule
		\multicolumn{8}{c}{\textbf{DRISHTI-OC} ($\text{MDF}=0.0072$)} \\
		\midrule
		Original         & 0.9532 & 0.9970 & 0.9965 & 0.9751 & 0.9031 & 0.8334 & 0.1666 \\
		Downsampled by 2 & 0.9513 & 0.9971 & 0.9965 & 0.9742 & 0.9010 & 0.8321 & 0.1676 \\
		Downsampled by 3 & 0.9508 & 0.9979 & 0.9964 & 0.9939 & 0.9006 & 0.8316 & 0.1684 \\
		Downsampled by 4 & 0.9505 & 0.9970 & 0.9963 & 0.9738 & 0.9001 & 0.8312 & 0.1691 \\
		\midrule
		\multicolumn{8}{c}{\textbf{DRISHTI-OD} ($\text{MDF}=0.0045$)} \\
		\midrule
		Original         & 0.9648 & 0.9985 & 0.9973 & 0.9817 & 0.9610 & 0.9277 & 0.0723 \\
		Downsampled by 2 & 0.9620 & 0.9986 & 0.9973 & 0.9803 & 0.9604 & 0.9264 & 0.0736 \\
		Downsampled by 3 & 0.9614 & 0.9985 & 0.9972 & 0.9800 & 0.9597 & 0.9249 & 0.0751 \\
		Downsampled by 4 & 0.9609 & 0.9985 & 0.9972 & 0.9797 & 0.9590 & 0.9237 & 0.0763 \\
		\bottomrule
		\end{tabular}
	}
	\label{tab:reaffirm}
\end{table}

\subsection{Experiment II: Image Complexity as a Guide for Network Depth}
In this experiment, we investigated the suitability of image complexity as a guideline in choosing a shallow or a deep network for segmentation. The assumption here was that training a deep network on moderate hardware would necessitate downsampling of the input images. To evaluate the impact of this, we used the datasets CHASE-DB1, which has high image complexity, and a combination of ISIC-2016 (training set) and PH2 (test set), which have low complexity. In both cases, we trained our LW-Net on the original images and our LS-Net on the 4-times downsampled images. For testing, to be able to compare with the original ground truth, the output of LS-Net was upsampled the same way as in Experiment I. The results of Experiment II (Table \ref{BDCOMPVSNETCOMP}) show that when the image complexity is high, the shallow LW-Net outperforms the deep LS-Net, despite having 10 times fewer layers. Conversely, when the image complexity is low, the deep network outperforms the shallow network even when trained on 4-times downsampled images and tested on the original ground truth.

\begin{table}
	\centering
	\caption{Performance of our shallow LW-Net and deep LS-Net on the high-complexity CHASE-DB1 dataset and the low-complexity ISIC-2016 and PH2 datasets.}
	\resizebox{\textwidth}{!}{
	\begin{tabular}{@{}lccccccc@{}}
		\toprule
		\textbf{Network} & \textbf{Se} & \textbf{Sp} & \textbf{A} & \textbf{BA} & \textbf{D} & \textbf{J} & \textbf{E} \\
		\midrule
		\multicolumn{8}{c}{\textbf{CHASE-DB1} ($\text{MDF}=0.1967$)} \\
		\midrule
		LW-Net & 0.8200 & 0.9845 & 0.9732 & 0.9022 & 0.8073 & 0.6775 & 0.3225 \\
		LS-Net & 0.7948 & 0.9786 & 0.9661 & 0.8867 & 0.7619 & 0.6169 & 0.3831 \\
		\midrule
		\multicolumn{8}{c}{\textbf{Trained on ISIC-2016 / Tested on PH2} ($\text{MDF}=0.0017 / 0.0049$)} \\
		\midrule
		LW-Net & 0.9115 & 0.8793 & 0.9290 & 0.8954 & 0.8789 & 0.7948 & 0.2052 \\
		LS-Net & 0.9601 & 0.9251 & 0.9673 & 0.9426 & 0.9503 & 0.9070 & 0.0930 \\
		\bottomrule
	\end{tabular}
	}
	\label{BDCOMPVSNETCOMP}
\end{table}

\section{Conclusion}
We have presented a framework to guide developers in making several critical macro-level neural network design choices for medical image segmentation based on image complexity measures. The proposed framework is independent of the segmentation task at hand and the image modalities used. This is possible because the design choices are based solely upon the information contained in the dataset. Extensive experiments on 10 different medical image segmentation benchmarks demonstrated the suitability of our framework. We conclude that the proposed image complexity measures help address the following critical issues in designing a neural network for medical image segmentation: 1) design and train neural networks for high-resolution medical images using generally available moderate compute resources, 2) minimizing the effects of downsampling the input images (usually to aid training) on segmentation performance, and 3) deciding on the depth of the architecture (number of layers/filters) for a given medical image segmentation task. We advocate that our framework complements NAS approaches and can be employed at the macro-level stage in conjunction with NAS for micro-level architectural optimization.

{\small
\bibliographystyle{IEEEtran}
\bibliography{egbib}
}

\end{document}